\documentclass[reprint, aps, pre, notitlepage, nofootinbib]{revtex4-2}

\usepackage{amsmath, amsfonts, amssymb, amsthm, mathtools}
\usepackage[utf8]{inputenc}
\usepackage[T1]{fontenc}
\usepackage{hyperref}
\usepackage{graphicx}
\usepackage[caption=false]{subfig}
\usepackage{lipsum}
\usepackage{xcolor}
\usepackage{empheq}

\theoremstyle{remark}

\theoremstyle{remark}

\theoremstyle{theorem}

\theoremstyle{theorem}

\newcommand{\LT}[1]{\widetilde{#1}}

\newcommand{\FPT}{\mathcal{T}}

\begin{document}
	
\title{First-passage functionals of Brownian motion in logarithmic potentials and heterogeneous diffusion}
\author{Mattia Radice}
\email[Corresponding author: ]{mradice@pks.mpg.de}
\affiliation{Max Planck Institute for the Physics of Complex Systems, N\"{o}thnitzer Str. 38, 01187 Dresden, Germany}
\begin{abstract}
We study the statistics of random functionals $\mathcal{Z}=\int_{0}^{\mathcal{T}}[x(t)]^{\gamma-2}dt$, where $x(t)$ is the trajectory of a one-dimensional Brownian motion with diffusion constant $D$ under the effect of a logarithmic potential $V(x)=V_0\ln(x)$. The trajectory starts from a point $x_0$ inside an interval entirely contained in the positive real axis, and the motion is evolved up to the first-exit time $\mathcal{T}$ from the interval. We compute explicitly the PDF of $\mathcal{Z}$ for $\gamma=0$, and its Laplace transform for $\gamma\neq0$, which can be inverted for particular combinations of $\gamma$ and $V_0$. Then we consider the dynamics in $(0,\infty)$ up to the first-passage time to the origin, and obtain the exact distribution for $\gamma>0$ and $V_0>-D$. By using a mapping between Brownian motion in logarithmic potentials and heterogeneous diffusion, we extend this result to functionals measured over trajectories generated by $\dot{x}(t)=\sqrt{2D}[x(t)]^{\theta}\eta(t)$, where $\theta<1$ and $\eta(t)$ is a Gaussian white noise. We also emphasize how the different interpretations that can be given to the Langevin equation affect the results. Our findings are illustrated by numerical simulations, with good agreement between data and theory.
\end{abstract}

\maketitle

\section{Introduction}
\begin{figure}
	\centering
	\includegraphics[width=\columnwidth]{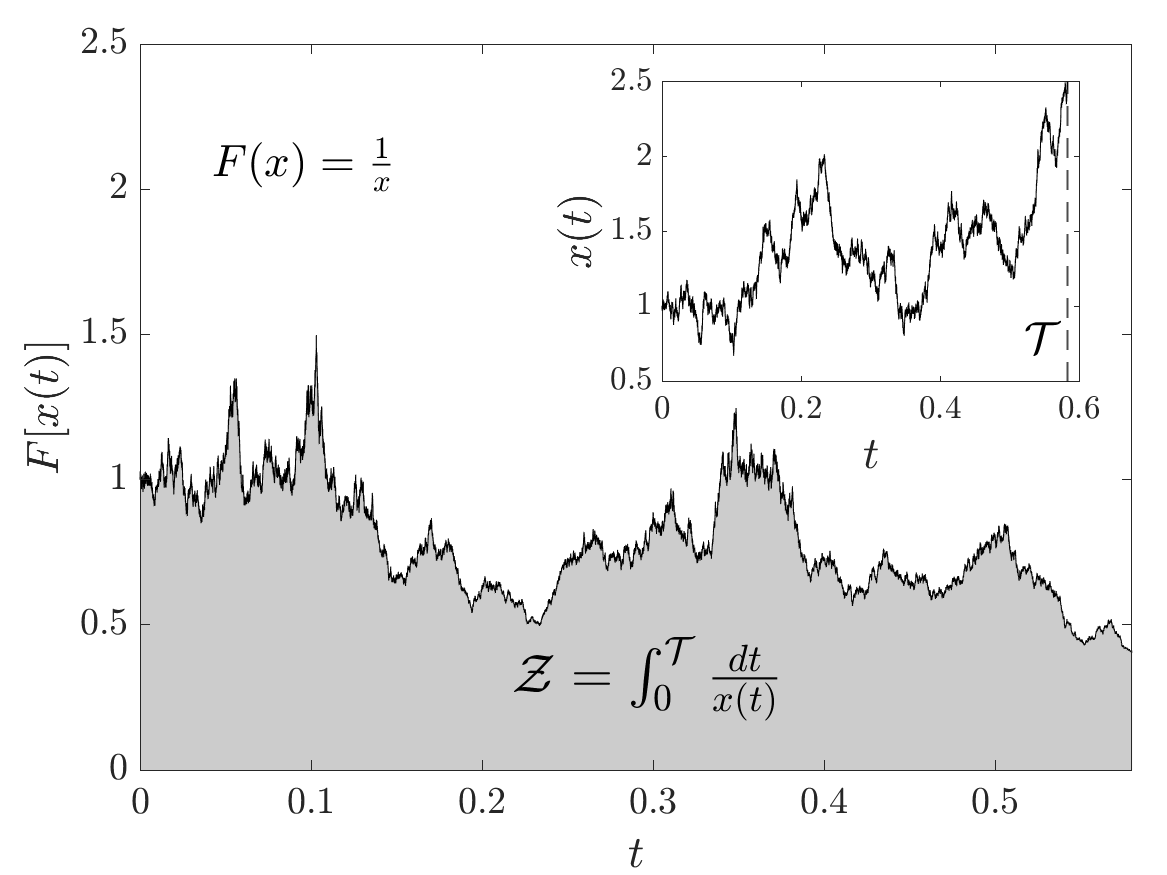}
	\caption{Example of first-passage functional $\mathcal{Z}=\int_{0}^{\mathcal{T}}F[x(t)]dt$ for Brownian motion in a potential $V(x)=V_0\ln(x)$ diffusing in $\Omega=(\frac12,\frac52)$. Here $F(x)=1/x$ and $\mathcal{Z}$ thus corresponds to the area under the graph of $1/x(t)$, where $x(t)$ is the stochastic trajectory displayed in the inset. The trajectory starts from $x_0=1$ and the motion is evolved up to the first-exit time from $\Omega$. The diffusion constant is equal to one and $V_0=0.5$.}
\end{figure}
Consider the stochastic trajectory of a one-dimensional particle described by the Langevin equation
\begin{equation}\label{eq:SDE}
	\frac{d x(t)}{d t}=\mu[x(t)]+\sqrt{2D[x(t)]}\eta(t),
\end{equation}
where $\mu(x)=- V'(x)$ represents a deterministic force derived from an external time-independent potential $V(x)$, $\eta(t)$ is a Gaussian white noise with zero mean and autocorrelation $\langle\eta(t)\eta(t')\rangle=\delta(t-t')$ and $D(x)$ is the space-dependent diffusion coefficient. Suppose that the motion generated by \eqref{eq:SDE} starts from a point $x_0$ inside a given interval $\Omega$, and the first passage outside $\Omega$ occurs after a random time $\mathcal{T}$, which we call the \emph{first-passage time}. Define
\begin{equation}\label{eq:func_0}
	\mathcal{Z}=\int_{0}^{\FPT}F[x(t)]d t,
\end{equation}
where $ F(x)$ is, in principle, an arbitrary function that makes the integral convergent. Such a random variable is known as \emph{first-passage functional}. Quantities of this kind have been extensively studied in the case of free Brownian motion, i.e., for $\mu(x)=0$ and $D(x)=D$, for the simple reason that many problems may be formulated in terms of first-passage Brownian functionals \cite{Maj-2005}. Of course, generalizations of the problem have also been proposed, in which, for example, dynamics other than purely Brownian or the introduction of stochastic resetting mechanisms are considered \cite{Abu-2013,KeaMaj-2005,KeaMajMar-2007,KeaMar-2016,AbuDel-2017,AbuFur-2019,KeaMar-2021,SinPal-2022,MeeOsh-2022,Mee-2023,Abu-2023,PalPalPar-2023arX,DubPal-2023}.

In this paper we wish to consider a subclass of first-passage functionals, where the integral in \eqref{eq:func_0} is evaluated for $F(x)=x^{\gamma-2}$, with $\gamma\in\mathbb{R}$, over the trajectory of a Brownian particle with constant diffusion coefficient $D$ in a logarithmic potential $V(x)=V_0\ln (x/\kappa)$, where $\kappa$ is a length scale that we can conveniently set to one. This choice is motivated by the fact that many interesting problems can be mapped to the study of functionals of this kind. For example, for $\gamma=2$ one has $F(x)=1$ and thus $\mathcal{Z}$ simply corresponds to the first-passage time $\mathcal{T}$, which is a stochastic quantity relevant for a plethora of applications \cite{Red,Metzler2014}. For $\gamma=3$, $\mathcal{Z}$ is equivalent to the first-passage area $\mathcal{A}$, i.e., the area swept by the trajectory $x(t)$ in the $xt$ plane till the first-passage time. This quantity has attracted a lot of interest and was studied for instance in the case of Brownian motion \cite{KeaMaj-2005,KeaMar-2016,MajMee-2020}, Brownian motion with drift \cite{KeaMaj-2005,KeaMajMar-2007,AbuDel-2017}, Brownian motion with stochastic resetting and jump-diffusion processes \cite{Abu-2013,SinPal-2022, Abu-2023arXiv}, Orstein-Uhlenbeck process with and without resetting \cite{KeaMar-2021,Abu-2023,DubPal-2023}, Lévy processes \cite{AbuFur-2019}, with applications in queueing theory and combinatorics \cite{KeaMaj-2005}, percolation \cite{Kea-2004}, animal movements \cite{Rey-2010}, snow melt \cite{DubBan-2018} and DNA breathing dynamics \cite{BanGupSeg-2011}, to cite a few examples. Other nontrivial and interesting cases are $\gamma=\case32$, which is related to the oscillation period in the underdamped one-dimensional Sinai model \cite{DeaMaj-2001}, and $\gamma=\case12$, which is associated with the lifetime of a comet in the solar system \cite{Maj-2005,Hammersley1961}. Remarkably, in the case of free Brownian motion diffusing in $\Omega=(0,\infty)$, it is possible to obtain the distribution of $\mathcal{Z}$ for any $\gamma>0$ \cite{MajMee-2020}. It is natural to try to extend this result to more general situations, for example by adding the presence of an external driving force. The specific case of a logarithmic potential is interesting for several reasons: first, it has been extensively studied in the literature \cite{KesBar-2010,DecLutBar-2011,HirMukSch-2011,MarBehGer-2011,DecLutBar-2012,RayReu-2020} and recognized as a model naturally appearing in different contexts, such as stochastic thermodynamics \cite{RyaDieChv-2013,HolDieEin-2015,ParMor-2021,ParMor-2022}, vortex dynamics \cite{Bray-2000,Cha-2007}, long-range interacting systems \cite{BouDau-2005,Cha-2005,CamDauRuf-2009}, ion condensation on a long polyelectrolyte \cite{Man-1969}, sleep-wake transitions \cite{LoNunHav-2002}, DNA denaturation \cite{BanGupSeg-2011} and diffusion of cold atoms in optical lattices \cite{Castin1991,MarEllZol-1996,Lut-2004,DouGerRen-2006,KesBar-2012,VezBarBur-2019}; in particular, in the latter two cases the first-passage area \cite{BanGupSeg-2011} and the area under an excursion \cite{BarAghKes-2014,KesMedBar-2014}, namely a trajectory that begins and ends at the origin without crossing it at intermediate times, are of particular interest. Second, there exists a discrete counterpart, known as the Gillis random walk \cite{Gill-bias,OPRA}, which can be solved exactly and has been considered in some recent work \cite{PROA,AOPR,RAD-2022-Gill,ZodAllEva-2023}. This model is a critical case for the study of recurrence in stochastic processes \cite{Lam-1960,Lam-1963,OPRA,Hug-I}, with unique first-passage properties that are also recovered in the continuous system. Third, it has been shown that certain models of heterogeneous diffusion can be mapped to the dynamics of Brownian motion (with constant diffusion coefficient) in a logarithmic potential \cite{Lei-Bar,Eli-2021EPL,Eli-2021}. Hence, obtaining the distribution of $\mathcal{Z}$ in the latter case allows us to derive also the solution of the problem in the case of a spatially-varying diffusion coefficient. We remark that heterogeneous diffusion has attracted a lot of interest in the statistical physics community \cite{CheCheMet-2013,CheMet-2014,CheCheMet-2014,BreLaw-2017,Lei-Bar,WanCheKan-2021,Sin-2022,SanDomKoc-2022} and not only, as situations where $D$ is nonconstant are ubiquitous: examples include contexts related to biology \cite{KuhIhaHyy-2011,PieHeyBra-2016,BerMak-2017,DosDorCol-2020}, finance \cite{Oks}, solute transport in heterogeneous media \cite{DenCorSch2004} and Richardson diffusion in turbulence \cite{Ric-1926}.

The outline of the paper is the following: in the next section we use the method of \cite{Maj-2005} to write a backward evolution equation for the Laplace-transformed probability density function of $\mathcal{Z}$, when evaluated along a trajectory generated by {\color{red}Eq.} \eqref{eq:SDE}. Then in section \ref{s:Results} we summarize the main results for the particular case of a logarithmic potential and a constant diffusion coefficient. As a corollary, we also obtain the distribution of $\mathcal{Z}$ when $x(t)$ is generated by $\dot{x}(t)=\sqrt{2D}x^{\theta}\eta(t)$, with $\theta<1$, which is a model for heterogeneous diffusion. In sections \ref{s:finite_interval}, \ref{s:InfInt} and \ref{s:semi-axis} we derive the results by providing detailed calculations. Finally, in section \ref{s:Concl} we draw our conclusions.

\section{Backward equation for the probability density function}\label{s:FP}
Let us call $p(z,x_0)$ the probability density function (PDF) of $\mathcal{Z}$, knowing that the trajectory started from $x_0\in\Omega$. The idea is to derive a backward evolution equation for the Laplace transform of the PDF, which corresponds to the expected value of $e^{-w\mathcal{Z}}$, where $w$ is the Laplace variable:
\begin{equation}\label{eq:LT}
	\LT{p}(w,x_0)=\int_{0}^{\infty}e^{-wz}p(z,x_0)d z=\mathbb{E}\left(e^{-w\mathcal{Z}}\right).
\end{equation}
Here the expected value is taken over all realizations that start from $ x_0 $ and leave $\Omega$ for the first time at $ \mathcal{T} $. To do this, one can rewrite Eq. \eqref{eq:func_0} as \cite{Maj-2005}
\begin{align}
	\mathcal{Z}&=\int_{0}^{d t}F[x(t')]d t'+\int_{d t}^{\mathcal{T}}F[x(t')]d t'\\
	&=F(x_0)d t+\int_{d t}^{\mathcal{T}}F[x(t')]d t'+o(d t),
\end{align}
and note that the second integral at the right-hand side (rhs) corresponds to the definition of $\mathcal{Z}$, but for a trajectory that starts from a random position $x(d t)=x_0+d x(0)$. Hence by using $\LT{p}(w,x_0)=\mathbb{E}\left(e^{-w\mathcal{Z}}\right)$, we have
\begin{equation}\label{eq:LT_first_app}
	\LT{p}(w,x_0)=\langle e^{-wF(x_0)d t}\LT{p}(w,x_0+d x)\rangle+o(d t),
\end{equation}
where the average at the rhs is taken over all possible $x(d t)$, viz., over all possible $d x(0)$. According to Eq. \eqref{eq:SDE}, for any $t$ the displacement $dx(t)=x(t+dt)-x(t)$ is given by
\begin{equation}\label{eq:SDE_diff}
	d x(t)=\mu[x(t)]d t+\sqrt{2D(x^*)}d W(t),
\end{equation}
where $d W(t)$ is the increment of a Wiener process of variance $d t$ and, more importantly, $x^*$ is a point between $x(t)$ and $x(t+d t)$. The choice of the point depends on the interpretation given to the Langevin equation \eqref{eq:SDE}, and different choices lead to different solutions \cite{WesBulLin-1979,Gar,Sok-2010,ManMcClin-2012,VacAntTal-2015}. In other words, if we set $x^*=\alpha x(t+d t)+(1-\alpha)x(t)$, with $0\leq \alpha\leq1$, the value of $\alpha$ determines the ``rule'' to integrate \eqref{eq:SDE}, and the choice is often motivated by physical reasons. The interpretations considered most significant in the physics literature are those of It\^{o} $(\alpha=0)$, Stratonovich $(\alpha=\case12)$ and H\"{a}nggi-Klimontovich $(\alpha=1)$ \cite{Ito-1944,Stra-1966,Han-1982,Kli-1990}. In our case, it is useful to make the nonanticipating choice $\alpha=0$ (It\^{o}). Nevertheless, we are not bound to consider exclusively the It\^{o} interpretation, as any other interpretation can be recovered by inserting an additional drift term dependent on $\alpha$. In other words, \eqref{eq:SDE_diff} is equivalent to
\begin{equation}
	d x(t)=\mu_\alpha[x(t)]d t+\sqrt{2D[x(t)]}d W(t),
\end{equation}
where
\begin{equation}
	\mu_\alpha(x)=\mu(x)+\alpha D'(x).
\end{equation}
Now, from It\^{o} formula \cite{Gar} we can write $d\LT{p}(w,x_0)=\LT{p}(w,x_0+dx)-\LT{p}(w,x_0)$ as
\begin{align}\label{key}
	d\LT{p}(w,x_0)=& \left[D(x_0)\frac{\partial^2\LT{p}(w,x_0)}{\partial x_0^2}+\mu_\alpha(x_0)\frac{\partial\LT{p}(w,x_0)}{\partial x_0}\right]d t\nonumber\\
	&+\sqrt{2D(x_0)}\frac{\partial\LT{p}(w,x_0)}{\partial x_0}d W,
\end{align}
thus by inserting this in Eq. \eqref{eq:LT_first_app}, taking the average over $d W$ and discarding terms that are $o(d t)$, we obtain
\begin{equation}\label{eq:Backw}
	D(x_0)\frac{\partial^2\LT{p}(w,x_0)}{\partial x_0^2}+\mu_\alpha(x_0)\frac{\partial\LT{p}(w,x_0)}{\partial x_0}- wF(x_0)\LT{p}(w,x_0)=0,
\end{equation}
which is the backward evolution equation for $ \LT{p}(w,x_0) $, to be accompanied by the appropriate boundary conditions and the normalization condition $\LT{p}(0,x_0)=1$. If $D(x)$ is always bigger than zero in $\Omega$, we can define
\begin{equation}
	\mathcal{N}(x_0)=\exp\left[\int_{a}^{x_0}\frac{\mu_\alpha(x)}{2D(x)}d x\right],
\end{equation}
where the lower bound of integration can be any point of $\Omega$, and then \eqref{eq:Backw} can be written as
\begin{equation}\label{key}
	\frac{1}{\mathcal{N}(x_0)}\left[\frac{\partial^2}{\partial x_0^2}-V_{\mathrm{eff}}(w,x_0)\right]\mathcal{N}(x_0)\LT{p}(w,x_0)=0,
\end{equation}
where 
\begin{equation}\label{key}
	V_{\mathrm{eff}}(w,x_0)=\frac{wF(x_0)}{D(x_0)}+\frac{1}{\mathcal{N}(x_0)}\frac{d^2\mathcal{N}(x_0)}{d x_0^2}
\end{equation}
We can therefore set
\begin{equation}\label{eq:p_resc_psi}
	\LT{p}(w,x_0)=\frac{\psi(w,x_0)}{\mathcal{N}(x_0)},
\end{equation}
to obtain a simpler equation for $ \psi(w,x_0) $:
\begin{equation}\label{eq:psi_pot}
	\left[\frac{\partial^2}{\partial x_0^2}-V_{\mathrm{eff}}(w,x_0)\right]\psi(w,x_0)=0.
\end{equation}
Note that the normalization condition $\LT{p}(0,x_0)=1$ imposes $\psi(0,x_0)=\mathcal{N}(x_0)$.

\section{Summary of the main results}\label{s:Results}
For $ V(x)=V_0\ln x $, with $-\infty<V_0<\infty$, and $D(x)=D$, Eq. \eqref{eq:psi_pot} simplifies to
\begin{equation}\label{eq:FP_log}
	\frac{\partial^2\psi(w,x_0)}{\partial x_0^2}-\left[\frac{wF(x_0)}{D}+\frac{\beta^2-1}{4x_0^2}\right]\psi(w,x_0)=0,
\end{equation}
where we have introduced the parameter
\begin{equation}
	\beta=1+\frac{V_0}{D}.
\end{equation}
We start by considering the dynamics in an interval $\Omega=(a,b)$, with $0<a<b$. Then we generalize to intervals of the kind $\Omega=(0,a)$ and $\Omega=(b,\infty)$. Finally, we will consider the problem in $\Omega=(0,\infty)$. In the last scenario, we will also provide the solution when the dynamics is generated by a Langevin equation of the kind
\begin{equation}
	\frac{d x(t)}{d t}=\sqrt{2D}x^{\theta}\eta(t),
\end{equation}
with $\theta<1$, and underline how it depends on different interpretations. To simplify the following formulas, it is convenient to define for $\gamma\neq0$ the exponent
\begin{equation}
	\nu=\frac{\beta}{\gamma}=\frac{1}{\gamma}+\frac{V_0}{\gamma D},
\end{equation}
and use the notation $\hat{q}$ to indicate the scaled variable
\begin{equation}\label{eq:q_hat}
	\hat{q}=\sqrt{\frac{wq^\gamma}{\gamma^2D}},
\end{equation}
where $w$ will be the Laplace variable.

\subsection{Finite intervals left-bounded by a positive number}
Consider $\Omega=(a,b)$, with $0<a<x_0<b$. For $\gamma\neq0$, the Laplace transform $\LT{p}(w,x_0)$ is given by
\begin{equation}\label{eq:p_vs_H}
	\LT{p}(w,x_0)=\left(\frac{x_0}{a}\right)^{\beta/2}\frac{\mathcal{H}_\nu(\hat{x}_0,\hat{b})}{\mathcal{H}_\nu(\hat{a},\hat{b})}+\left(\frac{x_0}{b}\right)^{\beta/2}\frac{\mathcal{H}_\nu(\hat{a},\hat{x}_0)}{\mathcal{H}_\nu(\hat{a},\hat{b})},
\end{equation}
where $\mathcal{H}_\nu(\hat{x},\hat{y})$ is defined as
\begin{equation}\label{eq:H}
	\mathcal{H}_\nu(\hat{x},\hat{y})= I_\nu(2\hat{x})K_\nu(2\hat{y})-I_\nu(2\hat{y})K_\nu(2\hat{x}).
\end{equation}
Here $I_\nu(z)$ and $K_\nu(z)$ are the modified Bessel functions of the first and second kind, respectively \cite{NIST}. Similarly, for $\gamma=0$ we define
\begin{equation}\label{eq:H_g0}
	H(x,y)=\sinh\left[\ln\left(\frac xy\right)\sqrt{\frac{w}{D}+\frac{\beta^2}{4}}\right],
\end{equation}
and have
\begin{equation}\label{eq:p_vs_H_g0}
	\LT{p}(w,x_0)=\left(\frac{x_0}{a}\right)^{\beta/2}\frac{H(x_0,b)}{H(a,b)}+\left(\frac{x_0}{b}\right)^{\beta/2}\frac{H(a,x_0)}{H(a,b)},
\end{equation}
which can be inverted, yielding the PDF
\begin{widetext}
	\begin{equation}\label{eq:p_in_ab_g0}
		p(z,x_0)=\frac{2\pi e^{-\tfrac{1}{4}\beta^2Dz}}{\ln^2(b/a)}\sum_{n=1}^{\infty}(-1)^{n+1}ne^{-\tfrac{n^2\pi^2Dz}{\ln^2(b/a)}}\left\{\left(\frac{x_0}{a}\right)^{\beta/2}\sin\left[\frac{\ln(b/x_0)}{\ln(b/a)}n\pi\right]+\left(\frac{x_0}{b}\right)^{\beta/2}\sin\left[\frac{\ln(x_0/a)}{\ln(b/a)}n\pi\right]\right\}.
	\end{equation}
\end{widetext} 
Furthermore, if we consider the set of trajectories that leave $\Omega=(a,b)$ from $b$, which has probability
\begin{equation}\label{eq:E_b}
	\mathcal{E}_b(a)=\begin{dcases}
		\frac{1-(x_0/a)^{\beta}}{1-(b/a)^{\beta}}&\text{if } \beta\neq0\\
		\frac{\ln(x_0/a)}{\ln(b/a)}&\text{if } \beta=0,
	\end{dcases}
\end{equation}
then the distribution of $\mathcal{Z}$ measured only on those trajectories has the normalized density
\begin{equation}\label{eq:p_b}
	\LT{p}(w,x_0)=\frac{1}{\mathcal{E}_b(a)}\times\begin{dcases}
	\left(\frac{x_0}{a}\right)^{\beta/2}\frac{\mathcal{H}_\nu(\hat{a},\hat{x}_0)}{\mathcal{H}_\nu(\hat{a},\hat{b})}&\gamma\neq0\\
		\left(\frac{x_0}{a}\right)^{\beta/2}\frac{H(a,x_0)}{H(a,b)}&\gamma=0,
	\end{dcases}
\end{equation}

Analogously, the probability $\mathcal{E}_a(b)$ of leaving from $a$ can be obtained from Eq. \eqref{eq:E_b}, and the corresponding normalized density from Eq. \eqref{eq:p_b}, by exchanging $a$ and $b$.

\subsection{Finite intervals left-bounded by the origin, or infinite intervals left-bounded by a positive number}
Now take $\Omega=(0,r)$ or $\Omega=(r,\infty)$, with $r>0$. There is a correspondence between the solutions in the two cases. More precisely:
\begin{enumerate}
	\item When $\beta>0$ and $\Omega=(0,r)$, the functional $\mathcal{Z}$ is well-defined only for $\gamma>0$. In this case, the Laplace transform $\LT{p}(w,x_0)$ is given by
	\begin{align}\label{eq:p_in_0b_g>0b>0}
		\LT{p}(w,x_0)=&\frac{2\hat{x}_0^\nu}{\Gamma(\nu)}K_\nu(2\hat{x}_0)\left[1-\frac{I_\nu(2\hat{x}_0)K_\nu(2\hat{r})}{I_\nu(2\hat{r})K_\nu(2\hat{x}_0)}\right]\nonumber\\
		&+\left(\frac{x_0}{r}\right)^{\beta/2}\frac{I_\nu(2\hat{x}_0)}{I_\nu(2\hat{r})}.
	\end{align}
	Nevertheless, if we examine only the set of trajectories that leave from $r$, which has probability $\mathcal{E}_R=(x_0/r)^\beta$, then $\mathcal{Z}$ is well-defined for any $\gamma$, and the corresponding normalized conditional PDF is
	\begin{equation}\label{eq:p_a_0a_b>0}
		\LT{p}(w,x_0)=\begin{dcases}
			\left(\frac{r}{x_0}\right)^{\beta/2}\frac{I_\nu(2\hat{x}_0)}{I_\nu(2\hat{r})}&\text{for }\gamma>0\\
			\left(\frac{r}{x_0}\right)^{\beta/2}\frac{K_\nu(2\hat{x}_0)}{K_\nu(2\hat{r})}&\text{for }\gamma<0,
		\end{dcases}
	\end{equation}
	whereas for $\gamma=0$ the conditional PDF is given explicitly by
	\begin{equation}\label{eq:p_a_0a_b>0_g0}
		p(z,x_0)=\sqrt{\frac{\ln^2(r/x_0)}{4\pi Dz^3}}e^{-\tfrac{\left[D\beta z-|\ln(r/x_0)|\right]^2}{4Dz}}.
	\end{equation}
	Similarly, if we take now $\Omega=(r,\infty)$ and exchange $\beta\to-\beta$ and $\gamma\to-\gamma$, the set of trajectories that leave the interval in a finite time has probability $\mathcal{E}=(x_0/r)^\beta$ and the corresponding normalized conditional PDF is given again by Eqs. \eqref{eq:p_a_0a_b>0} and \eqref{eq:p_a_0a_b>0_g0}.
	\item When $\beta\leq0$ and $\Omega=(0,r)$, a trajectory leaves the interval from $r$ with probability one, thus $\mathcal{Z}$ is well-defined for any $\gamma$. The Laplace transform $\LT{p}(w,x_0)$ is given by
	\begin{equation}\label{eq:p_a_0a_b<0}
		\LT{p}(w,x_0)=\begin{dcases}
			\left(\frac{x_0}{r}\right)^{\beta/2}\frac{I_\nu(2\hat{x}_0)}{I_\nu(2\hat{r})}&\text{for }\gamma>0\\
			\left(\frac{x_0}{r}\right)^{\beta/2}\frac{K_\nu(2\hat{x}_0)}{K_\nu(2\hat{r})}&\text{for }\gamma<0,
		\end{dcases}
	\end{equation}
	and for $\gamma=0$ the PDF is given explicitly by
	\begin{equation}\label{eq:p_a_0a_b<0_g0}
		p(z,x_0)=\sqrt{\frac{\ln^2(r/x_0)}{4\pi Dz^3}}e^{-\tfrac{\left[D\beta z+|\ln(r/x_0)|\right]^2}{4Dz}}.
	\end{equation}
	Similarly, if we take now $\Omega=(r,\infty)$ and exchange $\beta\to-\beta$ and $\gamma\to-\gamma$, the set of trajectories that leave the interval in a finite time has probability one and the corresponding PDF is given again by Eqs. \eqref{eq:p_a_0a_b<0} and \eqref{eq:p_a_0a_b<0_g0}.
\end{enumerate}

\subsection{Positive real axis}
For the positive real axis $\Omega=(0,\infty)$, the functional $\mathcal{Z}$ is well-defined only when both $\beta$ and $\gamma$ are positive. However, in this case the PDF can be computed explicitly. By defining
\begin{equation}\label{eq:ZD_def}
	Z_D=\frac{x_0^\gamma}{\gamma^2D},
\end{equation}
the PDF can be written as
\begin{equation}
	p(z,x_0)=\frac{Z_D^{\nu}}{\Gamma(\nu)}z^{-1-\nu}e^{-Z_D/z},
\end{equation}
where $\Gamma(\nu)$ is the Euler Gamma function. The result for free Brownian motion \cite{MajMee-2020} is recovered by setting $V_0=0$, i.e., by putting $\beta=1$, which yields the exponent $\nu=1/\gamma$.

As a corollary, consider now $\mathcal{Z}$ evaluated on trajectories generated by the Langevin equation
\begin{equation}
	\frac{dx(t)}{dt}=\sqrt{2D}x^{\theta}\eta(t),
\end{equation}
with $\theta<1$, that we may interpret with any $0\leq\alpha\leq1$. For $\gamma>2\theta$ and $0\leq\alpha\leq\case12$, the PDF of $\mathcal{Z}$ is
\begin{equation}\label{eq:distr_diff_coeff}
	g(z,x_0)=\frac{K_D^{\nu_\alpha}}{\Gamma(\nu_\alpha)}z^{-1-\nu_\alpha}e^{-K_D/z},
\end{equation}
where
\begin{equation}
	K_D=\frac{x_0^{\gamma-2\theta}}{(\gamma-2\theta)^2D},\quad\nu_\alpha=\frac{1-2\alpha\theta}{\gamma-2\theta}.
\end{equation}
The same applies to $\case12<\alpha\leq1$, if we add the condition $\theta<\case1{2\alpha}$. By way of illustration, the first-passage time density is recovered from Eq. \eqref{eq:distr_diff_coeff} by setting $\gamma=2$:
\begin{equation}
	p(t,x_0)=\left[\frac{x_0^{2(1-\theta)}}{(1-\theta)^2 4Dt}\right]^{\nu_\alpha}\frac{e^{-\tfrac{x_0^{2(1-\theta)}}{(1-\theta)^2 4Dt}}}{\Gamma(\nu_\alpha)t^{1+\nu_\alpha}},
\end{equation}
with $\nu_\alpha=(1-2\alpha\theta)/(2-2\theta)$, which agrees perfectly with recent results \cite{DosMenAnt-2022}.

In the following, we go into the details of the derivation and present plots in which we compare our findings with numerical simulations.

\section{Finite intervals left-bounded by a positive number}\label{s:finite_interval}
In this section we deal with intervals of the kind $\Omega=(a,b)$. This case can be treated for any value of $\gamma$, but we must distinguish $\gamma\neq0$ and $\gamma=0$.

Let us begin with $\gamma\neq0$. Eq. \eqref{eq:FP_log} can be brought back to the modified Bessel equation:
\begin{equation}\label{eq:Bessel_mod}
	z^2f''(z)+zf'(z)-(z^2+\nu^2)f(z)=0.
\end{equation}
To see this, we make the ansatz $ \psi(w,x_0)=x_0^\rho \varphi(\lambda x_0^\sigma) $, where $ \lambda $ depends on $ w $ and arrive at the following equation for $\varphi(z)$:
\begin{align}\label{key}
	0=&z^2\varphi''(z)+\frac{2\rho-1+\sigma}{\sigma} z\varphi'(z)\nonumber\\
	&-\left[\frac{wz^{\gamma/\sigma}}{D\sigma^2\lambda^{\gamma/\sigma}}-\frac{4\rho(\rho-1)+1-\beta^2}{4\sigma^2}\right]\varphi(z),
\end{align}
where $z=\lambda x_0^\sigma$. Then, by choosing
\begin{equation}\label{key}
	\rho=\frac12,\quad\sigma=\frac{\gamma}{2},\quad\lambda=\frac{2}{\gamma}\sqrt{\frac{w}{D}},
\end{equation}
we obtain the modified Bessel equation \eqref{eq:Bessel_mod}, with $\nu=\beta/\gamma$, which admits the general solution
\begin{equation}
	\varphi(2\hat{x}_0)=c_1I_\nu(2\hat{x}_0)+c_2K_\nu(2\hat{x}_0),
\end{equation}
where $c_1$ and $c_2$ are coefficients that depend on $a$, $b$ and $w$, and we recall
\begin{equation}
	\hat{x}_0=\sqrt{\frac{wx_0^\gamma}{\gamma^2D}},
\end{equation}
see Eq. \eqref{eq:q_hat}. The function $\psi(w,x_0)$ is thus of the form $\psi(w,x_0)=\sqrt{x_0}\varphi(2\hat{x}_0)$ and by recalling Eq. \eqref{eq:p_resc_psi}, we must have $\LT{p}(w,x_0)=x_0^{(\beta-1)/2} \psi(w,x_0)$. Therefore the general solution is
\begin{equation}\label{key}
	\LT{p}(w,x_0)=x_0^{\beta/2}\left[c_1I_\nu(2\hat{x}_0)+c_2K_\nu(2\hat{x}_0)\right].
\end{equation}
To determine the correct boundary conditions, we just note that when the starting point of the trajectory is close to one of the boundaries, the first-passage time tends to zero, and so does the integral in Eq. \eqref{eq:func_0}. Hence $\LT{p}(w,x_0)=\mathbb{E}(e^{-w\mathcal{Z}})$ must be equal to one for $x_0$ equal to $a$ or $b$:
\begin{equation}
	\LT{p}(w,a)=\LT{p}(w,b)=1.
\end{equation}
Then $c_1$ and $c_2$ can be determined from the simple linear system
\begin{equation}
	\mathcal{M}(a,b)\mathbf{c}=\mathbf{1},
\end{equation}
where
\begin{equation}
	\mathbf{c}=\begin{bmatrix}c_1\\c_2\end{bmatrix}\quad\mathbf{1}=\begin{bmatrix}1\\1\end{bmatrix},
\end{equation}
and
\begin{equation}
	\mathcal{M}(a,b)=\begin{bmatrix}
		a^{\beta/2}I_\nu(2\hat{a})&a^{\beta/2}K_\nu(2\hat{a})\\
		b^{\beta/2}I_\nu(2\hat{b})&b^{\beta/2}K_\nu(2\hat{b})
	\end{bmatrix}.
\end{equation}
The solution can be finally written as
\begin{equation}\label{eq:p_vs_H_II}
	\LT{p}(w,x_0)=\left(\frac{x_0}{a}\right)^{\beta/2}\frac{\mathcal{H}_\nu(\hat{x}_0,\hat{b})}{\mathcal{H}_\nu(\hat{a},\hat{b})}+\left(\frac{x_0}{b}\right)^{\beta/2}\frac{\mathcal{H}_\nu(\hat{a},\hat{x}_0)}{\mathcal{H}_\nu(\hat{a},\hat{b})},
\end{equation}
with
\begin{equation}
	\mathcal{H}_\nu(\hat{x},\hat{y})=I_\nu(2\hat{x})K_\nu(2\hat{y})-I_\nu(2\hat{y})K_\nu(2\hat{x}).
\end{equation}
One can verify that $\LT{p}(w,x_0)$ satisfies the normalization condition $\LT{p}(0,x_0)=1$.

Now we consider the case $\gamma=0$, which corresponds to
\begin{equation}
	\mathcal{Z}=\int_{0}^{\mathcal{T}}\frac{d t}{[x(t)]^2}.
\end{equation}
Equation \eqref{eq:psi_pot} reads
\begin{equation}\label{key}
	\frac{\partial^2\psi(w,x_0)}{\partial x_0^2}-\left[\frac{w}{D}+\frac{\beta^2-1}{4}\right]\frac{\psi(w,x_0)}{x_0^2}=0,
\end{equation}
and now we seek solutions of the form $\psi(w,x_0)=\varphi(\lambda\ln x_0)$. The corresponding equation for $\varphi(z)$
\begin{equation}\label{key}
	\lambda^2\varphi''(z)-\lambda \varphi'(z)-\left[\frac wD+\frac{\beta^2-1}{4}\right]\varphi(z)=0,
\end{equation}
is just a second order linear ordinary differential equation with constant coefficients. The characteristic roots are
\begin{equation}\label{key}
	r_\pm=\frac{1}{2\lambda}\left[1\pm2\sqrt{\frac{w}{D}+\frac{\beta^2}{4}}\right],
\end{equation} 
and the solution can be thus written as
\begin{equation}\label{key}
	\varphi(z)=e^{z/2\lambda}\left[c_1\cosh(kz/\lambda)+c_2\sinh(kz/\lambda)\right],
\end{equation}
where
\begin{equation}\label{key}
	k=\sqrt{\frac{w}{D}+\frac{\beta^2}{4}}.
\end{equation}
By using $ \psi(w,x_0)=\varphi(\lambda\ln x_0)$, we see that the value of $\lambda$ is arbitrary, so we can set it to one. Finally, recalling $\LT{p}(w,x_0)=x_0^{(\beta-1)/2} \psi(w,x_0)$, we have
\begin{equation}\label{key}
	\LT{p}(w,x_0)=x_0^{\beta/2}\left[c_1\cosh(k\ln x_0)+c_2\sinh(k\ln x_0)\right],
\end{equation}
where $c_1$ and $c_2$ have to be determined once again in such a way that the boundary conditions $\LT{p}(w,a)=1$ and $\LT{p}(w,b)=1$ are satisfied. Similarly to the previous case, we have to solve an equation of the kind
\begin{equation}
	M(a,b)\mathbf{c}=\mathbf{1},
\end{equation}
but this time with
\begin{equation}
	M(a,b)=\begin{bmatrix}
		a^{\beta/2}\cosh(k\ln a)&a^{\beta/2}\sinh(k\ln a)\\
		b^{\beta/2}\cosh(k\ln b)&b^{\beta/2}\sinh(k\ln b)
	\end{bmatrix}.
\end{equation}
Once the coefficients have been determined, we find that the solution can be written as
\begin{equation}\label{eq:p_vs_H_g0_II}
	\LT{p}(w,x_0)=\left(\frac{x_0}{a}\right)^{\beta/2}\frac{H(x_0,b)}{H(a,b)}+\left(\frac{x_0}{b}\right)^{\beta/2}\frac{H(a,x_0)}{H(a,b)},
\end{equation}
which has the same structure as Eq. \eqref{eq:p_vs_H}, with $\mathcal{H}_\nu(\hat{x},\hat{y})$ replaced by
\begin{align}
	H(x,y)&=\cosh(k\ln x)\sinh(k\ln y)-\sinh(k\ln x)\cosh(k\ln y)\nonumber\\
	&=\sinh\left[k\ln(y/x)\right].
\end{align}

\subsection{Conditioning on leaving the interval from a given boundary}\label{s:Cond}
\begin{figure*}
	\subfloat{
		\includegraphics[width=.31\linewidth]{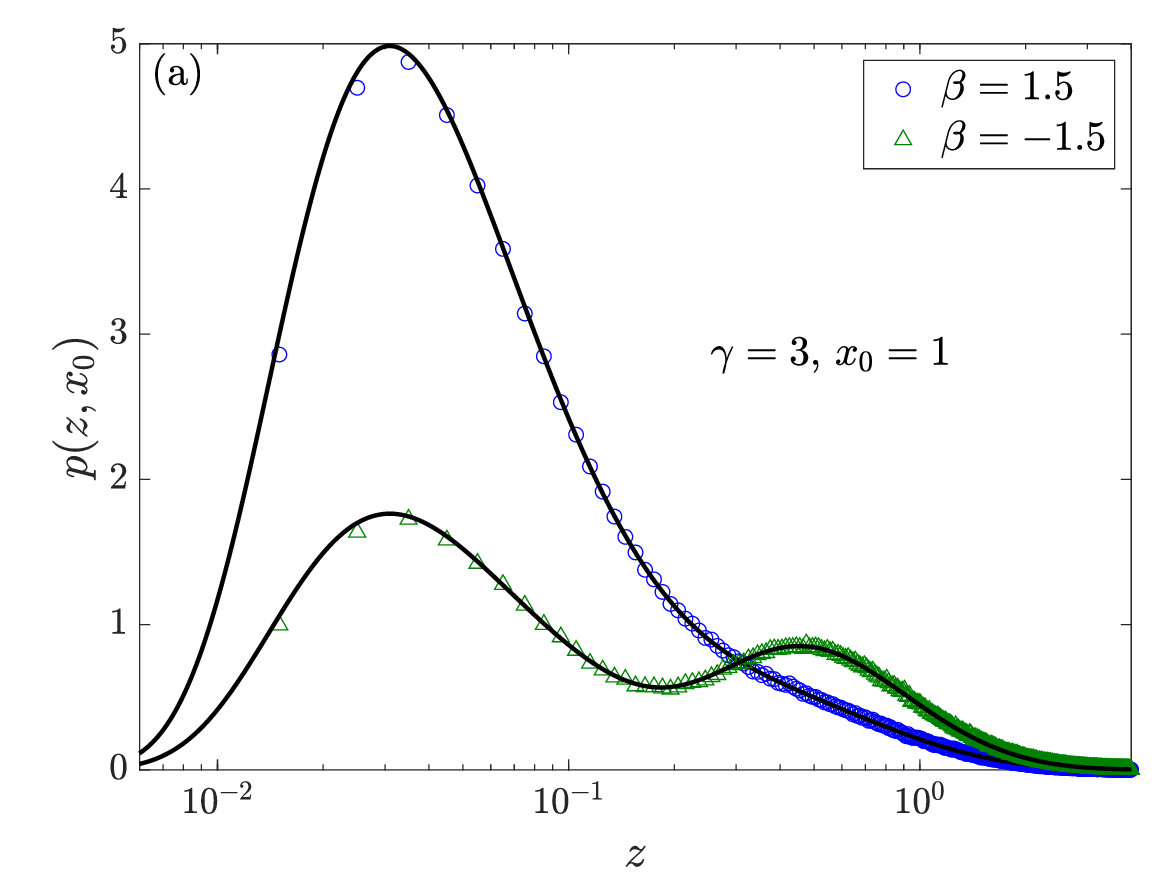}
	}\,
	\subfloat{
		\includegraphics[width=.31\linewidth]{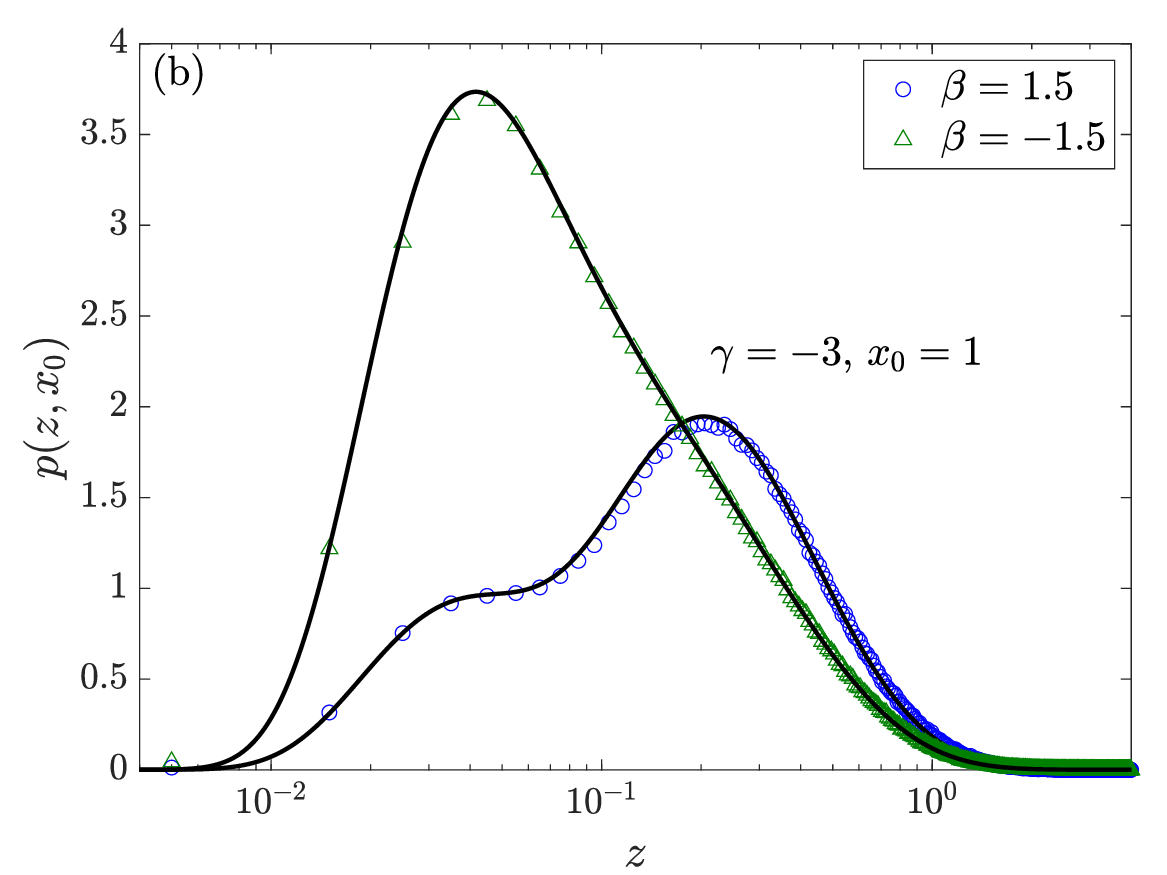}
	}\,
	\subfloat{
		\includegraphics[width=.31\linewidth]{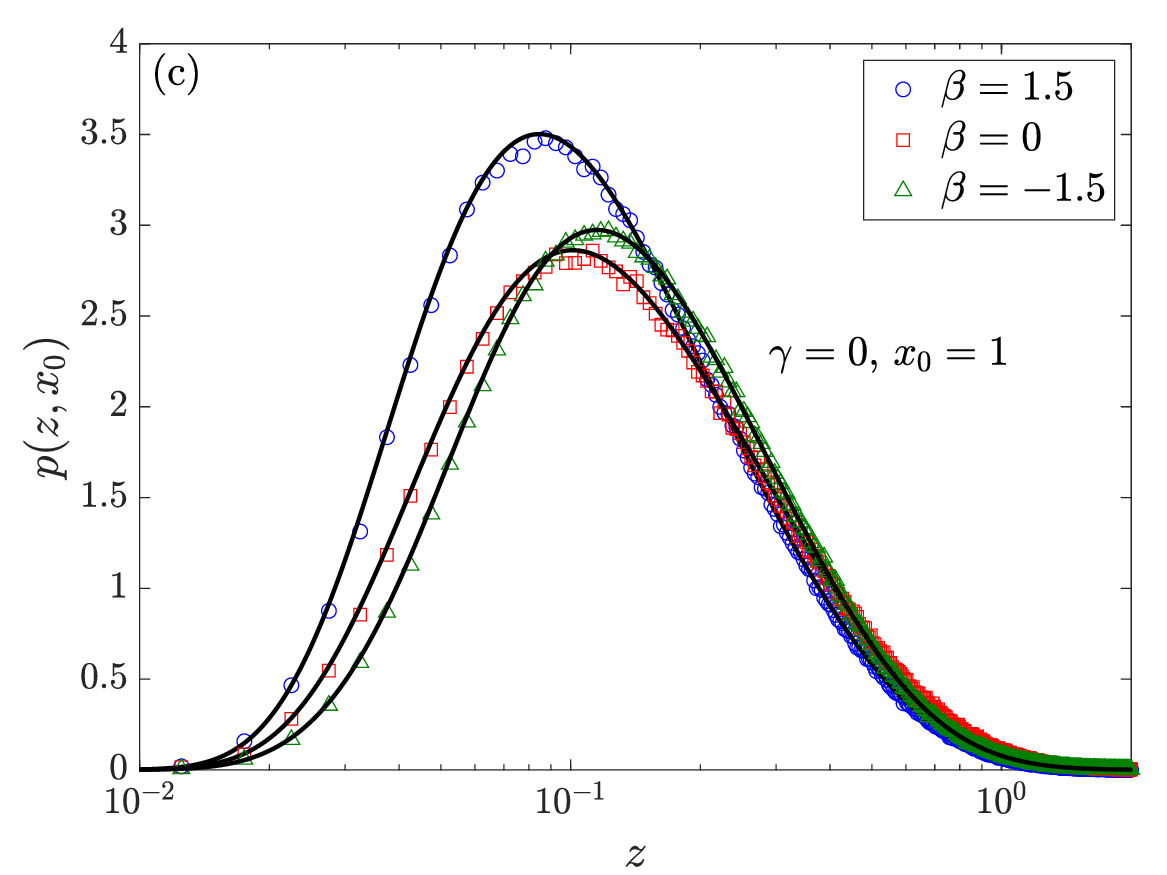}
	}
	\caption{PDF of the functionals (a) $\mathcal{Z}=\int_{0}^{\mathcal{T}}x(t)dt$, (b) $\mathcal{Z}=\int_{0}^{\mathcal{T}}[x(t)]^{-5}dt$ and (c) $\mathcal{Z}=\int_{0}^{\mathcal{T}}[x(t)]^{-2}dt$, for trajectories contained in $\Omega=(\case12,\case52)$. The motion starts from $x_0=1$ and the diffusion coefficient is set to one. The solid black curves are the theoretical predictions, which in (a)-(b) are obtained as $p(z,x_0)=p_a(z,b,x_0)+p_b(z,a,x_0)$, see Eq. \eqref{eq:p_inab_cond_gneq0}, while in (c) are given by Eq. \eqref{eq:p_in_ab_g0}. For (a)-(b) we have chosen $\beta=\case32$ (blue circles) and $\beta=-\case32$ (green triangles) to have $\nu=\pm\case12$. For (c) we have taken $\beta=\case32$ (blue circles), $\beta=0$ (red squares) and $\beta=-\case32$ (green triangles). The data have been obtained by measuring $\mathcal{Z}$ over $10^6$ walks evolved with small time step $\Delta t=10^{-5}$ up to the first-exit time from $\Omega$.}
	\label{fig:PDF_inab}
\end{figure*}
The structure of \eqref{eq:p_vs_H_II} and \eqref{eq:p_vs_H_g0_II} allows us to easily solve the problem with the additional condition that $x(t)$ leaves the interval from a chosen boundary. This request is relevant, for instance, in extreme value theory \cite{ZoiRosMaj-2009,MajRosZoi-2010,Maj-2010,MajPalSch-2020,HarGod-2019,HolBar-2020,AOPR}, where the probability of leaving the interval from a given boundary is related to the statistics of the maximum or the minimum of the process.

Let us take $s>0$ and $q>0$, and assume $\mathrm{min}\{s,q\}<x_0<\mathrm{max}\{s,q\}$. Define $p_s(z,q,x_0)$ as the PDF of the functional $\mathcal{Z}$ under the condition that $x(\mathcal{T})=s$, which corresponds to requiring the process to leave the interval from $s$.
Note that, according to this definition,
\begin{equation}\label{eq:cond_ps}
	\int_{0}^{\infty}p_s(z,q,x_0)dz=\mathcal{E}_s(q),
\end{equation}
where $\mathcal{E}_s(q)$ is the \emph{splitting probability}, namely, the probability of leaving the interval from $s$. The Laplace transform
\begin{equation}
	\LT{p}_s(w,q,x_0)=\int_{0}^{\infty}e^{-wz}p_s(z,q,x_0)d z,
\end{equation}
must thus satisfy the following boundary conditions:
\begin{enumerate}
	\item[a)] $\LT{p}_s(w,q,s)=1$: just as the case considered previously, if the dynamics starts close to the boundary $s$, the first-passage time and thus also $\mathcal{Z}$ tend to zero, hence $\mathbb{E}(e^{-w\mathcal{Z}})\to 1$;
	\item[b)] $\LT{p}_s(w,q,q)=0$: if the dynamics starts instead close to the other boundary $q$, the probability of leaving from $s$ tends to zero, and so does the integral in Eq. \eqref{eq:cond_ps}, from which it follows that also $p_s(z,q,x_0)$ and its Laplace transform must vanish.
\end{enumerate}
From \eqref{eq:p_vs_H_II} and \eqref{eq:p_vs_H_g0_II}, we see that the solutions we found previously are written as the sum of two terms. Taking for example \eqref{eq:p_vs_H_II}, it is easy to verify that
\begin{equation}
	\LT{p}(w,x_0)=\underbrace{\left(\frac{x_0}{a}\right)^{\beta/2}\frac{\mathcal{H}_\nu(\hat{x}_0,\hat{b})}{\mathcal{H}_\nu(\hat{a},\hat{b})}}_{\LT{p}_a(w,b,x_0)}+\underbrace{\left(\frac{x_0}{b}\right)^{\beta/2}\frac{\mathcal{H}_\nu(\hat{a},\hat{x}_0)}{\mathcal{H}_\nu(\hat{a},\hat{b})}}_{\LT{p}_b(w,a,x_0)},
\end{equation}
and the same is true for Eq. \eqref{eq:p_vs_H_g0_II}. Therefore
\begin{equation}\label{eq:p_in_ab_cond}
	\LT{p}_s(w,q,x_0)=\begin{dcases}
	\left(\frac{x_0}{s}\right)^{\beta/2}\frac{\mathcal{H}_\nu(\hat{x}_0,\hat{q})}{\mathcal{H}_\nu(\hat{s},\hat{q})}&\text{if } \gamma\neq0\\
		\left(\frac{x_0}{s}\right)^{\beta/2}\frac{H(x_0,q)}{H(s,q)}&\text{if } \gamma=0.
	\end{dcases}
\end{equation}
The splitting probability $\mathcal{E}_s(q)$ can be computed by using the results of Appendix \ref{s:Bessel}. We find
\begin{equation}\label{eq:E_s(q)}
	\mathcal{E}_s(q)=\begin{dcases}
		\frac{1-(x_0/q)^{\beta}}{1-(s/q)^{\beta}}&\text{if } \beta\neq0\\
		\frac{\ln(x_0/q)}{\ln(s/q)}&\text{if } \beta=0,
	\end{dcases}
\end{equation}
and consequently, the complementary probability of leaving from the other boundary is $\mathcal{E}_q(s)=1-\mathcal{E}_s(q)$.

Before continuing, let us illustrate the results of this section more explicitly. When $\gamma\neq0$, the Laplace transform of Eq. \eqref{eq:p_in_ab_cond} can be inverted for some particular values of the exponent $\nu$. For example, when $\nu=\pm\case12$, one can write the modified Bessel functions in terms of elementary functions \cite{NIST}
\begin{align}
	I_{\frac12}(z)&=\sqrt{\frac{2}{\pi z}}\sinh(z)\\
	I_{-\frac12}(z)&=\sqrt{\frac{2}{\pi z}}\cosh(z)\\
	K_{\frac12}(z)=K_{-\frac12}(z)&=\sqrt{\frac{\pi}{2 z}}e^{-z},
\end{align}
which makes the inversion particularly easy. Indeed, the function $\mathcal{H}_\nu(x,y)$ becomes
\begin{equation}
	\mathcal{H}_{\frac12}(x,y)=\mathcal{H}_{-\frac12}(x,y)=\frac{\sinh(2x-2y)}{\sqrt{xy}},
\end{equation}
thus
\begin{equation}
	\LT{p}_s(w,q,x_0)=\begin{dcases}
		\frac{\sinh(\ell_q\sqrt{w})}{\sinh(L\sqrt{w})}&\text{if } \nu=\frac12\\
		\left(\frac{x_0}{s}\right)^{\beta}\frac{\sinh(\ell_q\sqrt{w})}{\sinh(L\sqrt{w})}&\text{if } \nu=-\frac12,
	\end{dcases}
\end{equation}
where we have defined
\begin{align}
	\ell_q&=2\frac{|x_0^{\gamma/2}-q^{\gamma/2}|}{\sqrt{D\gamma^2}}\\
	L&=2\frac{|s^{\gamma/2}-q^{\gamma/2}|}{\sqrt{D\gamma^2}}.
\end{align}
The poles of $\LT{p}_s(w,q,x_0)$ are $w_n=-(n\pi/L)^2$, so the inversion yields
\begin{widetext}
\begin{equation}\label{eq:p_inab_cond_gneq0}
		p_s(z,q,x_0)=\frac{2\pi}{L^2}\sum_{n=1}^{\infty}(-1)^{n+1}ne^{-\tfrac{n^2\pi^2z}{L^2}}\times\begin{dcases}
			\sin\left(\frac{\ell_q}{L}n\pi\right)&\text{if }\nu=\frac12\\
			\left(\frac{x_0}{s}\right)^{\beta}\sin\left(\frac{\ell_q}{L}n\pi\right)&\text{if }\nu=-\frac12.
		\end{dcases}
\end{equation}
From $p(z,x_0)=p_s(z,q,x_0)+p_q(z,s,x_0)$ it is then possible to obtain the full distribution. Recall that $\nu=\pm\case12$ implies $\beta=\pm\case\gamma2$, hence the validity of this result is limited to those cases. When $\gamma=0$ instead, $\LT{p}_s(z,q,x_0)$ can be inverted for any value of $\beta$. The poles are now $w_n=-D[n\pi/\ln(q/s)]^2-D\beta^2/4$, so we get
\begin{equation}
	p_s(z,q,x_0)=\frac{2\pi D}{\ln^2(q/s)}\left(\frac{x_0}{s}\right)^{\beta/2}e^{-\frac14\beta^2Dz}\sum_{n=1}^{\infty}(-1)^{n+1}ne^{-\tfrac{n^2\pi^2Dz}{\ln^2(a/b)}}\sin\left[\frac{|\ln(q/x_0)|}{|\ln(q/s)|}n\pi\right],
\end{equation}
and from $p(z,x_0)=p_s(z,q,x_0)+p_q(z,s,x_0)$, with $s=a$ and $q=b$, we obtain Eq. \eqref{eq:p_in_ab_g0}.
\end{widetext}

In Fig. \ref{fig:PDF_inab} we show examples of $p(z,x_0)$ for $\gamma= 3$, $\gamma=-3$ and $\gamma=0$. Note that in the first two cases, the condition $\nu=\pm\case12$ requires us to choose $\beta=\pm\case32$. The datasets are obtained by measuring $\mathcal{Z}$ over trajectories in $\Omega=(\case12,\case52)$ starting from $x_0=1$ and evolved up to the first-exit time. The details on the numerical simulations are given in Appendix \ref{s:Num_sim}. We find that the comparison between the theoretical densities and the numerical results is good for any combination of the parameters. Interestingly, while the cases with $\nu=\case12$ display a unimodal distribution, for $\nu=-\case12$ the distribution can become bimodal, as shown in panel (a).

\section{Finite intervals left-bounded by the origin and infinite intervals left-bounded by a positive number}\label{s:InfInt}
We now want to generalize the treatment to intervals of the type $\Omega=(0,b)$ or $\Omega=(a,\infty)$. These cases may be interpreted as the limit for $a\to0$ or $b\to\infty$ of the results of Sec. \ref{s:finite_interval}, and both introduce difficulties not present before. For example, studying the problem in $(0,b)$ allows the trajectory to hit the origin, so there may be values of $\gamma$ for which the integral defining $\mathcal{Z}$ does not converge, see Eq. \eqref{eq:func_0}. For $(a,\infty)$, since the motion occurs in an infinite domain under the action of an external potential, there may be realizations for which the first-passage time $\mathcal{T}$ is not finite. For these reasons, it will be also necessary to rediscuss the boundary conditions that the solutions must satisfy.

\subsection{Finite intervals of the kind $\Omega=(0,b)$}
When we consider diffusion in a logarithmic potential $V(x)=V_0\ln(x)$, the first-passage problem to the origin must be treated with special attention. Indeed, the nature of this point depends on the relative magnitude of the potential $V_0$ with respect to the diffusion constant $D$, which in our discussion is measured by the parameter $\beta$. A detailed analysis following Feller's classification scheme can be found in Ref. \cite{MarBehGer-2011}, according to which the origin is an \emph{exit} boundary for $\beta\geq2$, a \emph{regular} boundary for $0<\beta<2$ and an \emph{entrance} boundary for $\beta\leq0$. Exit and regular boundaries are both \emph{accessible}; entrance boundaries are \emph{inaccessible} \cite{Eth}, meaning that they can not be reached in finite time from the interior of the state space (in our case, from any point inside $\Omega$). We can see this by evaluating the splitting probability $\mathcal{E}_a(b)$ in the limit $a\to0$, see Eq. \eqref{eq:E_s(q)}:
\begin{equation}\label{eq:E_b_in0b}
	\lim_{a\to0}\mathcal{E}_a(b)\equiv\mathcal{E}_L=\begin{dcases}
		1-\left(\frac{x_0}{b}\right)^\beta&\text{if }\beta>0\\
		0&\text{if }\beta\leq0.
	\end{dcases}
\end{equation}
Hence when $\beta\leq0$ the probability of hitting the origin vanishes, and a trajectory will leave $\Omega$ from $b$ with probability one. From a physical point of view, this can be motivated by noting that for $\beta\leq0$ there is a strong repulsive potential, with $V_0\geq D$, pushing the particle away from the origin. As a consequence, the boundary condition $\LT{p}(w,0)=1$ is no longer correct: even if the process starts very close to the origin, it is immediately pushed inside $\Omega$ and the motion goes on up to the first-passage to $b$. Thus, contrarily to what we had previously, $\mathcal{T}$ does not vanish in this case. The fact that the trajectory may or may not leave the origin affects the value of $\gamma$ we can choose so that the integral in Eq. \eqref{eq:func_0} will be convergent. Therefore in the following we differentiate the treatment depending on the sign of $\gamma$.

\subsubsection{Functionals with $\gamma>0$ }
When $\gamma>0$, the limit $a\to0$ corresponds to the limit $\hat{a}\to0$. From the results in Appendix \ref{s:Bessel}, we have
\begin{equation}\label{key}
	\mathcal{H}_\nu(\hat{x},\hat{y})\sim-I_{|\nu|}(2\hat{y})K_{|\nu|}(2\hat{x}),\quad \hat{x}\to0,
\end{equation}
thus if we evaluate $\LT{p}_b(z,a,x_0)$ as $a\to0$, see Eq. \eqref{eq:p_in_ab_cond}, we obtain
\begin{equation}
	\lim_{a\to0}\LT{p}_b(w,a,x_0)\equiv\LT{p}_b(w,x_0)=\left(\frac{x_0}{b}\right)^{\beta/2}\frac{I_{|\nu|}(2\hat{x}_0)}{I_{|\nu|}(2\hat{b})},
\end{equation}
which is the contribution of trajectories hitting $b$, whereas the contribution of trajectories hitting the origin behaves for small $a$ as
\begin{equation}\label{eq:p_a_in_0a_asy}
	\LT{p}_a(w,b,x_0)\sim-\left(\frac{x_0}{a}\right)^{\beta/2}\frac{\mathcal{H}_{\nu}(\hat{x}_0,\hat{b})}{I_{|\nu|}(2\hat{b})K_{|\nu|}(2\hat{a})},
\end{equation}
whose limit depends on the sign of $\beta$. We can use \cite{NIST}
\begin{equation}\label{key}
	K_{|\nu|}(2z)\sim\frac12\Gamma(|\nu|)z^{-|\nu|},\quad z\to0,
\end{equation}
to see that for $\beta>0$, since we have $|\nu|=\nu=\beta/\gamma$, both terms give a nonvanishing contribution in the limit, and the solution \eqref{eq:p_vs_H_II} converges thus to the function
\begin{align}
	\LT{p}(w,x_0)=&\frac{2\hat{x}_0^\nu}{\Gamma(\nu)}K_\nu(2\hat{x}_0)\left[1-\frac{I_\nu(2\hat{x}_0)K_\nu(2\hat{b})}{I_\nu(2\hat{b})K_\nu(2\hat{x}_0)}\right]\nonumber\\
	&+\left(\frac{x_0}{b}\right)^{\beta/2}\frac{I_\nu(2\hat{x}_0)}{I_\nu(2\hat{b})},
\end{align}
which is normalized, since $\LT{p}(0,x_0)=1$, and furthermore satisfies the boundary conditions $\LT{p}(w,0)=1$ and $\LT{p}(w,b)=1$. For $\beta\leq0$ instead, the rhs of \eqref{eq:p_a_in_0a_asy} vanishes, which is consistent with the fact that the origin is an entrance boundary. Thus in the limit $a\to0$ only the contribution of $\LT{p}_b(w,a,x_0)$ survives and the solution converges to
\begin{equation}\label{eq:p_in_0b_g>0b<0}
	\LT{p}(w,x_0)=\left(\frac{x_0}{b}\right)^{\beta/2}\frac{I_{{-\nu}}(2\hat{x}_0)}{I_{-\nu}(2\hat{b})}.
\end{equation}
We remark that, even if this expression originates only from $\LT{p}_b(w,a,x_0)$, it actually represent the full distribution, as can be seen from the fact that it satisfies the normalization condition $\LT{p}(0,x_0)=1$. At the boundaries, we have $\LT{p}(w,b)=1$, as expected, whereas for $x_0=0$ we obtain
\begin{equation}\label{eq:p_bc_b<0g>0}
	\LT{p}(w,0)=\frac{\hat{b}^{-\nu}}{\Gamma(1-\nu)I_{-\nu}(2\hat{b})},
\end{equation}
which is always smaller than one for $b>0$, as one can verify by using the series expansion of the modified Bessel function. Hence, consistently with the fact that the origin is an entrance boundary when $\beta\leq0$, the functional $\mathcal{Z}$ is strictly positive even when measured on trajectories that start very close to $x=0$.

\subsubsection{Functionals with $\gamma<0$ }
For $\gamma<0$, the limit $a\to0$ is equivalent to the limit $\hat{a}\to\infty$, which yields (see Appendix \ref{s:Bessel})
\begin{equation}
	\mathcal{H}_\nu(\hat{x},\hat{y})\sim I_\nu(2\hat{x})K_\nu(2\hat{y}),\quad \hat{x}\to\infty,
\end{equation}
then as $a\to0$ ($\hat{a}\to\infty$)
\begin{equation}
	\lim_{a\to0}\LT{p}_b(w,a,x_0)\equiv\LT{p}_b(w,x_0)=\left(\frac{x_0}{b}\right)^{\beta/2}\frac{K_\nu(2\hat{x}_0)}{K_\nu(2\hat{b})},
\end{equation}
whereas
\begin{equation}
	\LT{p}_a(w,b,x_0)\sim\left(\frac{x_0}{a}\right)^{\beta/2}\frac{\mathcal{H}_\nu(\hat{x}_0,\hat{b})}{I_\nu(2\hat{a})K_\nu(2\hat{b})},
\end{equation}
which vanishes when $\hat{a}\to\infty$, due to the exponential divergence of $I_\nu(2\hat{a})$. Hence, differently from the case $\gamma>0$, the contribution of the walks that hit the origin vanishes \emph{independently} of the value of $\beta$, and the only relevant term is
\begin{equation}\label{eq:p_in_0b_g<0}
	\LT{p}_b(w,x_0)=\left(\frac{x_0}{b}\right)^{\beta/2}\frac{K_\nu(2\hat{x}_0)}{K_\nu(2\hat{b})}.
\end{equation}
We note that this function satisfies $\LT{p}_b(w,b)=1$, but vanishes for any $\nu$ in the limit $x_0\to0$, due to the behavior of $K_{\nu}(z)$ at infinity \cite{NIST}
\begin{equation}
	K_\nu(z)\sim\sqrt{\frac{\pi}{2z}}e^{-z}\left[1+O\left(\frac1z\right)\right],\quad z\to\infty.
\end{equation}
Moreover, by computing $\LT{p}_b(0,x_0)$ we obtain
\begin{equation}\label{eq:p0x_in0b}
	\LT{p}_b(0,x_0)=\begin{dcases}\left(\frac{x_0}{b}\right)^\beta&\text{if }\beta>0\\
			1&\text{if }\beta\leq0,
	\end{dcases}
\end{equation}
which corresponds to the splitting probability $\mathcal{E}_R=1-\mathcal{E}_L$ of leaving $\Omega$ from $b$, see Eq. \eqref{eq:E_b_in0b} for the expression of $\mathcal{E}_L$. We can interpret this result as follows: when $\beta\leq0$, as we mentioned earlier, the origin is an entrance boundary, hence it can not be reached from the interior of $\Omega$ and all the trajectories starting from $x_0>0$ leave the interval from $b$. Therefore, the functional $\mathcal{Z}$ can be measured over each trajectory and Eq. \eqref{eq:p_in_0b_g<0} describes the full distribution, that is, $\LT{p}(w,x_0)=\LT{p}_b(w,x_0)$. On the other hand, when $\beta>0$, a trajectory can leave the interval from any of the two boundaries, but those that leave $\Omega$ from the origin yield a diverging $\mathcal{Z}$ for $\gamma<0$. In other words, $\mathcal{Z}$ is not well-defined if we allow the particle to hit the origin. As we can deduce from the fact that it is normalized to $\mathcal{E}_R$, Eq. \eqref{eq:p_in_0b_g<0} in this case describes the distribution of $\mathcal{Z}$ measured on the set of trajectories that leave the interval from $b$, namely, it is the conditional distribution. We can then define $\LT{p}(w,x_0)=\LT{p}_b(w,x_0)/\mathcal{E}_R$, so that $\LT{p}(w,x_0)$ always denotes the normalized PDF. In summary, we have
\begin{equation}\label{eq:p_in_0b_norm_g<0}
	\LT{p}(w,x_0)=\begin{dcases}
		\left(\frac{b}{x_0}\right)^{\beta/2}\frac{K_\nu(2\hat{x}_0)}{K_\nu(2\hat{b})}&\text{if }\beta>0\\
		\left(\frac{x_0}{b}\right)^{\beta/2}\frac{K_\nu(2\hat{x}_0)}{K_\nu(2\hat{b})}&\text{if }\beta\leq0.
	\end{dcases}
\end{equation}

The fact that $\LT{p}(w,x_0)$ tends to zero as $x_0$ approaches the origin means that $\mathcal{Z}$ is diverging, so contributions from paths passing near $x=0$ can be expected to cause a heavy-tailed decay of the distribution. Indeed, by expanding in powers of $w$, we find
\begin{widetext}
	\begin{equation}
		\LT{p}(w,x_0)\sim\begin{dcases}
			1+C_{\nu} w^{|\nu|}+\dots&\text{for }0<|\nu|<1\\
			1-\langle\mathcal{Z}\rangle w+C_{\nu}w^{|\nu|}+\dots&\text{for }1<|\nu|<2,
		\end{dcases}
	\end{equation}
\end{widetext}
with logarithmic corrections appearing for $\nu=0$ and $\nu=\pm1$. By using Tauberian arguments \cite{Fell-II}, we can conclude that the PDF is characterized by a power-law decay as $p(z,x_0)\sim z^{-1-|\nu|}$.
This can be shown even more explicitly if we consider $\nu=\pm\case12$, for which the inversion can be carried out easily. Indeed, in both cases we have
\begin{equation}
	\LT{p}(w,x_0)=e^{-2\sqrt{\tfrac{w}{\gamma^2D}}(x_0^{\gamma/2}-b^{\gamma/2})},
\end{equation}
and the inversion yields
\begin{equation}
	p(z,x_0)=\frac{x_0^{\gamma/2}-b^{\gamma/2}}{\sqrt{\pi \gamma^2D z^3}}e^{-\tfrac{(x_0^{\gamma/2}-b^{\gamma/2})^2}{\gamma^2 Dz}},
\end{equation}
which indeed decays as $p(z,x_0)\sim z^{-3/2}$.

\subsubsection{The case $\gamma=0$}
We now analyze the particular case $\gamma=0$. For any $\beta$, the limit $a\to0$ of $\LT{p}_b(w,a,x_0)$ yields
\begin{equation}
	\lim_{a\to0}\LT{p}_b(w,a,x_0)\equiv\LT{p}_b(w,x_0)=\left(\frac{x_0}{a}\right)^{k+\beta/2},
\end{equation}
with
\begin{equation}
	k=\sqrt{\frac{w}{D}+\frac{\beta^2}{4}},
\end{equation}
whereas $\LT{p}_a(w,b,x_0)$ vanishes in the limit. So we are in the same situation as the case $\gamma<0$: for $\beta>0$, there is a positive probability $\mathcal{E}_L=1-\mathcal{E}_R$ that a trajectory hits the origin, yielding a diverging $\mathcal{Z}$, hence the distribution must be measured only on the walks that leave $\Omega$ from $b$. For $\beta\leq0$ instead, a trajectory leaves from $b$ with probability one, hence $\LT{p}_b(w,x_0)$ corresponds to the full distribution. In both cases, we can set $\LT{p}(w,x_0)=\LT{p}_b(w,x_0)/\mathcal{E}_R$ and write
\begin{equation}\label{eq:p_in_0b_g0}
	\LT{p}(w,x_0)=\exp\left[-\left(\sqrt{\frac wD+\frac{\beta^2}{4}}-\frac{|\beta|}2\right)\ln\left(\frac{b}{x_0}\right)\right],
\end{equation}
which satisfies $\LT{p}(0,x_0)=1$ and $\LT{p}(w,b)=1$ and vanishes for $x_0\to0$. The inverse transform of Eq. \eqref{eq:p_in_0b_g0} is
\begin{equation}\label{eq:p_in_0b_g0_real}
	p(z,x_0)=\frac{\ln(b/x_0)}{\sqrt{4\pi Dz^3}}e^{-\tfrac{\left[D|\beta| z-\ln(b/x_0)\right]^2}{4Dz}}.
\end{equation}
We see that for $z\to0$ the PDF goes to zero as
\begin{equation}\label{key}
	p(z,x_0)\sim \frac{\ln(b/x_0)}{\sqrt{4\pi Dz^3}}\exp\left[-\frac{\ln^2(b/x_0)}{4Dz}\right],\quad z\to0,
\end{equation}
while for $z\to\infty$
\begin{equation}\label{key}
	p(z,x_0)\sim \frac{\ln(b/x_0)}{\sqrt{4\pi Dz^3}}\exp\left(-\frac{D\beta^2z}{4}\right),\quad z\to\infty.
\end{equation}
Hence, for $\beta\neq0$ there is an exponential cut-off ensuring the convergence of all moments,
while for $\beta=0$ we observe a pure power-law decay $p(z,x_0)\sim z^{-3/2}$ as $z\to\infty$.

\begin{figure}
	\centering
	\includegraphics[width=\columnwidth]{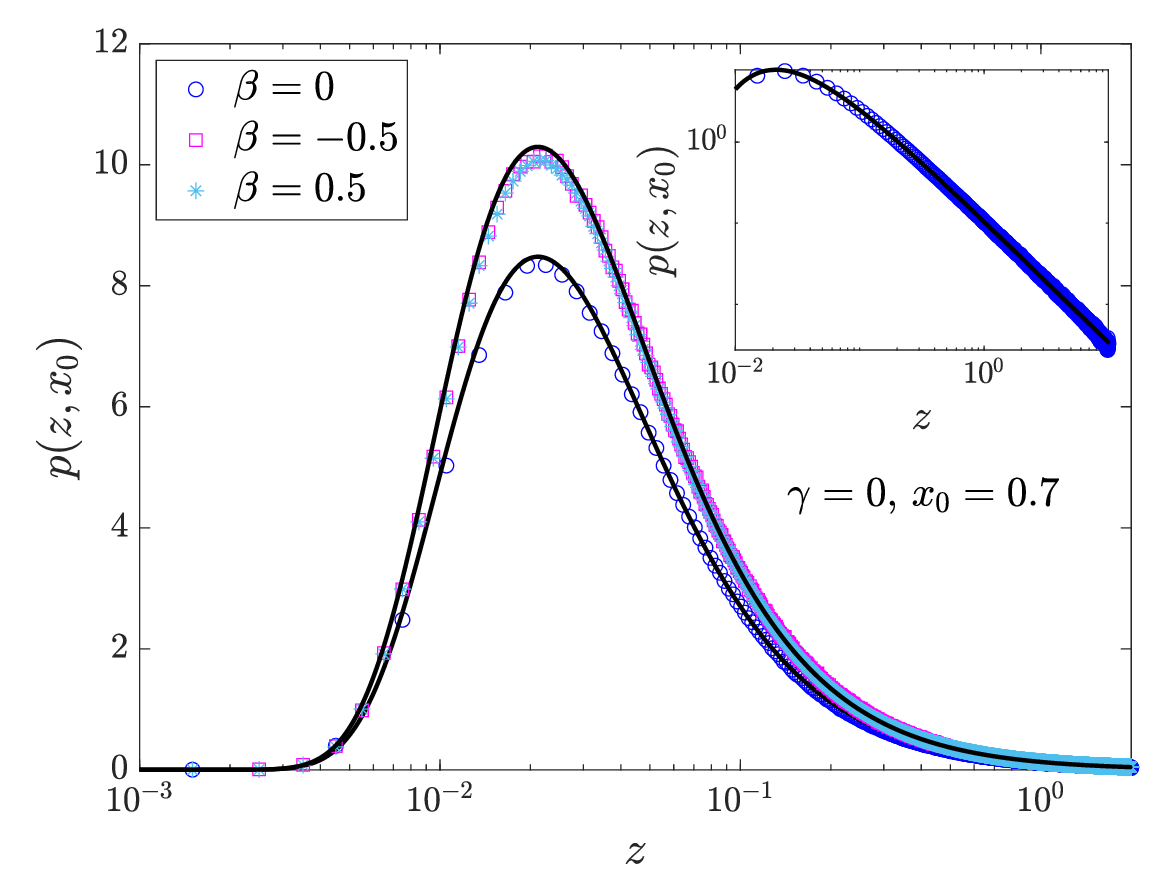}
	\caption{PDF of the functional $\mathcal{Z}=\int_{0}^{\mathcal{T}}[x(t)]^{-2}dt$ for trajectories in $\Omega=(0,1)$. The starting point is $x_0=0.7$ and the diffusion coefficient is set to one. The solid black curves are the theoretical predictions given by Eq. \eqref{eq:p_in_0b_g0_real}, the symbols represent the numerical distributions obtained from simulations. We have considered $\beta=-0.5$ (magenta squares), $\beta=0$ (blue circles) and $\beta=0.5$ (turquoise asterisks), with good agreement in all cases. Note that data corresponding to the same $|\beta|$ overlap, confirming the symmetry in $\beta$ of the PDF. In the inset, a plot in logarithmic scale of the case $\beta=0$, which shows that the simulations also capture the heavy tail of the distribution. The number of simulations is $10^8$ for $\beta=0.5$ and $10^7$ in the other cases. The trajectories are evolved up to the first-passage time with small time step $\Delta t=10^{-5}$.}
	\label{fig:PDF_g0_bb_v2}
\end{figure}

The PDF is displayed in Fig. \ref{fig:PDF_g0_bb_v2} and compared to the results obtained from numerical simulations, showing good agreement. The chosen values of $\beta$ cover all the cases: for $\beta=0.5$ the theoretical result of $\eqref{eq:p_in_0b_g0_real}$ corresponds to the full distribution, while for $\beta=0$ and $\beta=-0.5$ it is the PDF of $\mathcal{Z}$ measured on walks that leave the interval from $b$, normalized dividing by $\mathcal{E}_R$. Note that $p(z,x_0)$ only depends on the sign of $\beta$, hence the PDFs of the data with $\beta=0.5$ and $\beta=-0.5$ are described by the same theoretical curve, as we observe in the figure.

\subsection{Infinite intervals of the kind $\Omega=(a,\infty)$}
The case of infinite intervals marks a difference with the previous one in that the particle is not guaranteed to leave $\Omega$ in a finite time. Indeed, if we consider $\mathcal{E}_a(b)$ given by \eqref{eq:E_s(q)}, and take the limit $b\to\infty$, we get
\begin{equation}\label{eq:E_a}
	\lim_{b\to\infty}\mathcal{E}_a(b)\equiv\mathcal{E}=\begin{dcases}
			1 &\text{if } \beta\geq0\\
		\left(\frac{x_0}{a}\right)^\beta&\text{if }\beta<0,
	\end{dcases}
\end{equation}
meaning that for negative values of $\beta$, there is a nonzero probability $1-\mathcal{E}$ to observe an infinite first-passage time. Note that if we take also the limit $a\to0$, then $\mathcal{E}$ converges to $0$ for $\beta<0$, i.e., the set of trajectories that do not leave $\Omega=(0,\infty)$ has probability one, as is known \cite{Bray-2000}. To avoid having to deal with generalized functionals of the form
\begin{equation}
	\mathcal{Z}=\int_{0}^{\infty}[x(t)]^{\gamma-2}dt,
\end{equation}
in the following we restrict ourselves to the case where actually $\mathcal{T}<\infty$. Note that it would not be appropriate to speak about \emph{first-passage functionals} if the first-passage time is not finite.

\subsubsection{Functionals with $\gamma>0$}
When $\gamma$ has positive sign, the limit $b\to\infty$ corresponds to the limit $\hat{b}\to\infty$. Since
\begin{equation}
	\mathcal{H}_\nu(\hat{x},\hat{y})\sim-I_\nu(2\hat{y})K_\nu(2\hat{x}),\quad\hat{y}\to\infty,
\end{equation}
then as $b\to\infty$ the function $\LT{p}_a(w,b,x_0)$, see Eq. \eqref{eq:p_in_ab_cond}, converges to
\begin{equation}\label{eq:p_in_a00_g>0}
	\lim_{b\to\infty}\LT{p}_a(w,b,x_0)\equiv\LT{p}_a(w,x_0)=\left(\frac{x_0}{a}\right)^{\beta/2}\frac{K_\nu(2\hat{x}_0)}{K_\nu(2\hat{a})},
\end{equation}
which is the same result obtained for the problem in $(0,b)$ in the case $\gamma<0$, see Eq. \eqref{eq:p_in_0b_g<0}. The limiting function satisfies $\LT{p}_a(w,a)=1$ and vanishes for $x_0\to\infty$, i.e., $\LT{p}_a(w,\infty)=0$. The latter condition may be explained by the fact that if the motion starts very far from $a$ one observes a very large first-passage time, and thus we should expect larger and larger values of $\mathcal{Z}$, with $\LT{p}(w,x_0)=\langle e^{-w\mathcal{Z}}\rangle$ consequently vanishing. Regarding the normalization, we get $\LT{p}_a(0,x_0)=\mathcal{E}$, therefore we conclude that Eq. \eqref{eq:p_in_a00_g>0} is the PDF describing the full distribution of $\mathcal{Z}$ for $\beta\geq0$, while for $\beta<0$ it describes the distribution of $\mathcal{Z}$ measured only on the trajectories that actually leave $\Omega=(a,\infty)$ at some finite $\mathcal{T}$. Hence we define again the normalized distribution as
\begin{equation}
	\LT{p}(w,x_0)=\frac{\LT{p}_a(w,x_0)}{\mathcal{E}}.
\end{equation}
Note that this is equivalent to \eqref{eq:p_in_0b_norm_g<0}, hence the same considerations follow.

\subsubsection{Functionals with $\gamma<0$}
Here the limit $b\to\infty$ is equivalent to $\hat{b}\to0$. By using
\begin{equation}
	\mathcal{H}_\nu(\hat{x},\hat{y})\sim I_{|\nu|}(2\hat{x})K_{|\nu|}(2\hat{y}),\quad\hat{y}\to0,
\end{equation}
we see that as $b\to\infty$ ($\hat{b}\to0$), the Laplace transform $\LT{p}_a(w,b,x_0)$ goes to
\begin{equation}\label{eq:p_in_a00_g<0_b>0}
	\lim_{b\to\infty}\LT{p}_a(w,b,x_0)\equiv\LT{p}_a(w,x_0)=\left(\frac{x_0}{a}\right)^{\beta/2}\frac{I_{|\nu|}(2\hat{x}_0)}{I_{|\nu|}(2\hat{a})},
\end{equation}
which satisfies $\LT{p}_a(w,a)=1$, while $\LT{p}_a(0,x_0)=\mathcal{E}$, see the expression for $\mathcal{E}$ in Eq. \eqref{eq:E_a}. Therefore, when $\mathcal{E}=1$, i.e., for $\beta\geq0$, we have $\LT{p}(w,x_0)=\LT{p}_a(w,x_0)$, whereas in the opposite case we set $\LT{p}(w,x_0)=\LT{p}_a(w,x_0)/\mathcal{E}$, that is
\begin{equation}
	\LT{p}(w,x_0)=\begin{dcases}
		\left(\frac{x_0}{a}\right)^{\beta/2}\frac{I_{-\nu}(2\hat{x}_0)}{I_{-\nu}(2\hat{a})}&\beta\geq0\\
		\left(\frac{a}{x_0}\right)^{\beta/2}\frac{I_{\nu}(2\hat{x}_0)}{I_{\nu}(2\hat{a})}&\beta<0.
	\end{dcases}
\end{equation}
Remarkably, for $x_0\to\infty$, this function converges to
\begin{equation}
	\lim_{x_0\to\infty}\LT{p}(w,x_0)=\frac{\hat{a}^{|\nu|}}{\Gamma(1+|\nu|)I_{|\nu|}(2\hat{a})},
\end{equation}
which implies the convergence of all the moments even in the large-$x_0$ limit. 

\subsubsection{The case $\gamma=0$}
The limit $b\to\infty$ of \eqref{eq:p_in_ab_cond} yields
\begin{equation}
	\lim_{b\to\infty}\LT{p}_a(w,b,x_0)\equiv\LT{p}_a(w,x_0)=\left(\frac{a}{x_0}\right)^{k-\beta/2},
\end{equation}
which satisfies $\LT{p}_a(w,a)=1$ and vanishes in the limit $x_0\to\infty$, for the same reason of the case $\gamma>0$. We have once again $\LT{p}_a(0,x_0)=\mathcal{E}$, hence the normalized distribution is $\LT{p}(w,x_0)=\LT{p}_a(w,x_0)/\mathcal{E}$, which can be inverted, yielding
\begin{equation}
	p(z,x_0)=\frac{\ln(x_0/a)}{\sqrt{4\pi Dz^3}}e^{-\tfrac{\left[D|\beta|z-\ln(x_0/a)\right]^2}{4Dz}}.
\end{equation}
This result is equivalent to what we obtained in the interval $(0,b)$, therefore the considerations made in that case still hold.

\section{Positive real axis}\label{s:semi-axis}
\begin{figure*}
	\subfloat{
		\includegraphics[width=.48\linewidth]{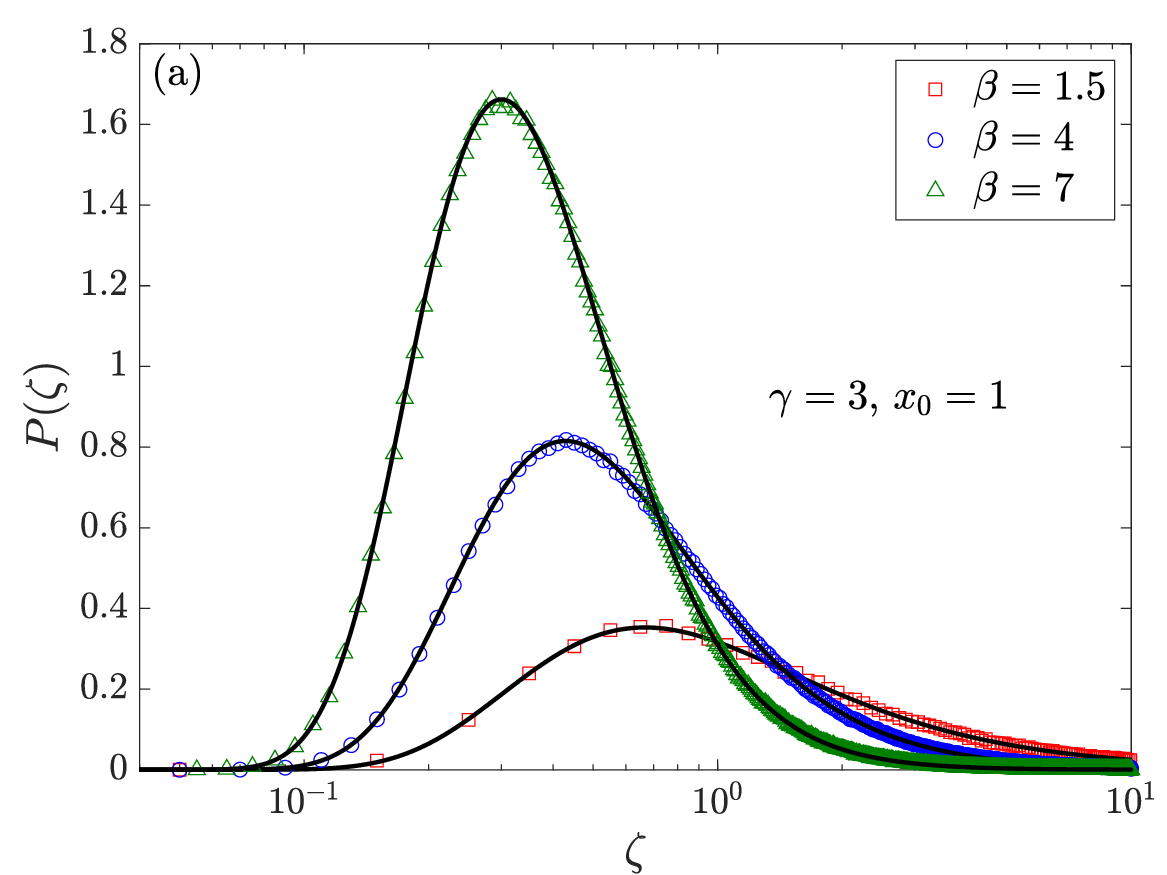}
	}\,
	\subfloat{
		\includegraphics[width=.48\linewidth]{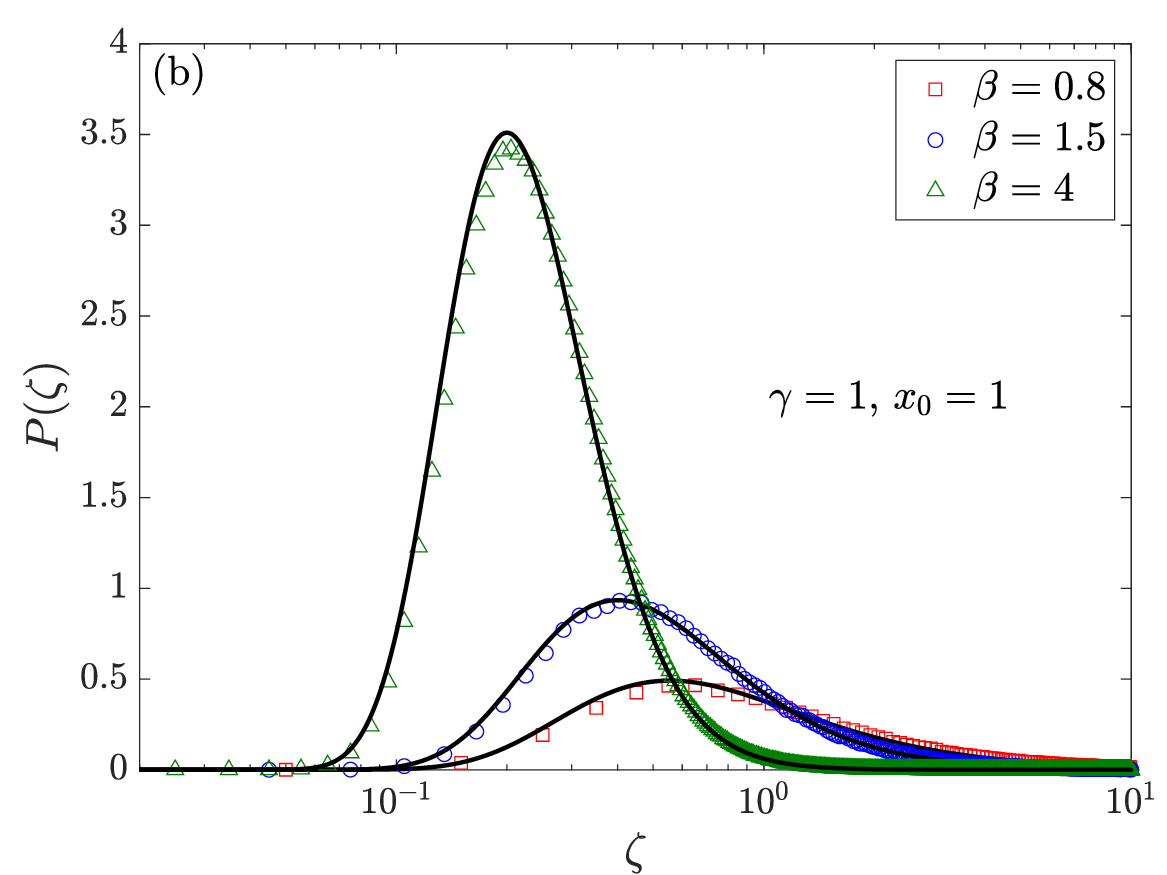}
	}
	\caption{PDF of the scaled variable $\mathcal{Z}/Z_D$, with $\mathcal{Z}$ representing the functionals (a) $\mathcal{Z}=\int_{0}^{\mathcal{T}}x(t)dt$ and (b) $\mathcal{Z}=\int_{0}^{\mathcal{T}}[x(t)]^{-1}dt$ for trajectories in the positive real axis starting from $x_0=1$ and evolved up to the first-passage time to the origin. The diffusion coefficient is set to one. The solid black curves are the theoretical predictions given by Eq. \eqref{eq:P_in0inf_scal}, the symbols represent the numerical distributions obtained from simulations. For both values of $\gamma$ we considered different values of $\beta$: (a) in the case $\gamma=3$ we choose $\beta=1.5$ (red squares), $\beta=4$ (blue circles) and $\beta=7$ (green triangles), (b) in the case $\gamma=1$ we take $\beta=0.8$ (red squares), $\beta=1.5$ (blue circles) and $\beta=4$ (green triangles). The number of simulations depends on $\beta$: for $\beta=0.8$ and $\beta=1.5$ it is $3\cdot10^5$, for $\beta=4$ and $\beta=7$ it is $3\cdot10^6$. All the trajectories are evolved with small time step $\Delta t=10^{-5}$.}
	\label{fig:PDF_in_0inf_log}
\end{figure*}
It is straightforward now to obtain the solution of the problem in the positive real axis $\Omega=(0,\infty)$. It should be clear that $\mathcal{Z}$ is well-defined only for $\beta>0$ and $\gamma>0$. Indeed, as we discussed previously, the condition $\beta>0$ is necessary for the first-passage time to be finite, while $\gamma>0$ ensures that $\mathcal{Z}$ does not diverge when a trajectory hits the origin. Hence we restrict to $\beta>0$ and $\gamma>0$, viz., $\nu>0$.

The solution can be obtained by simply considering the limit $b\to\infty$ of \eqref{eq:p_in_0b_g>0b>0}, which yields
\begin{equation}
	\LT{p}(w,x_0)=\frac{2\hat{x}^\nu_0}{\Gamma(\nu)}K_\nu(2\hat{x}_0).
\end{equation}
This is basically equivalent to the result obtained for free Brownian motion, see \cite{MajMee-2020}, which can be recovered by setting $V_0=0$, viz., $\beta=1$, yielding the exponent $\nu=1/\gamma$. By introducing a logarithmic potential, one thus obtains a generalized exponent $\nu=\beta/\gamma$. The Laplace transform can be inverted exactly by using \cite{Olv}
\begin{equation}\label{key}
	\frac{1}{2}t^{-\nu-1}e^{-y/t}=\frac{1}{2\pi i}\int_{\mathcal{B}} e^{st}\left(\frac{s}{y}\right)^{\nu/2}K_\nu(2\sqrt{ys})d s,
\end{equation}
valid for $\mathfrak{R}(y)>0$, yielding 
\begin{equation}\label{eq:p_in_0infty}
	p(z,x_0)=\frac{Z^{\nu}_D}{\Gamma(\nu)}z^{-1-\nu}e^{-Z_D/z}.
\end{equation}
We note that the PDF can be written in scaling form as $p(z,x_0)=P(\zeta)/Z_D$, where
\begin{equation}\label{eq:P_in0inf_scal}
	P(\zeta)=\frac{1}{\Gamma(\nu)}\zeta^{-1-\nu}e^{-1/\zeta},\quad \zeta=\frac{z}{Z_D}=\frac{\gamma^2D z}{x_0^\gamma}. 
\end{equation}
For $\zeta\to0$ the function $P(\zeta)$ vanishes displaying an essential singularity, while for $\zeta\to\infty$ we get a power-law decay $\mathcal{P}(\zeta)\sim\zeta^{-1-\nu}$. Therefore, the $m$-th moment of the distribution is finite only for $m<\nu$, in which case is equal to
\begin{equation}
	\langle\mathcal{Z}^m\rangle=\int_{0}^{\infty}z^mp(z,x_0)d z=\frac{\Gamma(\nu-m)}{\Gamma(\nu)}Z_D^m.
\end{equation}
Note that $\nu>m$ means $\beta>m\gamma$, hence for fixed $\gamma$ we can tune $\beta$ so that all moments up to the $m$-th are finite. On the other hand, for fixed $\beta$ the $m$-th moment is finite only if $\gamma<\beta/m$, therefore we can for instance have a finite $\langle\mathcal{T}\rangle$ but a diverging $\langle\mathcal{A}\rangle$ (first-passage area).

In Fig. \ref{fig:PDF_in_0inf_log} we present the distribution of the scaled variable $\mathcal{Z}/Z_D$ given by Eq. \eqref{eq:P_in0inf_scal}, and compare it with numerical data. We show the cases $\gamma=3$ and $\gamma=1$, each with three different values of $\beta$, chosen so that for both functionals a case with $0<\nu<1$ (infinite mean and variance), one with $1<\nu<2$ (finite mean, infinite variance) and one with $\nu>2$ (finite mean and variance) is displayed. The agreement between data and theory is evident in all cases.

\subsection{Heterogeneous diffusion}
\begin{figure*}
	\subfloat{
		\includegraphics[width=.48\linewidth]{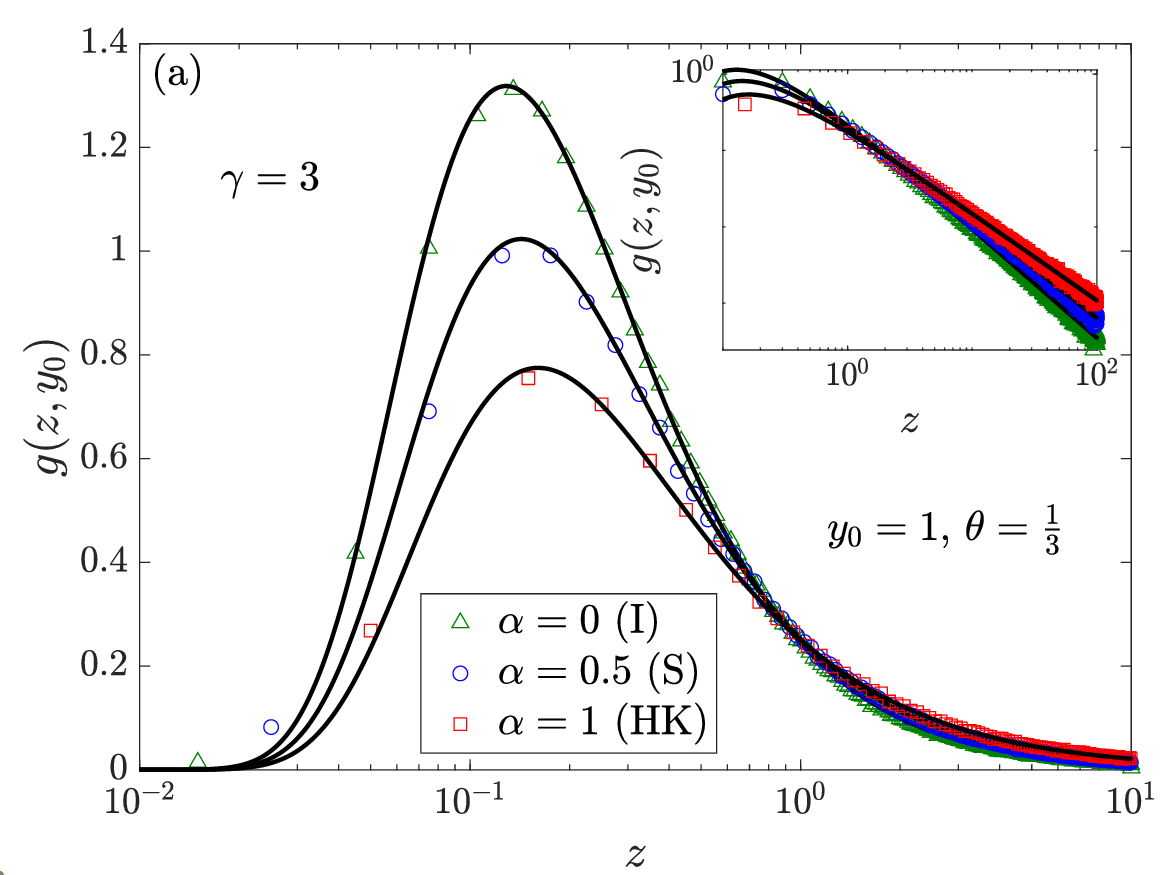}
	}\,
	\subfloat{
		\includegraphics[width=.48\linewidth]{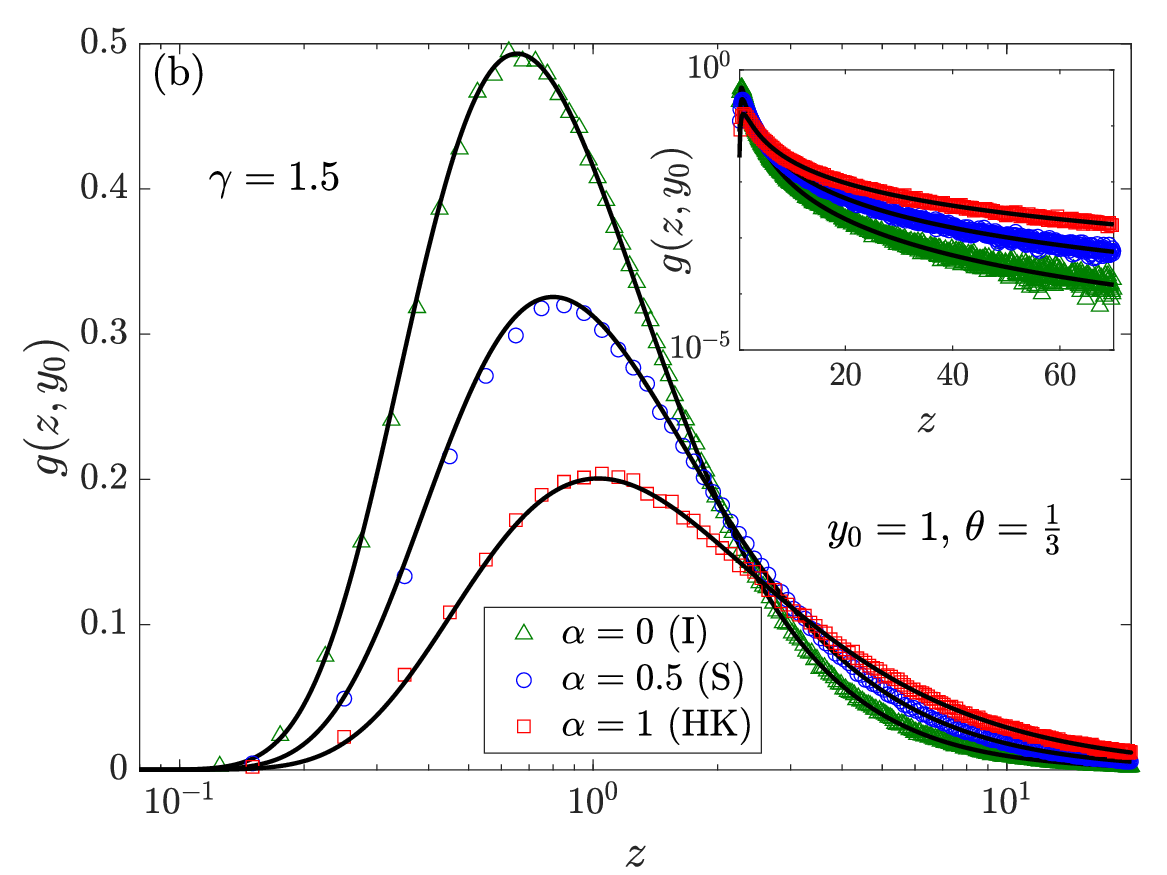}
	}
	\caption{PDF of (a) $\mathcal{Z}=\int_{0}^{\mathcal{T}}y(t)dt$ and (b) $\mathcal{Z}=\int_{0}^{\mathcal{T}}[y(t)]^{-1/2}dt$ for trajectories in the positive real axis generated by Eq. \eqref{eq:Lang_Diff_x} and evolved up to the first-passage time to the origin. The starting point is $y_0=1$ and the space-dependent diffusion coefficient is $D(x)=x^{2/3}$. The solid black curves are the theoretical predictions given by Eq. \eqref{eq:P_in0inf_Diff}, the symbols represent the numerical distributions obtained from simulations. For both values of $\gamma$ we considered three different interpretations, corresponding to It\^{o} ($\alpha=0$, green triangles), Stratonovich ($\alpha=\case12$, blue circles) and H\"{a}nggi-Klimontovich ($\alpha=1$, red squares). The insets, which are presented in log-log scale in (a) and semilog scale in (b), display the heavy-tails of the distributions. All datasets are obtained by evaluating $\mathcal{Z}$ over $10^6$ trajectories, with small time step $\Delta t=10^{-4}$. The integration scheme used to integrate Eq. \eqref{eq:Lang_Diff_x} is the It\^{o} scheme, and the desired interpretation is obtained by adding the appropriate drift $\mu_{\alpha}(x)=\alpha D'(x)$.}
	\label{fig:PDF_Diff}
\end{figure*}
We now extend the results of this section to the case where $\mathcal{Z}$ is measured over stochastic trajectories generated by
\begin{equation}\label{eq:Lang_Diff_x}
	\frac{dy(t)}{dt}=\sqrt{2D}[y(t)]^{\theta}\eta(t),
\end{equation}
with $\theta<1$. As discussed in Sec. \ref{s:FP}, different interpretations can be assigned to this Langevin equation, and we will see how the results are affected by the interpretation. One possible approach to this problem would be to write down Eq. \eqref{eq:psi_pot} for $D(x)=Dx^{2\theta}$ and $\mu_{\alpha}(x)=\alpha D'(x)$, and then solve the resulting equation, namely,
\begin{equation}
	\left[\frac{\partial^2}{\partial y_0^2}-\frac{wF(y_0)}{Dy_0^{2\theta}}-\frac{\alpha\theta(\alpha\theta-1)}{y_0^2}\right]\psi(w,y_0)=0,
\end{equation}
 accompanied by the appropriate boundary conditions. The solution may be then used to obtain the Laplace transform of the PDF. However, as it has already been pointed out \cite{Lei-Bar,Eli-2021,Eli-2021EPL}, there exists a close relation between Brownian motion in logarithmic potentials and heterogeneous diffusion which we may exploit to obtain the solution in a much more straightforward way.
 
 Let us call $x(t)$ a trajectory generated by
 \begin{equation}
 	\frac{dx(t)}{dt}=-\frac{V_0}{x(t)}+\sqrt{2D}\eta(t),
 \end{equation}
 with $x(0)=x_0>0$ and evolving in $\Omega=(0,\infty)$ till the first-passage time to the origin $\mathcal{T}$. Let $\theta<1$ and define the following transformation on the trajectory:
 \begin{equation}\label{eq:y_vs_x}
 	y(t)=\left[(1-\theta)x(t)\right]^\frac{1}{1-\theta}.
 \end{equation}
 By applying It\^{o} formula on $y(t)$, we see that the transformed trajectory evolves according to
 \begin{equation}
 	\frac{dy}{dt}=D\left(1-\beta+\beta\theta\right)y^{2\theta-1}+\sqrt{2D}y^{\theta}\eta(t),
 \end{equation}
 which is interpreted in It\^{o} scheme. We recall $\beta=1+V_0/D$. This is the Langevin equation of a system with a space-dependent diffusion coefficient $D(x)=Dx^{2\theta}$ and a drift term that may be written as
 \begin{equation}
 	\mu(x)=\left(\frac{1-\beta}{2\theta}+\frac\beta2\right)D'(x).
 \end{equation}
  Thus, by setting the coefficient in front of $D'(x)$ equal to $\alpha$, we obtain exactly the It\^{o} form of $\dot{y}=\sqrt{2D(y)}\eta(t)$ in the $\alpha$-interpretation. It is immediate to see that $\alpha=\case12$, which corresponds to Stratonovich interpretation, is recovered by setting $\beta=1$, i.e., by identifying $x(t)$ as free Brownian motion, with a free choice of $\theta<1$. This mapping between Brownian motion and heterogeneous diffusion with Stratonovich interpretation is well-known, see for instance Ref. \cite{CheCheMet-2013}. All other interpretations can be obtained by observing that the parameters $\alpha$, $\beta$ and $\theta$ are related by
 \begin{equation}\label{eq:thet_vs_alpha}
 	\theta=\frac{\beta-1}{\beta-2\alpha}.
 \end{equation}
This also means that for fixed $\alpha$ we can tune $\theta$ by changing the value of $\beta$ in the original model. One must recall, however, that since $\theta<1$, one is limited to take $\beta<2\alpha$ when $\alpha>\case12$ and $\beta>2\alpha$ when $\alpha<\case12$.
 
 It is clear that the first-passage time to the origin of the original trajectory $x(t)$, starting from $x_0$, is the same as the transformed trajectory $y(t)$, with the initial condition $y_0=[(1-\theta)x_0]^{1/(1-\theta)}$. Hence by using Eq. \eqref{eq:y_vs_x} we can write
 \begin{equation}
 	\int_{0}^{\mathcal{T}}[y(t)]^{\gamma-2}dt=\mathcal{C}\int_{0}^{\mathcal{T}}[x(t)]^{\gamma'-2}dt,
 \end{equation}
 with
 \begin{equation}
 	\gamma'=\frac{\gamma-2\theta}{1-\theta},\quad\mathcal{C}=(1-\theta)^{\frac{\gamma-2}{1-\theta}}.
 \end{equation}
 If the functional at the rhs has a proper distribution, with PDF $p(z,x_0)$, then the lhs has a proper distribution too, with PDF
 \begin{equation}
 	g(z,y_0)=\frac{1}{\mathcal{C}}p\left(\frac{z}{\mathcal{C}},\frac{y_0^{1-\theta}}{1-\theta}\right).
 \end{equation}
 We recall that this is true for $x(t)$ if both $\beta$ and $\gamma'$ are positive. The first condition is always met when $\alpha<\case12$, because $\theta<1$ implies $\beta>2\alpha>0$; when $\alpha>\case12$ instead, this must be added to the previous condition $\beta<2\alpha$, obtaining $0<\beta<2\alpha$, which implies that we are limited to $\theta<\case{1}{2\alpha}$. The condition $\gamma'>0$ is equivalent to
 \begin{equation}
 	\gamma>2\theta,
 \end{equation}
hence the lower bound for $\gamma$ is the exponent appearing in the expression of the diffusion coefficient. By using Eq. \eqref{eq:p_in_0infty}, we obtain
 \begin{equation}\label{eq:P_in0inf_Diff}
 	g(z,y_0)=\frac{K_D^{\nu_\alpha}}{\Gamma(\nu_\alpha)}z^{-1-\nu_\alpha}e^{-K_D/z},
 \end{equation}
 where
 \begin{align}
 	K_D&=\frac{y_0^{\gamma-2\theta}}{(\gamma-2\theta)^2D}\\
 	\nu_\alpha&=\frac{1-2\alpha\theta}{\gamma-2\theta}.
 \end{align}
Hence, the interpretation given to the Langevin equation strongly affects the distribution by changing the power-law decay exponent of the PDF.
 
The results are displayed in Fig. \ref{fig:PDF_Diff} for the cases $\gamma=3$ and $\gamma=\case32$, with a diffusion coefficient $D(x)=Dx^{2/3}$. For each $\gamma$, we choose three possible interpretations: $\alpha=0$ (It\^{o}), $\alpha=\case12$ (Stratonovich) and $\alpha=1$ (H\"{a}nggi-Klimontovich), corresponding to the exponents
\begin{align}
	\nu_I&=\frac{1}{\gamma-2\theta}\\
	\nu_S&=(1-\theta)\nu_I\\
	\nu_{HK}&=(1-2\theta)\nu_I.
\end{align}
Note that for $\theta>0$ we have $\nu_I>\nu_{S}>\nu_{HK}$, whereas the opposite happens for $\theta<0$. For the chosen values of $\theta$ and $\gamma$, we obtain in every case an heavy-tailed distribution: for the first-passage area we have $\nu_I=\case37$, $\nu_S=\case27$, and $\nu_{HK}=\case17$, while for $\gamma=1.5$ we get $\nu_I=\case65$, $\nu_S=\case45$ and $\nu_{HK}=\case25$. The agreement between theory and numerical results is generally good. The data can replicate all the features of the PDF, including the tails, see the insets in both panels. We remark that the numerical results have been obtained by measuring $\mathcal{Z}$ over trajectories generated by Eq. \eqref{eq:Lang_Diff_x}, hence they are independent of the method we discussed here.

\section{Conclusions and discussion}\label{s:Concl}
In this paper we have studied the statistical properties of random variables of the kind $\mathcal{Z}=\int_{0}^{\mathcal{T}}[x(t)]^{\gamma-2}dt$, where $x(t)$ is a one-dimensional trajectory of Brownian motion with diffusion constant $D$ evolving under the effect of a logarithmic potential $V(x)=V_0\ln(x)$ that can be either attractive or repulsive. The trajectory starts from $x_0$ inside a given interval $\Omega$ and leaves it for the first time at some random instant $\mathcal{T}$. We initially considered the problem for $\Omega=(a,b)$ entirely contained in the positive real axis, which can be treated for any $\gamma$ and any value of $V_0$. We then generalized to intervals of the kind $\Omega=(0,b)$ or $\Omega=(a,\infty)$. Both these generalizations introduce some limitations: in the former case, for $\gamma<0$ the functional $\mathcal{Z}$ is defined in terms of a divergent integral when measured on trajectories hitting the origin. In the latter case, the presence of a repulsive potential may prevent the particle to leave $\Omega$, which implies an infinite first-passage time. Interestingly, we have underlined that there is a correspondence between the solutions of the two cases, if we always restrict the study of $\mathcal{Z}$ on trajectories for which it is well-defined. Finally, we have computed exactly the density of $\mathcal{Z}$ when it is constructed on trajectories in $\Omega=(0,\infty)$, with $\gamma>0$ and $V_0>-D$. By using a close relation between Brownian motion in logarithmic potentials and heterogeneous diffusion, we have also obtained the distribution of $\mathcal{Z}$ measured on trajectories $x(t)$ generated by $\dot{x}=\sqrt{2D}x^{\theta}\eta(t)$, with $\theta<1$.

This work extends some previously known results regarding first-passage functionals of Brownian motion \cite{MajMee-2020}. By introducing a potential, we were able to study how it affects the statistical properties of $\mathcal{Z}$ for a fixed value of $\gamma$. We emphasize that the logarithmic potential has unique properties, which stem from the fact that it grows as a slowly varying function for $x\to\infty$, yielding a force that is proportional to $1/x$. As already noted in the literature, this causes both the drift term and the diffusion term in the Fokker-Planck equation to scale as $1/x^2$ \cite{KesBar-2010,DecLutBar-2012}. Therefore, the two effects (diffusion and drift) are comparable as long as the dynamics takes place away from the origin, and the system can be treated effectively as a perturbation of Brownian motion. Not surprisingly, the results we obtained in Sec. \ref{s:semi-axis} have the same functional form as those obtained for free Brownian motion \cite{MajMee-2020}. Nevertheless, the system is far from being trivial, as its behavior can be drastically modified by adjusting the parameters that govern the intensity of the drift and diffusion terms, namely the strength of the potential and the diffusion coefficient. This has consequences regarding for example the emergence of nonnormalizable steady states \cite{KesBar-2010,DecLutBar-2011,DecLutBar-2012} or the recurrence properties, of which this system is a critical case study, as evident from the analysis of related discrete models \cite{Lam-1960,Lam-1963,OPRA,Hug-I}. For the problem considered in this paper, for instance, we found that in the case $\Omega=(0,\infty)$ the PDF has a power-law decay as $p(z,x_0)\sim z^{-1-\nu}$, with $\nu=(D+V_0)/(D\gamma)$, which means that the distribution has infinite variance for $V_0<D(2\gamma-1)$ and also infinite mean for $V_0<D(\gamma-1)$. 

Another interesting feature of Brownian motion in a logarithmic potential is that it is associated with heterogeneous diffusion, which is studied in many contexts. Our results can be easily extended to the case where the dynamics is generated by $\dot{x}=\sqrt{2D(x)}\eta$, with $D(x)=Dx^{2\theta}$ and $\theta<1$, as we have done for $\Omega=(0,\infty)$. In this context, a key role is played by the interpretation given to the Langevin equation, and we have seen how the value of the interpretation parameter $\alpha$ contributes, along with the exponent $\theta$, to determine the statistics of $\mathcal{Z}$ for a given value of $\gamma$.

Finally, let us remark that the densities of $\mathcal{Z}$ over trajectories in $\Omega=(0,\infty)$ for logarithmic potentials and heterogeneous diffusion, given in Eq. \eqref{eq:p_in_0infty} and Eq. \eqref{eq:P_in0inf_Diff} respectively, have the same structure, viz, in both cases we obtain the PDF of an Inverse-Gamma random variable \cite{Hof}. This fact is strictly connected to the property of selfsimilarity, which is shared by both models, as shown in Refs. \cite{Eli-2021,Eli-2021EPL}. There the author proves that for any selfsimilar diffusion process the first-passage time to the origin has Inverse-Gamma statistics. We have found that the same statistics describes also $\mathcal{Z}=\int_{0}^{\mathcal{T}}[x(t)]^{\gamma-2}dt$, if it exists. Although this observation could be deduced from scaling arguments, at least with regard to the asymptotic behavior for large $z$ \cite{KeaMaj-2005,MajMee-2020}, it was not trivial to determine how the entire distribution changes with $\gamma$.

As a future perspective, one can ask how the different functionals studied in this article are correlated. The correlation can be measured, for example, by computing the joint probability distribution between two observables $\mathcal{Z}_1$ and $\mathcal{Z}_2$ evaluated for two different $\gamma$. In particular, the case where one of them corresponds to the first-passage time may be particularly relevant, so that information on the correlation between spatial and temporal variables can be obtained directly. This type of joint distribution is indeed useful for comprehensively quantifying the properties of stochastic search processes, as recently observed and studied in \cite{KliBarVoi-2022}.

\appendix
\begin{widetext}
\section{Some properties of the function $\mathcal{H}_\nu(x,y)$}\label{s:Bessel}
\subsection{Expansion in powers of $w$}
Here we consider
\begin{equation}\label{eq:App_H_def}
	\mathcal{H}_\nu(\hat{x},\hat{y})=I_\nu(2\hat{x})K_\nu(2\hat{y})-I_\nu(2\hat{y})K_\nu(2\hat{x}), 
\end{equation}
where $\nu\equiv\beta/\gamma$ and the notation $\hat{q}$ indicates
\begin{equation}
	\hat{q}=\sqrt{\frac{wq^\gamma}{\gamma^2D}}.
\end{equation}
We wish to compute the power series expansion up to first order in $w$, i.e., up to $\hat{q}^2$, which can be used to compute the splitting probabilities or the first moment. Here the modified Bessel function of the first kind $I_\nu(z)$ is defined as
\begin{equation}\label{key}
	I_\nu(z)=\left(\frac{z}{2}\right)^{\nu}\sum_{k=0}^{\infty}\frac{(z^2/4)^k}{k!\Gamma(k+\nu+1)},
\end{equation}
while $K_\nu(z)$ denotes the modified Bessel function of the second kind.

When $\nu$ is non-integer, we can use
\begin{equation}\label{eq:K_vs_I}
	K_\nu(z)=\frac{\pi}{2}\frac{I_{-\nu}(z)-I_\nu(z)}{\sin(\pi\nu)},
\end{equation}
to write
\begin{equation}\label{eq:app_H_nu_nonint}
	\mathcal{H}_\nu(\hat{x},\hat{y})=\frac{\pi}{2\sin(\pi\nu)}\left[I_\nu(2\hat{x})I_{-\nu}(2\hat{y})-I_\nu(2\hat{y})I_{-\nu}(2\hat{x})\right].
\end{equation}
Note that we have the symmetry $\mathcal{H}_\nu(\hat{x},\hat{y})=\mathcal{H}_{-\nu}(\hat{x},\hat{y})$. Hence, recalling that $\nu=\beta/\gamma$, the results do not change under a change of sign in $\beta$, whereas a change of sign in $\gamma$ yields
\begin{equation}
	\mathcal{H}_\nu\left(\sqrt{\frac{wx_0^\gamma}{D\gamma^2}},\sqrt{\frac{wy_0^\gamma}{D\gamma^2}}\right)\to\mathcal{H}_\nu\left(\sqrt{\frac{wx_0^{-\gamma}}{D\gamma^2}},\sqrt{\frac{wy_0^{-\gamma}}{D\gamma^2}}\right).
\end{equation}
By using the definition of $I_\nu(z)$, we find
\begin{equation}\label{key}
	\mathcal{H}_\nu(\hat{x},\hat{y})\sim f_0(x,y;\nu)+\frac{w}{D\gamma^2}f_1(x,y;\nu),
\end{equation}
where $f_0(x,y;\nu)$ and $f_1(x,y;\nu)$ are
\begin{align}
	f_0(x,y;\nu)=&\frac{x^{\gamma\nu}-y^{\gamma\nu}}{2\nu(xy)^{\gamma\nu/2}}\\
	f_1(x,y;\nu)=&\frac{x^{\gamma\nu}g_\nu(x,y)-y^{\gamma\nu}g_\nu(y,x)}{2\nu(xy)^{\gamma\nu/2}},
\end{align}
with
\begin{equation}
	g_\nu(x,y)=\frac{x^\gamma}{1+\nu}+\frac{y^\gamma}{1-\nu}.
\end{equation}
Note the property $g_{-\nu}(x,y)=g_\nu(y,x)$, from which it follows $f_i(x,y;-\nu)=-f_i(x,y;\nu)$.

When $\nu$ is instead an integer, we use the properties of the modified Bessel functions \cite{NIST}
\begin{align}
	I_{-\nu}(z)&=I_\nu(z)+\frac{2}{\pi}\sin(\pi\nu)K_\nu(z)\label{eq:App:Bess_p1}\\
	K_{-\nu}(z)&=K_\nu(z),\label{eq:App:Bess_p2}
\end{align}
to verify that we have again the symmetry $\mathcal{H}_\nu(\hat{x},\hat{y})=\mathcal{H}_{-\nu}(\hat{x},\hat{y})$. For $\nu=n$, with $n=0,1,2,\dots$, the modified Bessel function of the second kind can be expanded as \cite{NIST}
\begin{align}\label{key}
	K_0(2z)=&-\left[\ln(z)+\gamma_{E}\right]I_0(2z)+z^2+\frac{1+\frac12}{(2!)^2}z^4+\dots\\
	K_n(2z)=&\frac{z^{-n}}{2}\mathcal{G}_n(z)+(-1)^{n+1}\ln(z)I_n(2z)+(-1)^n\frac{z^n}{2}\mathcal{F}_n(z),\label{eq:K_exp_int}
\end{align}
where in the first line $\gamma_{E}$ is the Euler-Mascheroni constant, and in the second line
\begin{align}\label{key}
	\mathcal{G}_n(z)&=\sum_{k=0}^{n-1}\frac{(n-k-1)!}{k!}\left(-z^2\right)^k\\
	\mathcal{F}_n(z)&=\sum_{k=0}^{\infty}\frac{\psi(k+1)+\psi(k+n+1)}{k!(n+k)!}z^{2k}.
\end{align}
For negative integers, the corresponding expansion is still given by \eqref{eq:K_exp_int}, with $n$ replaced by its absolute value. Assuming from now on $n\geq0$, for $\mathcal{H}_n(\hat{x},\hat{y})$ we can write
\begin{equation}\label{key}
	\mathcal{H}_n(\hat{x},\hat{y})\sim f_0(x,y;n)+\frac{w}{D\gamma^2}f_1(x,y;n),
\end{equation}
so in each case we just need to identify the functions $f_0(x,y;n)$ and $f_1(x,y;n)$. When $n=0$, $\mathcal{H}_0(x,y)$ can be expanded as
\begin{align}\label{key}
	\mathcal{H}_0(x,y)&\sim\left(1+\hat{x}^2+\hat{y}^2\right)\frac\gamma2\ln\left(\frac xy\right)+\hat{y}^2-\hat{x}^2\\
	&\sim\frac\gamma2\ln\left(\frac xy\right)+\frac{w}{D\gamma^2}\left[(x^\gamma+y^\gamma)\frac\gamma2\ln\left(\frac xy\right)+y^\gamma-x^\gamma\right],
\end{align}
hence we have
\begin{align}\label{key}
	f_0(x,y;0)=&\frac{\gamma}{2}\ln\left(\frac xy\right)\\
	f_1(x,y;0)=&\frac{\gamma}{2}(x^\gamma+y^\gamma)\ln\left(\frac xy\right)+y^\gamma-x^\gamma.
\end{align}
In the case $n=1$, the expansion of $\mathcal{H}_1(\hat{x},\hat{y})$ is
\begin{align}\label{key}
	\mathcal{H}_1(x,y)&\sim \frac12\Bigg(\frac xy\Bigg)^{\gamma/2}\left(1-\frac{\hat{x}^2}{2}\right)-\frac12\Bigg(\frac yx\Bigg)^{\gamma/2}\left(1-\frac{\hat{y}^2}{2}\right)+\frac\gamma2(xy)^{\gamma/2}\ln\left(\frac yx\right)\\
	&\sim \frac{x^{\gamma}-y^\gamma}{2(xy)^{\gamma/2}}+\frac{w}{D\gamma^2}\left[\frac{x^{2\gamma}-y^{2\gamma}}{4(xy)^{\gamma/2}}+\frac\gamma2(xy)^{\gamma/2}\ln\left(\frac yx\right)\right],
\end{align}
therefore now
\begin{align}\label{key}
	f_0(x,y;1)=&\frac{x^\gamma-y^\gamma}{2(xy)^{\gamma/2}}\\
	f_1(x,y;1)=&\frac{x^{2\gamma}-y^{2\gamma}}{4(xy)^{\gamma/2}}+\frac\gamma2(xy)^{\gamma/2}\ln\left(\frac yx\right),
\end{align}
and the same holds for $n=-1$. Finally, for $\nu=2,3,\dots$, the expansion of $\mathcal{H}_n(\hat{x},\hat{y})$ is
\begin{equation}\label{key}
	\mathcal{H}_n(\hat{x},\hat{y})\sim\frac1{2n}\left(\frac xy\right)^{n\gamma /2}\left(1+\frac{\hat{x}^2}{n+1}-\frac{\hat{y}^2}{n-1}\right)-\frac1{2n}\left(\frac yx\right)^{n\gamma /2}\left(1-\frac{\hat{x}^2}{n-1}+\frac{\hat{y}^2}{n+1}\right)
\end{equation}
from which we find that $f_0(x,y;n)$ and $f_1(x,y;n)$ have the same expressions of the non-integer case
\begin{align}
	f_0(x,y;n)=&\frac{x^{n\gamma}-y^{n\gamma}}{2n(xy)^{n\gamma/2}}\\
	f_1(x,y;n)=&\frac{x^{n\gamma}g_n(x,y)-y^{n\gamma}g_n(y,x)}{2n(xy)^{n\gamma /2}},
\end{align}
with
\begin{equation}
	g_n(x,y)=\frac{x^\gamma}{1+n}+\frac{y^\gamma}{1-n},
\end{equation}
and the symmetry $f_i(x,y;n)=-f_i(x,y;-n)$ allows us to claim that the same holds for $n=-2,-3,\dots$
\subsection{Behavior for small and large values of the argument}
We now want to evaluate the behavior of $\mathcal{H}_\nu(\hat{x},\hat{y})$ when $\hat{x}\to0$ and $\hat{x}\to\infty$. Note from the definition \eqref{eq:App_H_def} that $\mathcal{H}_\nu(\hat{x},\hat{y})=-\mathcal{H}_\nu(\hat{y},\hat{x})$, hence the behavior of $\mathcal{H}_\nu(\hat{x},\hat{y})$ as $\hat{x}\to0$ ($\hat{x}\to\infty$) corresponds to the behavior of $-\mathcal{H}_\nu(\hat{x},\hat{y})$ as $\hat{y}\to0$ ($\hat{y}\to\infty$).

As we have shown before in this Appendix, $\mathcal{H}_\nu(\hat{x},\hat{y})$ has the symmetry $\mathcal{H}_\nu(\hat{x},\hat{y})=\mathcal{H}_{-\nu}(\hat{x},\hat{y})$, hence the results do not depend on the sign of $\nu$ and we can thus limit the study to the case $\nu\geq0$. When $z\to0$, the modified Bessel function of the first kind $I_\nu(z)$ behaves as $z^{\nu}$, whereas $K_\nu(z)$ diverges as $z^{-\nu}$ for $\nu>0$ or logarithmically for $\nu=0$. Keeping in mind the symmetry in $\nu$, we therefore have
\begin{equation}\label{key}
	\mathcal{H}_\nu(\hat{x},\hat{y})\sim-I_{|\nu|}(2\hat{y})K_{|\nu|}(2\hat{x}),\quad \hat{x}\to0.
\end{equation}
When $z\to\infty$, the leading-order behavior of both Bessel function is independent of $\nu$. In particular, $I_\nu(z)$ diverges and $K_\nu(z)$ vanishes, both exponentially. Therefore
\begin{equation}\label{key}
	\mathcal{H}_\nu(\hat{x},\hat{y})\sim I_\nu(2\hat{x})K_\nu(2\hat{y}),\quad \hat{x}\to\infty.
\end{equation}
Note that in the asymptotic expansions of $I_\nu(z)$ and $K_\nu(z)$ for large $z$ appear $\nu$-dependent coefficients, which, however, have the symmetry $c_\nu=c_{-\nu}$, see \cite{NIST}.

\section{Computation of the mean value}\label{s:Mean_value_App}
\subsection{Case $\gamma\neq0$}
Starting from the results of Appendix \ref{s:Bessel}, we can consider
\begin{equation}\label{key}
	\LT{p}(w,x_0)=\left(\frac{x_0}{a}\right)^{\beta/2}\frac{\mathcal{H}_\nu(\hat{x}_0,\hat{b})}{\mathcal{H}_\nu(\hat{a},\hat{b})}+\left(\frac{x_0}{b}\right)^{\beta/2}\frac{\mathcal{H}_\nu(\hat{a},\hat{x}_0)}{\mathcal{H}_\nu(\hat{a},\hat{b})},
\end{equation}
to compute the mean value of $\mathcal{Z}$, namely, the coefficient of the linear term in the series expansion in powers of $w$. We note that, while $\mathcal{H}_\nu(\hat{x},\hat{y})$ has the symmetry $\mathcal{H}_\nu(\hat{x},\hat{y})=\mathcal{H}_{-\nu}(\hat{x},\hat{y})$, the expression of $\LT{p}(w,x_0)$ contains the prefactors $(x_0/a)^{\beta/2}$ and $(x_0/b)^{\beta/2}$ that depend on the sign of $\beta$ and thus on the sign of $\nu$. In general, $\langle\mathcal{Z}\rangle$ can be written in terms of the functions $f_0(x,y;\nu)$ and $f_1(x,y;\nu)$, see \ref{s:Bessel} for their definitions, as
\begin{equation}\label{key}
	\langle \mathcal{Z}\rangle=\frac1{D\gamma^2}\left[\frac{f_1(a,b;\nu)}{f_0(a,b;\nu)}
	-\left(\frac{x_0}{a}\right)^{\gamma\nu/2}\frac{f_1(x_0,b;\nu)}{f_0(a,b;\nu)}-\left(\frac{x_0}{b}\right)^{\gamma\nu/2}\frac{f_1(a,x_0;\nu)}{f_0(a,b;\nu)}\right].
\end{equation}
Let us first take $|\nu|\neq0,1$. Then the term between square brackets in the previous equation reads
\begin{equation}\label{key}
	\left[\dots\right]=\frac{a^{\gamma\nu}(b^\gamma-x_0^\gamma)+b^{\gamma\nu}(x_0^\gamma-a^\gamma)+x_0^{\gamma\nu}(a^\gamma-b^\gamma)}{(1-\nu)(a^{\gamma\nu}-b^{\gamma\nu})},
\end{equation}
which yields
\begin{equation}\label{eq:Z_mean_nu_nonint}
	\langle\mathcal{Z}\rangle=\frac{Z_D}{1-\nu}\left[\frac{b^\gamma(x_0^{\gamma\nu}-a^{\gamma\nu})+a^{\gamma}(b^{\gamma\nu}-x_0^{\gamma\nu})}{x_0^\gamma(b^{\gamma\nu}-a^{\gamma\nu})}-1\right].
\end{equation}
When $\nu=\pm1$, the corresponding expression is
\begin{align}
	\left[\dots\right]=&\frac{a^{2\gamma}-b^{2\gamma}}{2(a^\gamma-b^\gamma)}-\left(\frac{x_0}{a}\right)^{\frac\gamma2(\pm1-1)}\frac{x_0^{2\gamma}-b^{2\gamma}}{2(a^\gamma-b^\gamma)}-\left(\frac{x_0}{b}\right)^{\frac\gamma2(\pm1-1)}\frac{a^{2\gamma}-x_0^{2\gamma}}{2(a^\gamma-b^\gamma)}\nonumber\\
	&+\frac{\gamma}{a^\gamma-b^\gamma}\left[(ab)^\gamma\ln\left(\frac ba\right)-\left(\frac{x_0}{a}\right)^{\frac\gamma2(\pm1-1)}(x_0b)^\gamma\ln\left(\frac b{x_0}\right)
	-\left(\frac{x_0}{b}\right)^{\frac\gamma2(\pm1-1)}(x_0a)^\gamma\ln\left(\frac {x_0}a\right)\right],\label{eq:sb_nu=1}
\end{align}
where the $\pm$ sign corresponds to the sign of $\nu$. For $\nu=1$, the term between square brackets in \eqref{eq:sb_nu=1} is the only non-vanishing term, and thus we obtain
\begin{equation}\label{key}
	\langle\mathcal{Z}\rangle=Z_D\left[\frac{b^\gamma(x_0^\gamma-a^\gamma)\ln(b/x_0)-a^\gamma(b^\gamma-x_0^\gamma)\ln(x_0/a)}{x_0^\gamma(b^\gamma-a^\gamma)}\right].
\end{equation}
On the contrary, for $\nu=-1$ the term between square brackets vanishes, while the remaining term yields
\begin{equation}\label{key}
	\langle\mathcal{Z}\rangle=\frac{Z_D}{2x_0^{2\gamma}}(x_0^\gamma-a^\gamma)(b^\gamma-x_0^\gamma),
\end{equation}
which corresponds to \eqref{eq:Z_mean_nu_nonint} for $\nu=-1$. Finally, the case $\nu=0$ is straightforward, and the corresponding mean value is
\begin{equation}\label{key}
	\langle\mathcal{Z}\rangle=Z_D\left[\frac{b^\gamma\ln(x_0/a)+a^\gamma\ln(b/x_0)}{x_0^\gamma\ln(b/a)}-1\right].
\end{equation}
For the sake of completeness, we also consider
\begin{equation}
	\LT{p}_s(w,q,x_0)=\left(\frac{x_0}{s}\right)^{\beta/2}\frac{\mathcal{H}_\nu(\hat{x}_0,\hat{q})}{\mathcal{H}_\nu(\hat{s},\hat{q})},
\end{equation}
and evaluate the coefficient of the linear term in the expansion in powers of $w$. In general, we can write
\begin{equation}
	\LT{p}_s(w,q,x_0)=\mathcal{E}_s(q)\left\{1+\frac{w}{D\gamma^2}\left[\frac{f_1(x_0,q;\nu)}{f_0(x_0,q;\nu)}-\frac{f_1(s,q;\nu)}{f_0(s,q;\nu)}\right]+o(w)\right\},
\end{equation}
where the splitting probability is given by
\begin{equation}
	\mathcal{E}_s(q)=\left(\frac{x_0}{s}\right)^{\beta/2}\frac{f_0(x_0,q;\nu)}{f_0(s,q;\nu)}.
\end{equation}
Then the conditional first moment is just
\begin{equation}
	\langle\mathcal{Z}_s(q)\rangle=\frac{1}{D\gamma^2}\left[\frac{f_1(s,q;\nu)}{f_0(s,q;\nu)}-\frac{f_1(x_0,q;\nu)}{f_0(x_0,q;\nu)}\right],
\end{equation}
and one can verify that the previous general expression for $\langle\mathcal{Z}\rangle$ can be obtained from $\langle\mathcal{Z}\rangle=\mathcal{E}_s(q)\langle\mathcal{Z}_s(q)\rangle+\mathcal{E}_q(s)\langle\mathcal{Z}_q(s)\rangle$. By skipping details, for $\nu\neq0,\pm1$ we find
\begin{equation}\label{key}
	\langle\mathcal{Z}_s(q)\rangle=\frac{1}{D\gamma^2}\left\{\frac{s^\gamma}{1-\nu}\left[\frac{1-(s/q)^{\gamma(\nu-1)}}{1-(s/q)^{\gamma\nu}}-\frac{1-(x_0/q)^{\gamma(\nu-1)}}{1-(x_0/q)^{\gamma\nu}}\left(\frac{x_0}{s}\right)^\gamma\right]
	+\frac{ q^\gamma}{1+\nu}\left[\frac{1-(s/q)^{\gamma(\nu+1)}}{1-(s/q)^{\gamma\nu}}-\frac{1-(x_0/q)^{\gamma(\nu+1)}}{1-(x_0/q)^{\gamma\nu}}\right]\right\},
\end{equation}
the cases $\nu=\pm1$ are both covered by
\begin{equation}\label{key}
	\langle\mathcal{Z}_s(q)\rangle=\frac{1}{D\gamma^2}\left\{\frac{s^\gamma-x_0^\gamma}{2}+\gamma q^\gamma\left[\frac{s^\gamma\ln(q/s)}{s^\gamma-q^\gamma}-\frac{x_0^\gamma\ln(q/x_0)}{x_0^\gamma-q^\gamma}\right]\right\},
\end{equation}
and $\nu=0$ yields
\begin{equation}\label{key}
	\langle\mathcal{Z}_s(q)\rangle=\frac{1}{D\gamma^2}\left[s^\gamma-x_0^\gamma+\frac{2(q^\gamma-s^\gamma)}{\gamma\ln(s/q)}-\frac{2(q^\gamma-x_0^\gamma)}{\gamma\ln(x_0/q)}\right].
\end{equation}
\subsection{Case $\gamma=0$}
For $\gamma=0$, we start by considering
\begin{equation}
	H(x,y)=\sinh\left[\ln\left(\frac{y}{x}\right)\sqrt{\frac{w}{D}+\frac{\beta^2}{4}}\right],
\end{equation}
for which we may write
\begin{equation}\label{eq:H_exp_app}
	H(x,y)=\begin{dcases}
		\sinh\left[\frac{|\beta|}{2}\ln\left(\frac{y}{x}\right)\right]+\frac{w\ln(y/x)}{|\beta|D}\cosh\left[\frac{|\beta|}{2}\ln\left(\frac{y}{x}\right)\right]+o(w)&\text{for }\beta\neq0\\
		\sqrt{\frac{w}{D}}\ln\left(\frac yx\right)+\frac{1}{6}\left(\frac wD\right)^{3/2}\ln^3\left(\frac{w}{D}\right)+o(w^{3/2})&\text{for }\beta=0.
	\end{dcases}
\end{equation}
Let us first evaluate the expansion of
\begin{equation}
	\LT{p}_s(w,q,x_0)=\left(\frac{x_0}{s}\right)^{\beta/2}\frac{H(x_0,q)}{H(s,q)},
\end{equation}
which is
\begin{equation}
	\LT{p}_s(w,q,x_0)=\mathcal{E}_s(q)\left[1+\langle\mathcal{Z}_s(q)\rangle w+o(w)\right].
\end{equation}
By using Eq. \eqref{eq:H_exp_app}, we see that for $\beta\neq0$ we have
\begin{equation}
	\mathcal{E}_s(q)=\left(\frac{x_0}{s}\right)^{\beta/2}\frac{\sinh\left[\frac{|\beta|}{2}\ln\left(\frac{x_0}{q}\right)\right]}{\sinh\left[\frac{|\beta|}{2}\ln\left(\frac{s}{q}\right)\right]}=\left(\frac{x_0}{s}\right)^{\beta/2-|\beta|/2}\frac{1-(x_0/q)^{|\beta|}}{1-(s/q)^{|\beta|}}=\frac{1-(x_0/q)^\beta}{1-(s/q)^\beta},
\end{equation}
and the conditional first moment is
\begin{align}
	\langle\mathcal{Z}_s(q)\rangle&=\frac{1}{|\beta|D}\left\{\ln\left(\frac{s}{q}\right)\coth\left[\frac{|\beta|}{2}\ln\left(\frac{s}{q}\right)\right]-\ln\left(\frac{x_0}{q}\right)\coth\left[\frac{|\beta|}{2}\ln\left(\frac{x_0}{q}\right)\right]\right\}\nonumber\\
	&=\frac{1}{|\beta|D}\left[\ln\left(\frac qs\right)\frac{1+(s/q)^{|\beta|}}{1-(s/q)^{|\beta|}}-\ln\left(\frac{q}{x_0}\right)\frac{1+(x_0/q)^{|\beta|}}{1-(x_0/q)^{|\beta|}}\right]\nonumber\\
	&=\frac{1}{\beta D}\left[\ln\left(\frac qs\right)\frac{1+(s/q)^{\beta}}{1-(s/q)^{\beta}}-\ln\left(\frac{q}{x_0}\right)\frac{1+(x_0/q)^{\beta}}{1-(x_0/q)^{\beta}}\right].
\end{align}
For $\beta=0$ instead, the splitting probability is
\begin{equation}
	\mathcal{E}_s(q)=\frac{\ln(x_0/q)}{\ln(s/q)},
\end{equation}
and the conditional first moment is given by
\begin{equation}
	\langle\mathcal{Z}_s(q)\rangle=\frac{1}{6D}\left[\ln^2\left(\frac{s}{q}\right)-\ln^2\left(\frac{x_0}{q}\right)\right]=\frac{1}{6D}\ln\left(\frac{x_0}{s}\right)\ln\left(\frac{q^2}{sx_0}\right).
\end{equation}
Now, from $\langle\mathcal{Z}\rangle=\mathcal{E}_s(q)\langle\mathcal{Z}_s(q)\rangle+\mathcal{E}_q(s)\langle\mathcal{Z}_q(s)\rangle$ and setting $s=a$ and $q=b$, we obtain
\begin{equation}
	\langle\mathcal{Z}\rangle=\begin{dcases}
		\frac{1}{2D}\ln\left(\frac{b}{x_0}\right)\ln\left(\frac{x_0}{a}\right)&\text{for }\beta=0\\
		\frac{1}{\beta D}\left[\frac{(b^\beta-x_0^\beta)\ln(x_0/a)-(x_0^\beta-a^\beta)\ln(b/x_0)}{b^\beta-a^\beta}\right]&\text{for }\beta\neq0.
	\end{dcases}
\end{equation}
\end{widetext}

\section{Details on numerical simulations}\label{s:Num_sim}
Here we illustrate the numerical scheme used to integrate the stochastic differential equation
\begin{equation}
	d x(t)=a(x)d t+b(x)d W(t).
\end{equation}
To obtain the results illustrated in this paper, in the case of diffusion in a logarithmic potential we used the weak order $2$ Runge-Kutta method \cite{Sauer2012,Klo-Pla}:
\begin{align}
	x_{n+1}=x_n&+\frac12\left[a(\Upsilon)+a(x_n)\right]\Delta t\nonumber\\
	&+\frac14\left[b(\Upsilon^+)+b(\Upsilon^-)+2b(x_n)\right]\Delta W_n\nonumber\\
	&+\frac14\left[b(\Upsilon^+)-b(\Upsilon^-)\right]\left[\frac{\Delta W_n^2-\Delta t}{\sqrt{\Delta t}}\right],
\end{align}
where
\begin{align}
	\Upsilon &=x_n+a(x_n)\Delta t+b(x_n)\Delta W_n\\
	\Upsilon^{\pm} &=x_n+a(x_n)\Delta t\pm b(x_n)\sqrt{\Delta t},
\end{align}
and $\Delta W_n$ are all independent and identically distributed random variables drawn from a common distribution $p(\Delta W)$, such that $\mathbb{E}(\Delta W)=0$ and $\mathbb{E}(\Delta W^2)=\Delta t$. For example, a popular choice is
\begin{equation}
	p(\Delta W)=\frac{1}{\sqrt{2\pi\Delta t}}\exp\left(-\frac{\Delta W^2}{2\Delta t}\right).
\end{equation}
We recall that a discrete-time approximation $x_n$ is said to converge weakly to $x(t)$ if for all polynomials $q(z)$ \cite{Sauer2012}:
\begin{equation}
	\lim_{\Delta t\to0}\mathbb{E}\left\{q(x_n)\right\}=\mathbb{E}\left\{q\left[x(t)\right]\right\}.
\end{equation}
In practice, weak convergence implies the convergence of all moments in the $\Delta t\to0$ limit. The \emph{order} of convergence $m$ is defined by the order of the error in the moments with the step size:
\begin{equation}
	\left|\mathbb{E}\left\{q(x_n)\right\}-\mathbb{E}\left\{q\left[x(t)\right]\right\}\right|=O\left(\Delta t^m\right),
\end{equation}
for sufficiently small $\Delta t$ \cite{Sauer2012}. In the case of heterogeneous diffusion instead, we considered the It\^{o} integration scheme and used the Euler-Maruyama method. By taking into account the possible interpretations of the Langevin equation, the method is implemented as
\begin{equation}
	x_{n+1}=x_n+\alpha D'(x_n)\Delta t+\sqrt{2D(x_n)}\Delta W_n.
\end{equation} 
Finally, to compute the functional defined by \eqref{eq:func_0} in the main text, we approximate
\begin{equation}
	\int_{0}^{\mathcal{T}}F[x(t)]d t\approx\sum_{i=0}^{N}\frac{F(x_{i+1})+F(x_i)}{2}\Delta t,
\end{equation}
where $\mathcal{T}$ is the first-passage time outside a given interval $\Omega$, and $N$ is the random number of steps needed for the approximated trajectory to exit from $\Omega$.

%


\begin{thebibliography}{99}%
	\makeatletter
	\providecommand \@ifxundefined [1]{%
		\@ifx{#1\undefined}
	}%
	\providecommand \@ifnum [1]{%
		\ifnum #1\expandafter \@firstoftwo
		\else \expandafter \@secondoftwo
		\fi
	}%
	\providecommand \@ifx [1]{%
		\ifx #1\expandafter \@firstoftwo
		\else \expandafter \@secondoftwo
		\fi
	}%
	\providecommand \natexlab [1]{#1}%
	\providecommand \enquote  [1]{``#1''}%
	\providecommand \bibnamefont  [1]{#1}%
	\providecommand \bibfnamefont [1]{#1}%
	\providecommand \citenamefont [1]{#1}%
	\providecommand \href@noop [0]{\@secondoftwo}%
	\providecommand \href [0]{\begingroup \@sanitize@url \@href}%
	\providecommand \@href[1]{\@@startlink{#1}\@@href}%
	\providecommand \@@href[1]{\endgroup#1\@@endlink}%
	\providecommand \@sanitize@url [0]{\catcode `\\12\catcode `\$12\catcode
		`\&12\catcode `\#12\catcode `\^12\catcode `\_12\catcode `\%12\relax}%
	\providecommand \@@startlink[1]{}%
	\providecommand \@@endlink[0]{}%
	\providecommand \url  [0]{\begingroup\@sanitize@url \@url }%
	\providecommand \@url [1]{\endgroup\@href {#1}{\urlprefix }}%
	\providecommand \urlprefix  [0]{URL }%
	\providecommand \Eprint [0]{\href }%
	\providecommand \doibase [0]{https://doi.org/}%
	\providecommand \selectlanguage [0]{\@gobble}%
	\providecommand \bibinfo  [0]{\@secondoftwo}%
	\providecommand \bibfield  [0]{\@secondoftwo}%
	\providecommand \translation [1]{[#1]}%
	\providecommand \BibitemOpen [0]{}%
	\providecommand \bibitemStop [0]{}%
	\providecommand \bibitemNoStop [0]{.\EOS\space}%
	\providecommand \EOS [0]{\spacefactor3000\relax}%
	\providecommand \BibitemShut  [1]{\csname bibitem#1\endcsname}%
	\let\auto@bib@innerbib\@empty
	\bibitem [{\citenamefont {Majumdar}(2005)}]{Maj-2005}%
	\BibitemOpen
	\bibfield  {author} {\bibinfo {author} {\bibfnamefont {S.~N.}\ \bibnamefont
			{Majumdar}},\ }\href@noop {} {\bibfield  {journal} {\bibinfo  {journal}
			{Curr. Sci.}\ }\textbf {\bibinfo {volume} {89}},\ \bibinfo {pages} {2076}
		(\bibinfo {year} {2005})}\BibitemShut {NoStop}%
	\bibitem [{\citenamefont {Abundo}(2013)}]{Abu-2013}%
	\BibitemOpen
	\bibfield  {author} {\bibinfo {author} {\bibfnamefont {M.}~\bibnamefont
			{Abundo}},\ }\href@noop {} {\bibfield  {journal} {\bibinfo  {journal}
			{Methodol. Comput. Appl. Probab.}\ }\textbf {\bibinfo {volume} {15}},\
		\bibinfo {pages} {85} (\bibinfo {year} {2013})}\BibitemShut {NoStop}%
	\bibitem [{\citenamefont {Kearney}\ and\ \citenamefont
		{Majumdar}(2005)}]{KeaMaj-2005}%
	\BibitemOpen
	\bibfield  {author} {\bibinfo {author} {\bibfnamefont {M.~J.}\ \bibnamefont
			{Kearney}}\ and\ \bibinfo {author} {\bibfnamefont {S.~N.}\ \bibnamefont
			{Majumdar}},\ }\href@noop {} {\bibfield  {journal} {\bibinfo  {journal} {J.
				Phys. A: Math. Gen.}\ }\textbf {\bibinfo {volume} {38}},\ \bibinfo {pages}
		{4097} (\bibinfo {year} {2005})}\BibitemShut {NoStop}%
	\bibitem [{\citenamefont {Kearney}\ \emph {et~al.}(2007)\citenamefont
		{Kearney}, \citenamefont {Majumdar},\ and\ \citenamefont
		{Martin}}]{KeaMajMar-2007}%
	\BibitemOpen
	\bibfield  {author} {\bibinfo {author} {\bibfnamefont {M.~J.}\ \bibnamefont
			{Kearney}}, \bibinfo {author} {\bibfnamefont {S.~N.}\ \bibnamefont
			{Majumdar}},\ and\ \bibinfo {author} {\bibfnamefont {R.~J.}\ \bibnamefont
			{Martin}},\ }\href@noop {} {\bibfield  {journal} {\bibinfo  {journal} {J.
				Phys. A: Math. Theor.}\ }\textbf {\bibinfo {volume} {40}},\ \bibinfo {pages}
		{F863} (\bibinfo {year} {2007})}\BibitemShut {NoStop}%
	\bibitem [{\citenamefont {Kearney}\ and\ \citenamefont
		{Martin}(2016)}]{KeaMar-2016}%
	\BibitemOpen
	\bibfield  {author} {\bibinfo {author} {\bibfnamefont {M.~J.}\ \bibnamefont
			{Kearney}}\ and\ \bibinfo {author} {\bibfnamefont {R.~J.}\ \bibnamefont
			{Martin}},\ }\href@noop {} {\bibfield  {journal} {\bibinfo  {journal} {J.
				Phys. A: Math. Theor.}\ }\textbf {\bibinfo {volume} {49}},\ \bibinfo {pages}
		{195001} (\bibinfo {year} {2016})}\BibitemShut {NoStop}%
	\bibitem [{\citenamefont {Abundo}\ and\ \citenamefont
		{Del~Vescovo}(2017)}]{AbuDel-2017}%
	\BibitemOpen
	\bibfield  {author} {\bibinfo {author} {\bibfnamefont {M.}~\bibnamefont
			{Abundo}}\ and\ \bibinfo {author} {\bibfnamefont {D.}~\bibnamefont
			{Del~Vescovo}},\ }\href@noop {} {\bibfield  {journal} {\bibinfo  {journal}
			{Methodol. Comput. Appl. Probab.}\ }\textbf {\bibinfo {volume} {19}},\
		\bibinfo {pages} {985} (\bibinfo {year} {2017})}\BibitemShut {NoStop}%
	\bibitem [{\citenamefont {Abundo}\ and\ \citenamefont
		{Furia}(2019)}]{AbuFur-2019}%
	\BibitemOpen
	\bibfield  {author} {\bibinfo {author} {\bibfnamefont {M.}~\bibnamefont
			{Abundo}}\ and\ \bibinfo {author} {\bibfnamefont {S.}~\bibnamefont {Furia}},\
	}\href@noop {} {\bibfield  {journal} {\bibinfo  {journal} {Methodol. Comput.
				Appl. Probab.}\ }\textbf {\bibinfo {volume} {21}},\ \bibinfo {pages} {1283}
		(\bibinfo {year} {2019})}\BibitemShut {NoStop}%
	\bibitem [{\citenamefont {Kearney}\ and\ \citenamefont
		{Martin}(2021)}]{KeaMar-2021}%
	\BibitemOpen
	\bibfield  {author} {\bibinfo {author} {\bibfnamefont {M.~J.}\ \bibnamefont
			{Kearney}}\ and\ \bibinfo {author} {\bibfnamefont {R.~J.}\ \bibnamefont
			{Martin}},\ }\href@noop {} {\bibfield  {journal} {\bibinfo  {journal} {J.
				Phys. A: Math. Theor.}\ }\textbf {\bibinfo {volume} {54}},\ \bibinfo {pages}
		{055002} (\bibinfo {year} {2021})}\BibitemShut {NoStop}%
	\bibitem [{\citenamefont {Singh}\ and\ \citenamefont
		{Pal}(2022)}]{SinPal-2022}%
	\BibitemOpen
	\bibfield  {author} {\bibinfo {author} {\bibfnamefont {P.}~\bibnamefont
			{Singh}}\ and\ \bibinfo {author} {\bibfnamefont {A.}~\bibnamefont {Pal}},\
	}\href@noop {} {\bibfield  {journal} {\bibinfo  {journal} {J. Phys. A: Math.
				Theor.}\ }\textbf {\bibinfo {volume} {55}},\ \bibinfo {pages} {234001}
		(\bibinfo {year} {2022})}\BibitemShut {NoStop}%
	\bibitem [{\citenamefont {Meerson}\ and\ \citenamefont
		{Oshanin}(2022)}]{MeeOsh-2022}%
	\BibitemOpen
	\bibfield  {author} {\bibinfo {author} {\bibfnamefont {B.}~\bibnamefont
			{Meerson}}\ and\ \bibinfo {author} {\bibfnamefont {G.}~\bibnamefont
			{Oshanin}},\ }\href@noop {} {\bibfield  {journal} {\bibinfo  {journal} {Phys.
				Rev. E}\ }\textbf {\bibinfo {volume} {105}},\ \bibinfo {pages} {064137}
		(\bibinfo {year} {2022})}\BibitemShut {NoStop}%
	\bibitem [{\citenamefont {Meerson}(2023)}]{Mee-2023}%
	\BibitemOpen
	\bibfield  {author} {\bibinfo {author} {\bibfnamefont {B.}~\bibnamefont
			{Meerson}},\ }\href@noop {} {\bibfield  {journal} {\bibinfo  {journal} {Phys.
				Rev. E}\ }\textbf {\bibinfo {volume} {107}},\ \bibinfo {pages} {064122}
		(\bibinfo {year} {2023})}\BibitemShut {NoStop}%
	\bibitem [{\citenamefont {Abundo}(2023{\natexlab{a}})}]{Abu-2023}%
	\BibitemOpen
	\bibfield  {author} {\bibinfo {author} {\bibfnamefont {M.}~\bibnamefont
			{Abundo}},\ }\href@noop {} {\bibfield  {journal} {\bibinfo  {journal} {Stoch.
				Anal. Appl.}\ }\textbf {\bibinfo {volume} {41}},\ \bibinfo {pages} {358}
		(\bibinfo {year} {2023}{\natexlab{a}})}\BibitemShut {NoStop}%
	\bibitem [{\citenamefont {Pal}\ \emph {et~al.}(2023)\citenamefont {Pal},
		\citenamefont {Pal}, \citenamefont {Park},\ and\ \citenamefont
		{Lee}}]{PalPalPar-2023arX}%
	\BibitemOpen
	\bibfield  {author} {\bibinfo {author} {\bibfnamefont {P.~S.}\ \bibnamefont
			{Pal}}, \bibinfo {author} {\bibfnamefont {A.}~\bibnamefont {Pal}}, \bibinfo
		{author} {\bibfnamefont {H.}~\bibnamefont {Park}},\ and\ \bibinfo {author}
		{\bibfnamefont {J.~S.}\ \bibnamefont {Lee}},\ }\href@noop {} {\bibfield
		{journal} {\bibinfo  {journal} {arXiv:2305.04562}\ } (\bibinfo {year}
		{2023})}\BibitemShut {NoStop}%
	\bibitem [{\citenamefont {Dubey}\ and\ \citenamefont
		{Pal}(2023)}]{DubPal-2023}%
	\BibitemOpen
	\bibfield  {author} {\bibinfo {author} {\bibfnamefont {A.}~\bibnamefont
			{Dubey}}\ and\ \bibinfo {author} {\bibfnamefont {A.}~\bibnamefont {Pal}},\
	}\href@noop {} {\bibfield  {journal} {\bibinfo  {journal} {arXiv:2304.05226}\
		} (\bibinfo {year} {2023})}\BibitemShut {NoStop}%
	\bibitem [{\citenamefont {Redner}(2001)}]{Red}%
	\BibitemOpen
	\bibfield  {author} {\bibinfo {author} {\bibfnamefont {S.}~\bibnamefont
			{Redner}},\ }\href@noop {} {\emph {\bibinfo {title} {A guide to first-passage
				processes}}}\ (\bibinfo  {publisher} {Cambridge University Press},\ \bibinfo
	{address} {Cambridge},\ \bibinfo {year} {2001})\BibitemShut {NoStop}%
	\bibitem [{\citenamefont {Metzler}\ \emph {et~al.}(2014)\citenamefont
		{Metzler}, \citenamefont {Oshanin},\ and\ \citenamefont
		{Redner}}]{Metzler2014}%
	\BibitemOpen
	\bibinfo {editor} {\bibfnamefont {R.}~\bibnamefont {Metzler}}, \bibinfo
	{editor} {\bibfnamefont {G.}~\bibnamefont {Oshanin}},\ and\ \bibinfo {editor}
	{\bibfnamefont {S.}~\bibnamefont {Redner}},\ eds.,\ \href@noop {} {\emph
		{\bibinfo {title} {First-passage phenomena and their applications}}}\
	(\bibinfo  {publisher} {World Scientific},\ \bibinfo {address} {Singapore},\
	\bibinfo {year} {2014})\BibitemShut {NoStop}%
	\bibitem [{\citenamefont {Majumdar}\ and\ \citenamefont
		{Meerson}(2020)}]{MajMee-2020}%
	\BibitemOpen
	\bibfield  {author} {\bibinfo {author} {\bibfnamefont {S.~N.}\ \bibnamefont
			{Majumdar}}\ and\ \bibinfo {author} {\bibfnamefont {B.}~\bibnamefont
			{Meerson}},\ }\href@noop {} {\bibfield  {journal} {\bibinfo  {journal} {J.
				Stat. Mech.}\ }\textbf {\bibinfo {volume} {2020}},\ \bibinfo {pages} {023202}
		(\bibinfo {year} {2020})}\BibitemShut {NoStop}%
	\bibitem [{\citenamefont {Abundo}(2023{\natexlab{b}})}]{Abu-2023arXiv}%
	\BibitemOpen
	\bibfield  {author} {\bibinfo {author} {\bibfnamefont {M.}~\bibnamefont
			{Abundo}},\ }\href@noop {} {\bibfield  {journal} {\bibinfo  {journal}
			{arXiv:2307.12154}\ } (\bibinfo {year} {2023}{\natexlab{b}})}\BibitemShut
	{NoStop}%
	\bibitem [{\citenamefont {Kearney}(2004)}]{Kea-2004}%
	\BibitemOpen
	\bibfield  {author} {\bibinfo {author} {\bibfnamefont {M.~J.}\ \bibnamefont
			{Kearney}},\ }\href@noop {} {\bibfield  {journal} {\bibinfo  {journal} {J.
				Phys. A: Math. Gen.}\ }\textbf {\bibinfo {volume} {37}},\ \bibinfo {pages}
		{8421} (\bibinfo {year} {2004})}\BibitemShut {NoStop}%
	\bibitem [{\citenamefont {Reynolds}(2010)}]{Rey-2010}%
	\BibitemOpen
	\bibfield  {author} {\bibinfo {author} {\bibfnamefont {A.~M.}\ \bibnamefont
			{Reynolds}},\ }\href@noop {} {\bibfield  {journal} {\bibinfo  {journal} {J.
				R. Soc. Interface}\ }\textbf {\bibinfo {volume} {7}},\ \bibinfo {pages}
		{1753} (\bibinfo {year} {2010})}\BibitemShut {NoStop}%
	\bibitem [{\citenamefont {Dubey}\ and\ \citenamefont
		{Bandyopadhyay}(2018)}]{DubBan-2018}%
	\BibitemOpen
	\bibfield  {author} {\bibinfo {author} {\bibfnamefont {A.}~\bibnamefont
			{Dubey}}\ and\ \bibinfo {author} {\bibfnamefont {M.}~\bibnamefont
			{Bandyopadhyay}},\ }\href@noop {} {\bibfield  {journal} {\bibinfo  {journal}
			{Eur. Phys. J. B}\ }\textbf {\bibinfo {volume} {91}},\ \bibinfo {pages} {276}
		(\bibinfo {year} {2018})}\BibitemShut {NoStop}%
	\bibitem [{\citenamefont {Bandyopadhyay}\ \emph {et~al.}(2011)\citenamefont
		{Bandyopadhyay}, \citenamefont {Gupta},\ and\ \citenamefont
		{Segal}}]{BanGupSeg-2011}%
	\BibitemOpen
	\bibfield  {author} {\bibinfo {author} {\bibfnamefont {M.}~\bibnamefont
			{Bandyopadhyay}}, \bibinfo {author} {\bibfnamefont {S.}~\bibnamefont
			{Gupta}},\ and\ \bibinfo {author} {\bibfnamefont {D.}~\bibnamefont {Segal}},\
	}\href@noop {} {\bibfield  {journal} {\bibinfo  {journal} {Phys. Rev. E}\
		}\textbf {\bibinfo {volume} {83}},\ \bibinfo {pages} {031905} (\bibinfo
		{year} {2011})}\BibitemShut {NoStop}%
	\bibitem [{\citenamefont {Dean}\ and\ \citenamefont
		{Majumdar}(2001)}]{DeaMaj-2001}%
	\BibitemOpen
	\bibfield  {author} {\bibinfo {author} {\bibfnamefont {S.~N.}\ \bibnamefont
			{Dean}}\ and\ \bibinfo {author} {\bibfnamefont {S.~N.}\ \bibnamefont
			{Majumdar}},\ }\href@noop {} {\bibfield  {journal} {\bibinfo  {journal} {J.
				Phys. A: Math. Gen.}\ }\textbf {\bibinfo {volume} {34}},\ \bibinfo {pages}
		{L697} (\bibinfo {year} {2001})}\BibitemShut {NoStop}%
	\bibitem [{\citenamefont {Hammersley}(1961)}]{Hammersley1961}%
	\BibitemOpen
	\bibfield  {author} {\bibinfo {author} {\bibfnamefont {J.~M.}\ \bibnamefont
			{Hammersley}},\ }in\ \href@noop {} {\emph {\bibinfo {booktitle} {Proceedings
				of the fourth Berkeley Symposium on Mathematical statistics and probability,
				vol. 3}}},\ \bibinfo {editor} {edited by\ \bibinfo {editor} {\bibfnamefont
			{J.}~\bibnamefont {Neyman}}}\ (\bibinfo  {publisher} {University of
		California press},\ \bibinfo {address} {Berkeley and Los Angeles},\ \bibinfo
	{year} {1961})\ pp.\ \bibinfo {pages} {17--78}\BibitemShut {NoStop}%
	\bibitem [{\citenamefont {Kessler}\ and\ \citenamefont
		{Barkai}(2010)}]{KesBar-2010}%
	\BibitemOpen
	\bibfield  {author} {\bibinfo {author} {\bibfnamefont {D.~A.}\ \bibnamefont
			{Kessler}}\ and\ \bibinfo {author} {\bibfnamefont {E.}~\bibnamefont
			{Barkai}},\ }\href@noop {} {\bibfield  {journal} {\bibinfo  {journal} {Phys.
				Rev. Lett.}\ }\textbf {\bibinfo {volume} {105}},\ \bibinfo {pages} {120602}
		(\bibinfo {year} {2010})}\BibitemShut {NoStop}%
	\bibitem [{\citenamefont {Dechant}\ \emph {et~al.}(2011)\citenamefont
		{Dechant}, \citenamefont {Lutz}, \citenamefont {Barkai},\ and\ \citenamefont
		{Kessler}}]{DecLutBar-2011}%
	\BibitemOpen
	\bibfield  {author} {\bibinfo {author} {\bibfnamefont {A.}~\bibnamefont
			{Dechant}}, \bibinfo {author} {\bibfnamefont {E.}~\bibnamefont {Lutz}},
		\bibinfo {author} {\bibfnamefont {E.}~\bibnamefont {Barkai}},\ and\ \bibinfo
		{author} {\bibfnamefont {D.~A.}\ \bibnamefont {Kessler}},\ }\href@noop {}
	{\bibfield  {journal} {\bibinfo  {journal} {J. Stat. Phys.}\ }\textbf
		{\bibinfo {volume} {145}},\ \bibinfo {pages} {1524} (\bibinfo {year}
		{2011})}\BibitemShut {NoStop}%
	\bibitem [{\citenamefont {Hirschberg}\ \emph {et~al.}(2011)\citenamefont
		{Hirschberg}, \citenamefont {Mukamel},\ and\ \citenamefont
		{Sch\"{u}tz}}]{HirMukSch-2011}%
	\BibitemOpen
	\bibfield  {author} {\bibinfo {author} {\bibfnamefont {O.}~\bibnamefont
			{Hirschberg}}, \bibinfo {author} {\bibfnamefont {D.}~\bibnamefont
			{Mukamel}},\ and\ \bibinfo {author} {\bibfnamefont {G.~M.}\ \bibnamefont
			{Sch\"{u}tz}},\ }\href@noop {} {\bibfield  {journal} {\bibinfo  {journal}
			{Phys. Rev. E}\ }\textbf {\bibinfo {volume} {84}},\ \bibinfo {pages} {041111}
		(\bibinfo {year} {2011})}\BibitemShut {NoStop}%
	\bibitem [{\citenamefont {Martin}\ \emph {et~al.}(2011)\citenamefont {Martin},
		\citenamefont {Behn},\ and\ \citenamefont {Germano}}]{MarBehGer-2011}%
	\BibitemOpen
	\bibfield  {author} {\bibinfo {author} {\bibfnamefont {E.}~\bibnamefont
			{Martin}}, \bibinfo {author} {\bibfnamefont {U.}~\bibnamefont {Behn}},\ and\
		\bibinfo {author} {\bibfnamefont {G.}~\bibnamefont {Germano}},\ }\href@noop
	{} {\bibfield  {journal} {\bibinfo  {journal} {Phys. Rev. E}\ }\textbf
		{\bibinfo {volume} {83}},\ \bibinfo {pages} {051115} (\bibinfo {year}
		{2011})}\BibitemShut {NoStop}%
	\bibitem [{\citenamefont {Dechant}\ \emph {et~al.}(2012)\citenamefont
		{Dechant}, \citenamefont {Lutz}, \citenamefont {Kessler},\ and\ \citenamefont
		{Barkai}}]{DecLutBar-2012}%
	\BibitemOpen
	\bibfield  {author} {\bibinfo {author} {\bibfnamefont {A.}~\bibnamefont
			{Dechant}}, \bibinfo {author} {\bibfnamefont {E.}~\bibnamefont {Lutz}},
		\bibinfo {author} {\bibfnamefont {D.~A.}\ \bibnamefont {Kessler}},\ and\
		\bibinfo {author} {\bibfnamefont {E.}~\bibnamefont {Barkai}},\ }\href@noop {}
	{\bibfield  {journal} {\bibinfo  {journal} {Phys. Rev. E}\ }\textbf {\bibinfo
			{volume} {85}},\ \bibinfo {pages} {051124} (\bibinfo {year}
		{2012})}\BibitemShut {NoStop}%
	\bibitem [{\citenamefont {Ray}\ and\ \citenamefont
		{Reuveni}(2020)}]{RayReu-2020}%
	\BibitemOpen
	\bibfield  {author} {\bibinfo {author} {\bibfnamefont {S.}~\bibnamefont
			{Ray}}\ and\ \bibinfo {author} {\bibfnamefont {S.}~\bibnamefont {Reuveni}},\
	}\href@noop {} {\bibfield  {journal} {\bibinfo  {journal} {J. Chem. Phys.}\
		}\textbf {\bibinfo {volume} {152}},\ \bibinfo {pages} {234110} (\bibinfo
		{year} {2020})}\BibitemShut {NoStop}%
	\bibitem [{\citenamefont {Ryabov}\ \emph {et~al.}(2013)\citenamefont {Ryabov},
		\citenamefont {Dierl}, \citenamefont {Chvosta}, \citenamefont {Einax},\ and\
		\citenamefont {Maass}}]{RyaDieChv-2013}%
	\BibitemOpen
	\bibfield  {author} {\bibinfo {author} {\bibfnamefont {A.}~\bibnamefont
			{Ryabov}}, \bibinfo {author} {\bibfnamefont {M.}~\bibnamefont {Dierl}},
		\bibinfo {author} {\bibfnamefont {P.}~\bibnamefont {Chvosta}}, \bibinfo
		{author} {\bibfnamefont {M.}~\bibnamefont {Einax}},\ and\ \bibinfo {author}
		{\bibfnamefont {P.}~\bibnamefont {Maass}},\ }\href@noop {} {\bibfield
		{journal} {\bibinfo  {journal} {J. Phys. A: Math. Theor.}\ }\textbf {\bibinfo
			{volume} {46}},\ \bibinfo {pages} {075002} (\bibinfo {year}
		{2013})}\BibitemShut {NoStop}%
	\bibitem [{\citenamefont {Holubec}\ \emph {et~al.}(2015)\citenamefont
		{Holubec}, \citenamefont {Dierl}, \citenamefont {Einax}, \citenamefont
		{Maass}, \citenamefont {Chvosta},\ and\ \citenamefont
		{Ryabov}}]{HolDieEin-2015}%
	\BibitemOpen
	\bibfield  {author} {\bibinfo {author} {\bibfnamefont {V.}~\bibnamefont
			{Holubec}}, \bibinfo {author} {\bibfnamefont {M.}~\bibnamefont {Dierl}},
		\bibinfo {author} {\bibfnamefont {M.}~\bibnamefont {Einax}}, \bibinfo
		{author} {\bibfnamefont {P.}~\bibnamefont {Maass}}, \bibinfo {author}
		{\bibfnamefont {P.}~\bibnamefont {Chvosta}},\ and\ \bibinfo {author}
		{\bibfnamefont {A.}~\bibnamefont {Ryabov}},\ }\href@noop {} {\bibfield
		{journal} {\bibinfo  {journal} {Phys. Scr.}\ }\textbf {\bibinfo {volume}
			{2015}},\ \bibinfo {pages} {014024} (\bibinfo {year} {2015})}\BibitemShut
	{NoStop}%
	\bibitem [{\citenamefont {Paraguass\'{u}}\ and\ \citenamefont
		{Morgado}(2021)}]{ParMor-2021}%
	\BibitemOpen
	\bibfield  {author} {\bibinfo {author} {\bibfnamefont {P.~V.}\ \bibnamefont
			{Paraguass\'{u}}}\ and\ \bibinfo {author} {\bibfnamefont {W.~A.~M.}\
			\bibnamefont {Morgado}},\ }\href@noop {} {\bibfield  {journal} {\bibinfo
			{journal} {J. Stat. Mech.}\ }\textbf {\bibinfo {volume} {2021}},\ \bibinfo
		{pages} {023205} (\bibinfo {year} {2021})}\BibitemShut {NoStop}%
	\bibitem [{\citenamefont {Paraguass\'{u}}\ and\ \citenamefont
		{Morgado}(2022)}]{ParMor-2022}%
	\BibitemOpen
	\bibfield  {author} {\bibinfo {author} {\bibfnamefont {P.~V.}\ \bibnamefont
			{Paraguass\'{u}}}\ and\ \bibinfo {author} {\bibfnamefont {W.~A.~M.}\
			\bibnamefont {Morgado}},\ }\href@noop {} {\bibfield  {journal} {\bibinfo
			{journal} {Physica A}\ }\textbf {\bibinfo {volume} {588}},\ \bibinfo {pages}
		{126576} (\bibinfo {year} {2022})}\BibitemShut {NoStop}%
	\bibitem [{\citenamefont {Bray}(2000)}]{Bray-2000}%
	\BibitemOpen
	\bibfield  {author} {\bibinfo {author} {\bibfnamefont {A.~J.}\ \bibnamefont
			{Bray}},\ }\href@noop {} {\bibfield  {journal} {\bibinfo  {journal} {Phys.
				Rev. E}\ }\textbf {\bibinfo {volume} {62}},\ \bibinfo {pages} {103} (\bibinfo
		{year} {2000})}\BibitemShut {NoStop}%
	\bibitem [{\citenamefont {Chavanis}\ and\ \citenamefont
		{Lemou}(2007)}]{Cha-2007}%
	\BibitemOpen
	\bibfield  {author} {\bibinfo {author} {\bibfnamefont {P.~H.}\ \bibnamefont
			{Chavanis}}\ and\ \bibinfo {author} {\bibfnamefont {M.}~\bibnamefont
			{Lemou}},\ }\href@noop {} {\bibfield  {journal} {\bibinfo  {journal} {Eur.
				Phys. J. B}\ }\textbf {\bibinfo {volume} {59}},\ \bibinfo {pages} {217}
		(\bibinfo {year} {2007})}\BibitemShut {NoStop}%
	\bibitem [{\citenamefont {Bouchet}\ and\ \citenamefont
		{Dauxois}(2005)}]{BouDau-2005}%
	\BibitemOpen
	\bibfield  {author} {\bibinfo {author} {\bibfnamefont {F.}~\bibnamefont
			{Bouchet}}\ and\ \bibinfo {author} {\bibfnamefont {T.}~\bibnamefont
			{Dauxois}},\ }\href@noop {} {\bibfield  {journal} {\bibinfo  {journal} {Phys.
				Rev. E}\ }\textbf {\bibinfo {volume} {72}},\ \bibinfo {pages} {045103(R)}
		(\bibinfo {year} {2005})}\BibitemShut {NoStop}%
	\bibitem [{\citenamefont {Chavanis}\ and\ \citenamefont
		{Lemou}(2005)}]{Cha-2005}%
	\BibitemOpen
	\bibfield  {author} {\bibinfo {author} {\bibfnamefont {P.~H.}\ \bibnamefont
			{Chavanis}}\ and\ \bibinfo {author} {\bibfnamefont {M.}~\bibnamefont
			{Lemou}},\ }\href@noop {} {\bibfield  {journal} {\bibinfo  {journal} {Phys.
				Rev. E}\ }\textbf {\bibinfo {volume} {72}},\ \bibinfo {pages} {061106}
		(\bibinfo {year} {2005})}\BibitemShut {NoStop}%
	\bibitem [{\citenamefont {Campa}\ \emph {et~al.}(2009)\citenamefont {Campa},
		\citenamefont {Dauxois},\ and\ \citenamefont {Ruffo}}]{CamDauRuf-2009}%
	\BibitemOpen
	\bibfield  {author} {\bibinfo {author} {\bibfnamefont {A.}~\bibnamefont
			{Campa}}, \bibinfo {author} {\bibfnamefont {T.}~\bibnamefont {Dauxois}},\
		and\ \bibinfo {author} {\bibfnamefont {S.}~\bibnamefont {Ruffo}},\
	}\href@noop {} {\bibfield  {journal} {\bibinfo  {journal} {Phys. Rep.}\
		}\textbf {\bibinfo {volume} {480}},\ \bibinfo {pages} {57} (\bibinfo {year}
		{2009})}\BibitemShut {NoStop}%
	\bibitem [{\citenamefont {Manning}(1969)}]{Man-1969}%
	\BibitemOpen
	\bibfield  {author} {\bibinfo {author} {\bibfnamefont {G.~S.}\ \bibnamefont
			{Manning}},\ }\href@noop {} {\bibfield  {journal} {\bibinfo  {journal} {J.
				Chem. Phys.}\ }\textbf {\bibinfo {volume} {51}},\ \bibinfo {pages} {924}
		(\bibinfo {year} {1969})}\BibitemShut {NoStop}%
	\bibitem [{\citenamefont {Lo}\ \emph {et~al.}(2002)\citenamefont {Lo},
		\citenamefont {Nunes~Amaral}, \citenamefont {Havlin}, \citenamefont {Ivanov},
		\citenamefont {Penzel}, \citenamefont {Peter},\ and\ \citenamefont
		{Stanley}}]{LoNunHav-2002}%
	\BibitemOpen
	\bibfield  {author} {\bibinfo {author} {\bibfnamefont {C.-C.}\ \bibnamefont
			{Lo}}, \bibinfo {author} {\bibfnamefont {L.~A.}\ \bibnamefont
			{Nunes~Amaral}}, \bibinfo {author} {\bibfnamefont {S.}~\bibnamefont
			{Havlin}}, \bibinfo {author} {\bibfnamefont {P.~C.}\ \bibnamefont {Ivanov}},
		\bibinfo {author} {\bibfnamefont {T.}~\bibnamefont {Penzel}}, \bibinfo
		{author} {\bibfnamefont {J.-H.}\ \bibnamefont {Peter}},\ and\ \bibinfo
		{author} {\bibfnamefont {H.~E.}\ \bibnamefont {Stanley}},\ }\href@noop {}
	{\bibfield  {journal} {\bibinfo  {journal} {Europhys. Lett.}\ }\textbf
		{\bibinfo {volume} {57}},\ \bibinfo {pages} {625} (\bibinfo {year}
		{2002})}\BibitemShut {NoStop}%
	\bibitem [{\citenamefont {Castin}\ \emph {et~al.}(1991)\citenamefont {Castin},
		\citenamefont {Dalibard},\ and\ \citenamefont
		{Cohen-Tannoudji}}]{Castin1991}%
	\BibitemOpen
	\bibfield  {author} {\bibinfo {author} {\bibfnamefont {Y.}~\bibnamefont
			{Castin}}, \bibinfo {author} {\bibfnamefont {J.}~\bibnamefont {Dalibard}},\
		and\ \bibinfo {author} {\bibfnamefont {C.}~\bibnamefont {Cohen-Tannoudji}},\
	}in\ \href@noop {} {\emph {\bibinfo {booktitle} {Light Induced Kinetic
				Effects on Atoms, Ions and Molecules}}},\ \bibinfo {editor} {edited by\
		\bibinfo {editor} {\bibfnamefont {L.}~\bibnamefont {Moi}}, \bibinfo {editor}
		{\bibfnamefont {S.}~\bibnamefont {Gozzini}}, \bibinfo {editor} {\bibfnamefont
			{C.}~\bibnamefont {Gabbanini}}, \bibinfo {editor} {\bibfnamefont
			{E.}~\bibnamefont {Arimondo}},\ and\ \bibinfo {editor} {\bibfnamefont
			{F.}~\bibnamefont {Strumia}}}\ (\bibinfo  {publisher} {ETS Editrice},\
	\bibinfo {address} {Pisa},\ \bibinfo {year} {1991})\ pp.\ \bibinfo {pages}
	{5--24}\BibitemShut {NoStop}%
	\bibitem [{\citenamefont {Marksteiner}\ \emph {et~al.}(1996)\citenamefont
		{Marksteiner}, \citenamefont {Ellinger},\ and\ \citenamefont
		{Zoller}}]{MarEllZol-1996}%
	\BibitemOpen
	\bibfield  {author} {\bibinfo {author} {\bibfnamefont {S.}~\bibnamefont
			{Marksteiner}}, \bibinfo {author} {\bibfnamefont {K.}~\bibnamefont
			{Ellinger}},\ and\ \bibinfo {author} {\bibfnamefont {P.}~\bibnamefont
			{Zoller}},\ }\href@noop {} {\bibfield  {journal} {\bibinfo  {journal} {Phys.
				Rev. A}\ }\textbf {\bibinfo {volume} {53}},\ \bibinfo {pages} {3409}
		(\bibinfo {year} {1996})}\BibitemShut {NoStop}%
	\bibitem [{\citenamefont {Lutz}(2004)}]{Lut-2004}%
	\BibitemOpen
	\bibfield  {author} {\bibinfo {author} {\bibfnamefont {E.}~\bibnamefont
			{Lutz}},\ }\href@noop {} {\bibfield  {journal} {\bibinfo  {journal} {Phys.
				Rev. Lett.}\ }\textbf {\bibinfo {volume} {93}},\ \bibinfo {pages} {190602}
		(\bibinfo {year} {2004})}\BibitemShut {NoStop}%
	\bibitem [{\citenamefont {Douglas}\ \emph {et~al.}(2006)\citenamefont
		{Douglas}, \citenamefont {Gergamini},\ and\ \citenamefont
		{Renzoni}}]{DouGerRen-2006}%
	\BibitemOpen
	\bibfield  {author} {\bibinfo {author} {\bibfnamefont {P.}~\bibnamefont
			{Douglas}}, \bibinfo {author} {\bibfnamefont {S.}~\bibnamefont {Gergamini}},\
		and\ \bibinfo {author} {\bibfnamefont {F.}~\bibnamefont {Renzoni}},\
	}\href@noop {} {\bibfield  {journal} {\bibinfo  {journal} {Phys. Rev. Lett.}\
		}\textbf {\bibinfo {volume} {96}},\ \bibinfo {pages} {110601} (\bibinfo
		{year} {2006})}\BibitemShut {NoStop}%
	\bibitem [{\citenamefont {Kessler}\ and\ \citenamefont
		{Barkai}(2012)}]{KesBar-2012}%
	\BibitemOpen
	\bibfield  {author} {\bibinfo {author} {\bibfnamefont {D.~A.}\ \bibnamefont
			{Kessler}}\ and\ \bibinfo {author} {\bibfnamefont {E.}~\bibnamefont
			{Barkai}},\ }\href@noop {} {\bibfield  {journal} {\bibinfo  {journal} {Phys.
				Rev. Lett.}\ }\textbf {\bibinfo {volume} {108}},\ \bibinfo {pages} {230602}
		(\bibinfo {year} {2012})}\BibitemShut {NoStop}%
	\bibitem [{\citenamefont {Vezzani}\ \emph {et~al.}(2019)\citenamefont
		{Vezzani}, \citenamefont {Barkai},\ and\ \citenamefont
		{Burioni}}]{VezBarBur-2019}%
	\BibitemOpen
	\bibfield  {author} {\bibinfo {author} {\bibfnamefont {A.}~\bibnamefont
			{Vezzani}}, \bibinfo {author} {\bibfnamefont {E.}~\bibnamefont {Barkai}},\
		and\ \bibinfo {author} {\bibfnamefont {R.}~\bibnamefont {Burioni}},\
	}\href@noop {} {\bibfield  {journal} {\bibinfo  {journal} {Phys. Rev. E}\
		}\textbf {\bibinfo {volume} {100}},\ \bibinfo {pages} {012108} (\bibinfo
		{year} {2019})}\BibitemShut {NoStop}%
	\bibitem [{\citenamefont {Barkai}\ \emph {et~al.}(2014)\citenamefont {Barkai},
		\citenamefont {Aghion},\ and\ \citenamefont {Kessler}}]{BarAghKes-2014}%
	\BibitemOpen
	\bibfield  {author} {\bibinfo {author} {\bibfnamefont {E.}~\bibnamefont
			{Barkai}}, \bibinfo {author} {\bibfnamefont {E.}~\bibnamefont {Aghion}},\
		and\ \bibinfo {author} {\bibfnamefont {D.~A.}\ \bibnamefont {Kessler}},\
	}\href@noop {} {\bibfield  {journal} {\bibinfo  {journal} {Phys. Rev. X}\
		}\textbf {\bibinfo {volume} {4}},\ \bibinfo {pages} {021036} (\bibinfo {year}
		{2014})}\BibitemShut {NoStop}%
	\bibitem [{\citenamefont {Kessler}\ \emph {et~al.}(2014)\citenamefont
		{Kessler}, \citenamefont {Medalion},\ and\ \citenamefont
		{Barkai}}]{KesMedBar-2014}%
	\BibitemOpen
	\bibfield  {author} {\bibinfo {author} {\bibfnamefont {D.~A.}\ \bibnamefont
			{Kessler}}, \bibinfo {author} {\bibfnamefont {S.}~\bibnamefont {Medalion}},\
		and\ \bibinfo {author} {\bibfnamefont {E.}~\bibnamefont {Barkai}},\
	}\href@noop {} {\bibfield  {journal} {\bibinfo  {journal} {J. Stat. Phys.}\
		}\textbf {\bibinfo {volume} {156}},\ \bibinfo {pages} {686} (\bibinfo {year}
		{2014})}\BibitemShut {NoStop}%
	\bibitem [{\citenamefont {Gillis}(1956)}]{Gill-bias}%
	\BibitemOpen
	\bibfield  {author} {\bibinfo {author} {\bibfnamefont {J.}~\bibnamefont
			{Gillis}},\ }\href@noop {} {\bibfield  {journal} {\bibinfo  {journal} {Quart.
				J. Math.}\ }\textbf {\bibinfo {volume} {7}},\ \bibinfo {pages} {144}
		(\bibinfo {year} {1956})}\BibitemShut {NoStop}%
	\bibitem [{\citenamefont {Onofri}\ \emph {et~al.}(2020)\citenamefont {Onofri},
		\citenamefont {Pozzoli}, \citenamefont {Radice},\ and\ \citenamefont
		{Artuso}}]{OPRA}%
	\BibitemOpen
	\bibfield  {author} {\bibinfo {author} {\bibfnamefont {M.}~\bibnamefont
			{Onofri}}, \bibinfo {author} {\bibfnamefont {G.}~\bibnamefont {Pozzoli}},
		\bibinfo {author} {\bibfnamefont {M.}~\bibnamefont {Radice}},\ and\ \bibinfo
		{author} {\bibfnamefont {R.}~\bibnamefont {Artuso}},\ }\href@noop {}
	{\bibfield  {journal} {\bibinfo  {journal} {J. Stat. Mech.}\ }\textbf
		{\bibinfo {volume} {2020}},\ \bibinfo {pages} {113201} (\bibinfo {year}
		{2020})}\BibitemShut {NoStop}%
	\bibitem [{\citenamefont {Pozzoli}\ \emph {et~al.}(2020)\citenamefont
		{Pozzoli}, \citenamefont {Radice}, \citenamefont {Onofri},\ and\
		\citenamefont {Artuso}}]{PROA}%
	\BibitemOpen
	\bibfield  {author} {\bibinfo {author} {\bibfnamefont {P.}~\bibnamefont
			{Pozzoli}}, \bibinfo {author} {\bibfnamefont {M.}~\bibnamefont {Radice}},
		\bibinfo {author} {\bibfnamefont {M.}~\bibnamefont {Onofri}},\ and\ \bibinfo
		{author} {\bibfnamefont {R.}~\bibnamefont {Artuso}},\ }\href@noop {}
	{\bibfield  {journal} {\bibinfo  {journal} {Entropy}\ }\textbf {\bibinfo
			{volume} {22}},\ \bibinfo {pages} {1431} (\bibinfo {year}
		{2020})}\BibitemShut {NoStop}%
	\bibitem [{\citenamefont {Artuso}\ \emph {et~al.}(2022)\citenamefont {Artuso},
		\citenamefont {Onofri}, \citenamefont {Pozzoli},\ and\ \citenamefont
		{Radice}}]{AOPR}%
	\BibitemOpen
	\bibfield  {author} {\bibinfo {author} {\bibfnamefont {R.}~\bibnamefont
			{Artuso}}, \bibinfo {author} {\bibfnamefont {M.}~\bibnamefont {Onofri}},
		\bibinfo {author} {\bibfnamefont {G.}~\bibnamefont {Pozzoli}},\ and\ \bibinfo
		{author} {\bibfnamefont {M.}~\bibnamefont {Radice}},\ }\href@noop {}
	{\bibfield  {journal} {\bibinfo  {journal} {J. Stat. Mech.}\ }\textbf
		{\bibinfo {volume} {2022}},\ \bibinfo {pages} {103209} (\bibinfo {year}
		{2022})}\BibitemShut {NoStop}%
	\bibitem [{\citenamefont {Radice}(2022)}]{RAD-2022-Gill}%
	\BibitemOpen
	\bibfield  {author} {\bibinfo {author} {\bibfnamefont {M.}~\bibnamefont
			{Radice}},\ }\href@noop {} {\bibfield  {journal} {\bibinfo  {journal} {J.
				Stat. Mech.}\ }\textbf {\bibinfo {volume} {2022}},\ \bibinfo {pages} {123206}
		(\bibinfo {year} {2022})}\BibitemShut {NoStop}%
	\bibitem [{\citenamefont {Zodage}\ \emph {et~al.}(2023)\citenamefont {Zodage},
		\citenamefont {Allen}, \citenamefont {Evans},\ and\ \citenamefont
		{Majumdar}}]{ZodAllEva-2023}%
	\BibitemOpen
	\bibfield  {author} {\bibinfo {author} {\bibfnamefont {A.}~\bibnamefont
			{Zodage}}, \bibinfo {author} {\bibfnamefont {R.~J.}\ \bibnamefont {Allen}},
		\bibinfo {author} {\bibfnamefont {M.}~\bibnamefont {Evans}},\ and\ \bibinfo
		{author} {\bibfnamefont {S.~N.}\ \bibnamefont {Majumdar}},\ }\href@noop {}
	{\bibfield  {journal} {\bibinfo  {journal} {J. Stat. Mech.}\ }\textbf
		{\bibinfo {volume} {2023}},\ \bibinfo {pages} {033211} (\bibinfo {year}
		{2023})}\BibitemShut {NoStop}%
	\bibitem [{\citenamefont {Lamperti}(1960)}]{Lam-1960}%
	\BibitemOpen
	\bibfield  {author} {\bibinfo {author} {\bibfnamefont {J.}~\bibnamefont
			{Lamperti}},\ }\href@noop {} {\bibfield  {journal} {\bibinfo  {journal} {J.
				Math. Anal. and Appl.}\ }\textbf {\bibinfo {volume} {1}},\ \bibinfo {pages}
		{314} (\bibinfo {year} {1960})}\BibitemShut {NoStop}%
	\bibitem [{\citenamefont {Lamperti}(1963)}]{Lam-1963}%
	\BibitemOpen
	\bibfield  {author} {\bibinfo {author} {\bibfnamefont {J.}~\bibnamefont
			{Lamperti}},\ }\href@noop {} {\bibfield  {journal} {\bibinfo  {journal} {J.
				Math. Anal. and Appl.}\ }\textbf {\bibinfo {volume} {7}},\ \bibinfo {pages}
		{127} (\bibinfo {year} {1963})}\BibitemShut {NoStop}%
	\bibitem [{\citenamefont {Hughes}(1995)}]{Hug-I}%
	\BibitemOpen
	\bibfield  {author} {\bibinfo {author} {\bibfnamefont {B.~D.}\ \bibnamefont
			{Hughes}},\ }\href@noop {} {\emph {\bibinfo {title} {Random walks in random
				environments. {V}olume 1: Random walks}}},\ Vol.~\bibinfo {volume} {1}\
	(\bibinfo  {publisher} {Clarendon press},\ \bibinfo {address} {Oxford},\
	\bibinfo {year} {1995})\BibitemShut {NoStop}%
	\bibitem [{\citenamefont {Leibovich}\ and\ \citenamefont
		{Barkai}(2019)}]{Lei-Bar}%
	\BibitemOpen
	\bibfield  {author} {\bibinfo {author} {\bibfnamefont {N.}~\bibnamefont
			{Leibovich}}\ and\ \bibinfo {author} {\bibfnamefont {E.}~\bibnamefont
			{Barkai}},\ }\href@noop {} {\bibfield  {journal} {\bibinfo  {journal} {Phys.
				Rev. E}\ }\textbf {\bibinfo {volume} {99}},\ \bibinfo {pages} {042138}
		(\bibinfo {year} {2019})}\BibitemShut {NoStop}%
	\bibitem [{\citenamefont {Eliazar}(2021{\natexlab{a}})}]{Eli-2021EPL}%
	\BibitemOpen
	\bibfield  {author} {\bibinfo {author} {\bibfnamefont {I.}~\bibnamefont
			{Eliazar}},\ }\href@noop {} {\bibfield  {journal} {\bibinfo  {journal} {EPL}\
		}\textbf {\bibinfo {volume} {146}},\ \bibinfo {pages} {40002} (\bibinfo
		{year} {2021}{\natexlab{a}})}\BibitemShut {NoStop}%
	\bibitem [{\citenamefont {Eliazar}(2021{\natexlab{b}})}]{Eli-2021}%
	\BibitemOpen
	\bibfield  {author} {\bibinfo {author} {\bibfnamefont {I.}~\bibnamefont
			{Eliazar}},\ }\href@noop {} {\bibfield  {journal} {\bibinfo  {journal} {J.
				Phys. A: Math. Theor.}\ }\textbf {\bibinfo {volume} {54}},\ \bibinfo {pages}
		{35LT01} (\bibinfo {year} {2021}{\natexlab{b}})}\BibitemShut {NoStop}%
	\bibitem [{\citenamefont {Cherstvy}\ \emph {et~al.}(2013)\citenamefont
		{Cherstvy}, \citenamefont {Chechkin},\ and\ \citenamefont
		{Metzler}}]{CheCheMet-2013}%
	\BibitemOpen
	\bibfield  {author} {\bibinfo {author} {\bibfnamefont {A.~G.}\ \bibnamefont
			{Cherstvy}}, \bibinfo {author} {\bibfnamefont {A.~V.}\ \bibnamefont
			{Chechkin}},\ and\ \bibinfo {author} {\bibfnamefont {R.}~\bibnamefont
			{Metzler}},\ }\href@noop {} {\bibfield  {journal} {\bibinfo  {journal} {New
				J. Phys.}\ }\textbf {\bibinfo {volume} {15}},\ \bibinfo {pages} {083039}
		(\bibinfo {year} {2013})}\BibitemShut {NoStop}%
	\bibitem [{\citenamefont {Cherstvy}\ and\ \citenamefont
		{Metzler}(2014)}]{CheMet-2014}%
	\BibitemOpen
	\bibfield  {author} {\bibinfo {author} {\bibfnamefont {A.~G.}\ \bibnamefont
			{Cherstvy}}\ and\ \bibinfo {author} {\bibfnamefont {R.}~\bibnamefont
			{Metzler}},\ }\href@noop {} {\bibfield  {journal} {\bibinfo  {journal} {Phys.
				Rev. E}\ }\textbf {\bibinfo {volume} {90}},\ \bibinfo {pages} {012134}
		(\bibinfo {year} {2014})}\BibitemShut {NoStop}%
	\bibitem [{\citenamefont {Cherstvy}\ \emph {et~al.}(2014)\citenamefont
		{Cherstvy}, \citenamefont {Chechkin},\ and\ \citenamefont
		{Metzler}}]{CheCheMet-2014}%
	\BibitemOpen
	\bibfield  {author} {\bibinfo {author} {\bibfnamefont {A.~G.}\ \bibnamefont
			{Cherstvy}}, \bibinfo {author} {\bibfnamefont {A.~V.}\ \bibnamefont
			{Chechkin}},\ and\ \bibinfo {author} {\bibfnamefont {R.}~\bibnamefont
			{Metzler}},\ }\href@noop {} {\bibfield  {journal} {\bibinfo  {journal} {J.
				Phys. A: Math. Theor.}\ }\textbf {\bibinfo {volume} {47}},\ \bibinfo {pages}
		{485002} (\bibinfo {year} {2014})}\BibitemShut {NoStop}%
	\bibitem [{\citenamefont {Bressloff}\ and\ \citenamefont
		{Lawley}(2017)}]{BreLaw-2017}%
	\BibitemOpen
	\bibfield  {author} {\bibinfo {author} {\bibfnamefont {P.~C.}\ \bibnamefont
			{Bressloff}}\ and\ \bibinfo {author} {\bibfnamefont {S.~D.}\ \bibnamefont
			{Lawley}},\ }\href@noop {} {\bibfield  {journal} {\bibinfo  {journal} {Phys.
				Rev. E}\ }\textbf {\bibinfo {volume} {95}},\ \bibinfo {pages} {060101}
		(\bibinfo {year} {2017})}\BibitemShut {NoStop}%
	\bibitem [{\citenamefont {Wang}\ \emph {et~al.}(2021)\citenamefont {Wang},
		\citenamefont {Cherstvy}, \citenamefont {Kantz}, \citenamefont {Metzler},\
		and\ \citenamefont {Sokolov}}]{WanCheKan-2021}%
	\BibitemOpen
	\bibfield  {author} {\bibinfo {author} {\bibfnamefont {W.}~\bibnamefont
			{Wang}}, \bibinfo {author} {\bibfnamefont {A.~G.}\ \bibnamefont {Cherstvy}},
		\bibinfo {author} {\bibfnamefont {H.}~\bibnamefont {Kantz}}, \bibinfo
		{author} {\bibfnamefont {R.}~\bibnamefont {Metzler}},\ and\ \bibinfo {author}
		{\bibfnamefont {I.~M.}\ \bibnamefont {Sokolov}},\ }\href@noop {} {\bibfield
		{journal} {\bibinfo  {journal} {Phys. Rev. E}\ }\textbf {\bibinfo {volume}
			{104}},\ \bibinfo {pages} {024105} (\bibinfo {year} {2021})}\BibitemShut
	{NoStop}%
	\bibitem [{\citenamefont {Singh}(2022)}]{Sin-2022}%
	\BibitemOpen
	\bibfield  {author} {\bibinfo {author} {\bibfnamefont {P.}~\bibnamefont
			{Singh}},\ }\href@noop {} {\bibfield  {journal} {\bibinfo  {journal} {Phys.
				Rev. E}\ }\textbf {\bibinfo {volume} {105}},\ \bibinfo {pages} {024113}
		(\bibinfo {year} {2022})}\BibitemShut {NoStop}%
	\bibitem [{\citenamefont {Sandev}\ \emph {et~al.}(2022)\citenamefont {Sandev},
		\citenamefont {Domazetoski}, \citenamefont {Kocarev}, \citenamefont
		{Metzler},\ and\ \citenamefont {Chechkin}}]{SanDomKoc-2022}%
	\BibitemOpen
	\bibfield  {author} {\bibinfo {author} {\bibfnamefont {T.}~\bibnamefont
			{Sandev}}, \bibinfo {author} {\bibfnamefont {V.}~\bibnamefont {Domazetoski}},
		\bibinfo {author} {\bibfnamefont {L.}~\bibnamefont {Kocarev}}, \bibinfo
		{author} {\bibfnamefont {R.}~\bibnamefont {Metzler}},\ and\ \bibinfo {author}
		{\bibfnamefont {A.}~\bibnamefont {Chechkin}},\ }\href@noop {} {\bibfield
		{journal} {\bibinfo  {journal} {J. Phys. A: Math. Theor.}\ }\textbf {\bibinfo
			{volume} {55}},\ \bibinfo {pages} {074003} (\bibinfo {year}
		{2022})}\BibitemShut {NoStop}%
	\bibitem [{\citenamefont {K\"{u}hn}\ \emph {et~al.}(2011)\citenamefont
		{K\"{u}hn}, \citenamefont {Ihalainen}, \citenamefont {Hyv\"{a}luoma},
		\citenamefont {Dross}, \citenamefont {Willman}, \citenamefont {Langowski},
		\citenamefont {Vihinen-Ranta},\ and\ \citenamefont
		{Timonen}}]{KuhIhaHyy-2011}%
	\BibitemOpen
	\bibfield  {author} {\bibinfo {author} {\bibfnamefont {T.}~\bibnamefont
			{K\"{u}hn}}, \bibinfo {author} {\bibfnamefont {T.~O.}\ \bibnamefont
			{Ihalainen}}, \bibinfo {author} {\bibfnamefont {J.}~\bibnamefont
			{Hyv\"{a}luoma}}, \bibinfo {author} {\bibfnamefont {N.}~\bibnamefont
			{Dross}}, \bibinfo {author} {\bibfnamefont {S.~F.}\ \bibnamefont {Willman}},
		\bibinfo {author} {\bibfnamefont {J.}~\bibnamefont {Langowski}}, \bibinfo
		{author} {\bibfnamefont {M.}~\bibnamefont {Vihinen-Ranta}},\ and\ \bibinfo
		{author} {\bibfnamefont {J.}~\bibnamefont {Timonen}},\ }\href@noop {}
	{\bibfield  {journal} {\bibinfo  {journal} {{PL}o{S} {ONE}}\ }\textbf
		{\bibinfo {volume} {6}},\ \bibinfo {pages} {e22962} (\bibinfo {year}
		{2011})}\BibitemShut {NoStop}%
	\bibitem [{\citenamefont {Pieprzyk}\ \emph {et~al.}(2016)\citenamefont
		{Pieprzyk}, \citenamefont {Heyes},\ and\ \citenamefont
		{Bra\'{n}ka}}]{PieHeyBra-2016}%
	\BibitemOpen
	\bibfield  {author} {\bibinfo {author} {\bibfnamefont {S.}~\bibnamefont
			{Pieprzyk}}, \bibinfo {author} {\bibfnamefont {D.~M.}\ \bibnamefont
			{Heyes}},\ and\ \bibinfo {author} {\bibfnamefont {A.~C.}\ \bibnamefont
			{Bra\'{n}ka}},\ }\href@noop {} {\bibfield  {journal} {\bibinfo  {journal}
			{Biomicrofluidics}\ }\textbf {\bibinfo {volume} {10}},\ \bibinfo {pages}
		{054118} (\bibinfo {year} {2016})}\BibitemShut {NoStop}%
	\bibitem [{\citenamefont {Berezhkovskii}\ and\ \citenamefont
		{Makarov}(2017)}]{BerMak-2017}%
	\BibitemOpen
	\bibfield  {author} {\bibinfo {author} {\bibfnamefont {A.~M.}\ \bibnamefont
			{Berezhkovskii}}\ and\ \bibinfo {author} {\bibfnamefont {D.~E.}\ \bibnamefont
			{Makarov}},\ }\href@noop {} {\bibfield  {journal} {\bibinfo  {journal} {J.
				Chem. Phys.}\ }\textbf {\bibinfo {volume} {147}},\ \bibinfo {pages} {201102}
		(\bibinfo {year} {2017})}\BibitemShut {NoStop}%
	\bibitem [{\citenamefont {dos Santos}\ \emph {et~al.}(2020)\citenamefont {dos
			Santos}, \citenamefont {Dornelas}, \citenamefont {Colombo},\ and\
		\citenamefont {Anteneodo}}]{DosDorCol-2020}%
	\BibitemOpen
	\bibfield  {author} {\bibinfo {author} {\bibfnamefont {M.~A.~F.}\
			\bibnamefont {dos Santos}}, \bibinfo {author} {\bibfnamefont
			{V.}~\bibnamefont {Dornelas}}, \bibinfo {author} {\bibfnamefont {E.~H.}\
			\bibnamefont {Colombo}},\ and\ \bibinfo {author} {\bibfnamefont
			{C.}~\bibnamefont {Anteneodo}},\ }\href@noop {} {\bibfield  {journal}
		{\bibinfo  {journal} {Phys. Rev. E}\ }\textbf {\bibinfo {volume} {102}},\
		\bibinfo {pages} {042139} (\bibinfo {year} {2020})}\BibitemShut {NoStop}%
	\bibitem [{\citenamefont {Oksendal}(2013)}]{Oks}%
	\BibitemOpen
	\bibfield  {author} {\bibinfo {author} {\bibfnamefont {B.}~\bibnamefont
			{Oksendal}},\ }\href@noop {} {\emph {\bibinfo {title} {Stochastic
				Differential Equations: An Introduction with Applications}}}\ (\bibinfo
	{publisher} {Springer science \& business media},\ \bibinfo {address}
	{Berlin},\ \bibinfo {year} {2013})\BibitemShut {NoStop}%
	\bibitem [{\citenamefont {Dentz}\ \emph {et~al.}(2004)\citenamefont {Dentz},
		\citenamefont {Cortis}, \citenamefont {Scher},\ and\ \citenamefont
		{Berkowitz}}]{DenCorSch2004}%
	\BibitemOpen
	\bibfield  {author} {\bibinfo {author} {\bibfnamefont {M.}~\bibnamefont
			{Dentz}}, \bibinfo {author} {\bibfnamefont {A.}~\bibnamefont {Cortis}},
		\bibinfo {author} {\bibfnamefont {H.}~\bibnamefont {Scher}},\ and\ \bibinfo
		{author} {\bibfnamefont {B.}~\bibnamefont {Berkowitz}},\ }\href@noop {}
	{\bibfield  {journal} {\bibinfo  {journal} {Adv. Water Resources}\ }\textbf
		{\bibinfo {volume} {27}},\ \bibinfo {pages} {155} (\bibinfo {year}
		{2004})}\BibitemShut {NoStop}%
	\bibitem [{\citenamefont {Richardson}(1926)}]{Ric-1926}%
	\BibitemOpen
	\bibfield  {author} {\bibinfo {author} {\bibfnamefont {L.~F.}\ \bibnamefont
			{Richardson}},\ }\href@noop {} {\bibfield  {journal} {\bibinfo  {journal}
			{Proc. R. Soc. Lond. A}\ }\textbf {\bibinfo {volume} {110}},\ \bibinfo
		{pages} {709} (\bibinfo {year} {1926})}\BibitemShut {NoStop}%
	\bibitem [{\citenamefont {West}\ \emph {et~al.}(1979)\citenamefont {West},
		\citenamefont {Bulsara}, \citenamefont {Lindenberg}, \citenamefont
		{Seshadri},\ and\ \citenamefont {Shuler}}]{WesBulLin-1979}%
	\BibitemOpen
	\bibfield  {author} {\bibinfo {author} {\bibfnamefont {B.~J.}\ \bibnamefont
			{West}}, \bibinfo {author} {\bibfnamefont {A.~R.}\ \bibnamefont {Bulsara}},
		\bibinfo {author} {\bibfnamefont {K.}~\bibnamefont {Lindenberg}}, \bibinfo
		{author} {\bibfnamefont {V.}~\bibnamefont {Seshadri}},\ and\ \bibinfo
		{author} {\bibfnamefont {K.~E.}\ \bibnamefont {Shuler}},\ }\href@noop {}
	{\bibfield  {journal} {\bibinfo  {journal} {Physica A}\ }\textbf {\bibinfo
			{volume} {97}},\ \bibinfo {pages} {211} (\bibinfo {year} {1979})}\BibitemShut
	{NoStop}%
	\bibitem [{\citenamefont {Gardiner}(2003)}]{Gar}%
	\BibitemOpen
	\bibfield  {author} {\bibinfo {author} {\bibfnamefont {C.~W.}\ \bibnamefont
			{Gardiner}},\ }\href@noop {} {\emph {\bibinfo {title} {Handbook of Stochastic
				Methods for Physics, Chemistry and the Natural Sciences}}}\ (\bibinfo
	{publisher} {Springer},\ \bibinfo {address} {Berlin},\ \bibinfo {year}
	{2003})\BibitemShut {NoStop}%
	\bibitem [{\citenamefont {Sokolov}(2010)}]{Sok-2010}%
	\BibitemOpen
	\bibfield  {author} {\bibinfo {author} {\bibfnamefont {I.~M.}\ \bibnamefont
			{Sokolov}},\ }\href@noop {} {\bibfield  {journal} {\bibinfo  {journal} {Chem.
				Phys.}\ }\textbf {\bibinfo {volume} {375}},\ \bibinfo {pages} {359} (\bibinfo
		{year} {2010})}\BibitemShut {NoStop}%
	\bibitem [{\citenamefont {Mannella}\ and\ \citenamefont
		{McClintock}(2012)}]{ManMcClin-2012}%
	\BibitemOpen
	\bibfield  {author} {\bibinfo {author} {\bibfnamefont {R.}~\bibnamefont
			{Mannella}}\ and\ \bibinfo {author} {\bibfnamefont {P.~V.~E.}\ \bibnamefont
			{McClintock}},\ }\href@noop {} {\bibfield  {journal} {\bibinfo  {journal}
			{Fluct. Noise Lett.}\ }\textbf {\bibinfo {volume} {11}},\ \bibinfo {pages}
		{1240010} (\bibinfo {year} {2012})}\BibitemShut {NoStop}%
	\bibitem [{\citenamefont {Vaccario}\ \emph {et~al.}(2015)\citenamefont
		{Vaccario}, \citenamefont {Antoine},\ and\ \citenamefont
		{Talbot}}]{VacAntTal-2015}%
	\BibitemOpen
	\bibfield  {author} {\bibinfo {author} {\bibfnamefont {G.}~\bibnamefont
			{Vaccario}}, \bibinfo {author} {\bibfnamefont {C.}~\bibnamefont {Antoine}},\
		and\ \bibinfo {author} {\bibfnamefont {J.}~\bibnamefont {Talbot}},\
	}\href@noop {} {\bibfield  {journal} {\bibinfo  {journal} {Phys. Rev. Lett.}\
		}\textbf {\bibinfo {volume} {115}},\ \bibinfo {pages} {240601} (\bibinfo
		{year} {2015})}\BibitemShut {NoStop}%
	\bibitem [{\citenamefont {It\^{o}}(1944)}]{Ito-1944}%
	\BibitemOpen
	\bibfield  {author} {\bibinfo {author} {\bibfnamefont {K.}~\bibnamefont
			{It\^{o}}},\ }\href@noop {} {\bibfield  {journal} {\bibinfo  {journal} {Proc.
				Imp. Acad.}\ }\textbf {\bibinfo {volume} {20}},\ \bibinfo {pages} {519}
		(\bibinfo {year} {1944})}\BibitemShut {NoStop}%
	\bibitem [{\citenamefont {Stratonovich}(1966)}]{Stra-1966}%
	\BibitemOpen
	\bibfield  {author} {\bibinfo {author} {\bibfnamefont {R.~L.}\ \bibnamefont
			{Stratonovich}},\ }\href@noop {} {\bibfield  {journal} {\bibinfo  {journal}
			{SIAM J. Control}\ }\textbf {\bibinfo {volume} {4}},\ \bibinfo {pages} {362}
		(\bibinfo {year} {1966})}\BibitemShut {NoStop}%
	\bibitem [{\citenamefont {H\"{a}nggi}(1982)}]{Han-1982}%
	\BibitemOpen
	\bibfield  {author} {\bibinfo {author} {\bibfnamefont {P.}~\bibnamefont
			{H\"{a}nggi}},\ }\href@noop {} {\bibfield  {journal} {\bibinfo  {journal}
			{Phys. Rev. A}\ }\textbf {\bibinfo {volume} {25}},\ \bibinfo {pages} {1130}
		(\bibinfo {year} {1982})}\BibitemShut {NoStop}%
	\bibitem [{\citenamefont {Klimontovich}(1990)}]{Kli-1990}%
	\BibitemOpen
	\bibfield  {author} {\bibinfo {author} {\bibfnamefont {Y.~L.}\ \bibnamefont
			{Klimontovich}},\ }\href@noop {} {\bibfield  {journal} {\bibinfo  {journal}
			{Physica A}\ }\textbf {\bibinfo {volume} {163}},\ \bibinfo {pages} {515}
		(\bibinfo {year} {1990})}\BibitemShut {NoStop}%
	\bibitem [{\citenamefont {Olver}\ \emph {et~al.}(2010)\citenamefont {Olver},
		\citenamefont {Lozier}, \citenamefont {Boisvert},\ and\ \citenamefont
		{Clark}}]{NIST}%
	\BibitemOpen
	\bibinfo {editor} {\bibfnamefont {F.~W.~J.}\ \bibnamefont {Olver}}, \bibinfo
	{editor} {\bibfnamefont {D.~W.}\ \bibnamefont {Lozier}}, \bibinfo {editor}
	{\bibfnamefont {R.~F.}\ \bibnamefont {Boisvert}},\ and\ \bibinfo {editor}
	{\bibfnamefont {C.~W.}\ \bibnamefont {Clark}},\ eds.,\ \href@noop {} {\emph
		{\bibinfo {title} {NIST Handbook of Mathematical Functions}}}\ (\bibinfo
	{publisher} {Cambridge University Press},\ \bibinfo {address} {Cambridge},\
	\bibinfo {year} {2010})\BibitemShut {NoStop}%
	\bibitem [{\citenamefont {dos Santos}\ \emph {et~al.}(2022)\citenamefont {dos
			Santos}, \citenamefont {Menon~Jr.},\ and\ \citenamefont
		{Anteneodo}}]{DosMenAnt-2022}%
	\BibitemOpen
	\bibfield  {author} {\bibinfo {author} {\bibfnamefont {M.~A.~F.}\
			\bibnamefont {dos Santos}}, \bibinfo {author} {\bibfnamefont
			{L.}~\bibnamefont {Menon~Jr.}},\ and\ \bibinfo {author} {\bibfnamefont
			{C.}~\bibnamefont {Anteneodo}},\ }\href@noop {} {\bibfield  {journal}
		{\bibinfo  {journal} {Phys. Rev. E}\ }\textbf {\bibinfo {volume} {106}},\
		\bibinfo {pages} {044113} (\bibinfo {year} {2022})}\BibitemShut {NoStop}%
	\bibitem [{\citenamefont {Zoia}\ \emph {et~al.}(2009)\citenamefont {Zoia},
		\citenamefont {Rosso},\ and\ \citenamefont {Majumdar}}]{ZoiRosMaj-2009}%
	\BibitemOpen
	\bibfield  {author} {\bibinfo {author} {\bibfnamefont {A.}~\bibnamefont
			{Zoia}}, \bibinfo {author} {\bibfnamefont {A.}~\bibnamefont {Rosso}},\ and\
		\bibinfo {author} {\bibfnamefont {S.~N.}\ \bibnamefont {Majumdar}},\
	}\href@noop {} {\bibfield  {journal} {\bibinfo  {journal} {Phys. Rev. Lett.}\
		}\textbf {\bibinfo {volume} {102}},\ \bibinfo {pages} {120602} (\bibinfo
		{year} {2009})}\BibitemShut {NoStop}%
	\bibitem [{\citenamefont {Majumdar}\ \emph {et~al.}(2010)\citenamefont
		{Majumdar}, \citenamefont {Rosso},\ and\ \citenamefont
		{Zoia}}]{MajRosZoi-2010}%
	\BibitemOpen
	\bibfield  {author} {\bibinfo {author} {\bibfnamefont {S.~N.}\ \bibnamefont
			{Majumdar}}, \bibinfo {author} {\bibfnamefont {A.}~\bibnamefont {Rosso}},\
		and\ \bibinfo {author} {\bibfnamefont {A.}~\bibnamefont {Zoia}},\ }\href@noop
	{} {\bibfield  {journal} {\bibinfo  {journal} {Phys. Rev. Lett.}\ }\textbf
		{\bibinfo {volume} {104}},\ \bibinfo {pages} {020602} (\bibinfo {year}
		{2010})}\BibitemShut {NoStop}%
	\bibitem [{\citenamefont {Majumdar}(2010)}]{Maj-2010}%
	\BibitemOpen
	\bibfield  {author} {\bibinfo {author} {\bibfnamefont {S.~N.}\ \bibnamefont
			{Majumdar}},\ }\href@noop {} {\bibfield  {journal} {\bibinfo  {journal}
			{Physica A}\ }\textbf {\bibinfo {volume} {389}},\ \bibinfo {pages} {4299}
		(\bibinfo {year} {2010})}\BibitemShut {NoStop}%
	\bibitem [{\citenamefont {Majumdar}\ \emph {et~al.}(2020)\citenamefont
		{Majumdar}, \citenamefont {Pal},\ and\ \citenamefont
		{Schehr}}]{MajPalSch-2020}%
	\BibitemOpen
	\bibfield  {author} {\bibinfo {author} {\bibfnamefont {S.~N.}\ \bibnamefont
			{Majumdar}}, \bibinfo {author} {\bibfnamefont {A.}~\bibnamefont {Pal}},\ and\
		\bibinfo {author} {\bibfnamefont {G.}~\bibnamefont {Schehr}},\ }\href@noop {}
	{\bibfield  {journal} {\bibinfo  {journal} {Phys. Rep.}\ }\textbf {\bibinfo
			{volume} {840}},\ \bibinfo {pages} {1} (\bibinfo {year} {2020})}\BibitemShut
	{NoStop}%
	\bibitem [{\citenamefont {Hartich}\ and\ \citenamefont
		{Godec}(2019)}]{HarGod-2019}%
	\BibitemOpen
	\bibfield  {author} {\bibinfo {author} {\bibfnamefont {D.}~\bibnamefont
			{Hartich}}\ and\ \bibinfo {author} {\bibfnamefont {A.}~\bibnamefont
			{Godec}},\ }\href@noop {} {\bibfield  {journal} {\bibinfo  {journal} {J.
				Phys. A: Math. Theor.}\ }\textbf {\bibinfo {volume} {52}},\ \bibinfo {pages}
		{244001} (\bibinfo {year} {2019})}\BibitemShut {NoStop}%
	\bibitem [{\citenamefont {H\"{o}ll}\ and\ \citenamefont
		{Barkai}(2020)}]{HolBar-2020}%
	\BibitemOpen
	\bibfield  {author} {\bibinfo {author} {\bibfnamefont {M.}~\bibnamefont
			{H\"{o}ll}}\ and\ \bibinfo {author} {\bibfnamefont {E.}~\bibnamefont
			{Barkai}},\ }\href@noop {} {\bibfield  {journal} {\bibinfo  {journal} {Phys.
				Rev. E}\ }\textbf {\bibinfo {volume} {102}},\ \bibinfo {pages} {042141}
		(\bibinfo {year} {2020})}\BibitemShut {NoStop}%
	\bibitem [{\citenamefont {Ethier}\ and\ \citenamefont {Kurtz}(1986)}]{Eth}%
	\BibitemOpen
	\bibfield  {author} {\bibinfo {author} {\bibfnamefont {S.~N.}\ \bibnamefont
			{Ethier}}\ and\ \bibinfo {author} {\bibfnamefont {T.~G.}\ \bibnamefont
			{Kurtz}},\ }\href@noop {} {\emph {\bibinfo {title} {Markov processes:
				characterization and convergence}}}\ (\bibinfo  {publisher} {Wiley},\
	\bibinfo {address} {New York},\ \bibinfo {year} {1986})\BibitemShut {NoStop}%
	\bibitem [{\citenamefont {Feller}(1971)}]{Fell-II}%
	\BibitemOpen
	\bibfield  {author} {\bibinfo {author} {\bibfnamefont {W.}~\bibnamefont
			{Feller}},\ }\href@noop {} {\emph {\bibinfo {title} {An introduction to
				probability theory and its applications}}},\ Vol.~\bibinfo {volume} {2}\
	(\bibinfo  {publisher} {Wiley},\ \bibinfo {address} {New York},\ \bibinfo
	{year} {1971})\BibitemShut {NoStop}%
	\bibitem [{\citenamefont {Olver}(2014)}]{Olv}%
	\BibitemOpen
	\bibfield  {author} {\bibinfo {author} {\bibfnamefont {P.~J.}\ \bibnamefont
			{Olver}},\ }\href@noop {} {\emph {\bibinfo {title} {Introduction to partial
				differential equations}}}\ (\bibinfo  {publisher} {Springer science \&
		business media},\ \bibinfo {address} {Berlin},\ \bibinfo {year}
	{2014})\BibitemShut {NoStop}%
	\bibitem [{\citenamefont {Hoff}(2009)}]{Hof}%
	\BibitemOpen
	\bibfield  {author} {\bibinfo {author} {\bibfnamefont {P.~D.}\ \bibnamefont
			{Hoff}},\ }\href@noop {} {\emph {\bibinfo {title} {A first course in Bayesian
				statistical methods}}}\ (\bibinfo  {publisher} {Springer-Verlag},\ \bibinfo
	{address} {New York},\ \bibinfo {year} {2009})\BibitemShut {NoStop}%
	\bibitem [{\citenamefont {Klinger}\ \emph {et~al.}(2022)\citenamefont
		{Klinger}, \citenamefont {Barbier-Chebbah}, \citenamefont {Voiturez},\ and\
		\citenamefont {B\'{e}nichou}}]{KliBarVoi-2022}%
	\BibitemOpen
	\bibfield  {author} {\bibinfo {author} {\bibfnamefont {J.}~\bibnamefont
			{Klinger}}, \bibinfo {author} {\bibfnamefont {A.}~\bibnamefont
			{Barbier-Chebbah}}, \bibinfo {author} {\bibfnamefont {R.}~\bibnamefont
			{Voiturez}},\ and\ \bibinfo {author} {\bibfnamefont {O.}~\bibnamefont
			{B\'{e}nichou}},\ }\href@noop {} {\bibfield  {journal} {\bibinfo  {journal}
			{Phys. Rev. E}\ }\textbf {\bibinfo {volume} {105}},\ \bibinfo {pages}
		{034116} (\bibinfo {year} {2022})}\BibitemShut {NoStop}%
	\bibitem [{\citenamefont {Sauer}(2012)}]{Sauer2012}%
	\BibitemOpen
	\bibfield  {author} {\bibinfo {author} {\bibfnamefont {T.}~\bibnamefont
			{Sauer}},\ }in\ \href@noop {} {\emph {\bibinfo {booktitle} {Handbook of
				Computational Finance}}},\ \bibinfo {editor} {edited by\ \bibinfo {editor}
		{\bibfnamefont {J.~C.}\ \bibnamefont {Duan}}, \bibinfo {editor}
		{\bibfnamefont {W.}~\bibnamefont {H\"{a}rdle}},\ and\ \bibinfo {editor}
		{\bibfnamefont {J.}~\bibnamefont {Gentle}}}\ (\bibinfo  {publisher} {Springer
		Berlin},\ \bibinfo {address} {Heidelberg},\ \bibinfo {year} {2012})\ pp.\
	\bibinfo {pages} {529--550}\BibitemShut {NoStop}%
	\bibitem [{\citenamefont {Kloeden}\ and\ \citenamefont
		{Platen}(1992)}]{Klo-Pla}%
	\BibitemOpen
	\bibfield  {author} {\bibinfo {author} {\bibfnamefont {P.~E.}\ \bibnamefont
			{Kloeden}}\ and\ \bibinfo {author} {\bibfnamefont {E.}~\bibnamefont
			{Platen}},\ }\href@noop {} {\emph {\bibinfo {title} {Numerical Solution of
				Stochastic Differential Equations}}}\ (\bibinfo  {publisher} {Springer
		Berlin},\ \bibinfo {address} {Heidelberg},\ \bibinfo {year}
	{1992})\BibitemShut {NoStop}%
\end{thebibliography}
\end{document}